\documentclass[12pt]{article}

\usepackage[dvipsnames]{xcolor}
\usepackage{jheppub, amsmath,amssymb,amsfonts,amsxtra, mathrsfs, makeidx,graphics,graphicx,amsthm,epsfig, bm,longtable,float, color,mathtools,xfrac,footnote,rotating, lscape, makecell, environ,mathtools, empheq, ytableau}
\usepackage{cancel}
\usepackage{tikz}

\DeclareMathOperator\Gh{\Gamma_h}
\def\Gh[#1] {
	\Gamma_h \left(#1\right)
	}
\DeclareMathOperator\Ge{\Gamma_e}
\def\Ge[#1] {
	\Gamma_e \left(#1\right)
	}

\DeclareMathOperator\At{\tilde{A}}
\DeclareMathOperator\Bt{\tilde{B}}

\newcommand\Ccancel[2][black]{
    \let\OldcancelColor\CancelColor
    \renewcommand\CancelColor{\color{#1}}
    \cancel{#2}
    \renewcommand\CancelColor{\OldcancelColor}
}

\title{
\begin{center}
A - BCD dualities
\end{center}
}

\author[a]{Antonio Amariti,}	
\author[b]{Fabio Mantegazza}
\author[c,d]{Simone Rota}
 \author[a,e]{Andrea Zanetti}	

\affiliation[a]{INFN, Sezione di Milano, Via Celoria 16, I-20133 Milano, Italy}
\affiliation[b]{Deutsches Elektronen-Synchrotron DESY, Notkestr. 85, 22607 Hamburg, Germany}
\affiliation[c]{SISSA, Via Bonomea 265, 34136 Trieste, Italy}
\affiliation[d]{INFN, Sezione di Trieste, Via Valerio 2, 34127 Trieste, Italy}
\affiliation[e]{Dipartimento di Fisica, Università degli studi di Milano, Via Celoria 16, I-20133}

\emailAdd{antonio.amariti@mi.infn.it}
\emailAdd{fabio.mantegazza@desy.de}
\emailAdd{srota@sissa.it}
\emailAdd{andrea.zanetti@mi.infn.it}

\preprint{DESY-25-143}

\abstract{In this paper we propose 4d and 3d dualities among special unitary gauge theories with fundamentals and antisymmetric flavors and symplectic or orthogonal gauge theories with fundamentals and two index tensor matter.
The various dualities originate from a conjectured 4d self-duality for $SU(N)$ with an antisymmetric and four fundamental flavors. While we  provide a proof of such self duality for $SU(4)$,  we focus on  baryonic deformations for the cases at higher ranks. The deformations give rise to RG flows, deforming the self duality into new types of  dualities, involving  $SU(N)$ and $USp(2M)$ gauge theories, where the precise value of $M$ depends on the baryonic deformation. 
We provide strong checks on the validity of these dualities, by proving the integral identities among  their superconformal index.
 By dimensional reduction on a circle, real mass flows and other deformations we then find  a rich set of new dualities in 3d.
These dualities are first conjectured  from localization, by the application of the duplication formula for the one loop determinants of the matter fields, and then they are  proved by using the tensor deconfinement technique. 
}

\begin{document}
\maketitle
\flushbottom
\allowdisplaybreaks 


%
%
%
%
%
\section{Introduction}
\label{sec:intro}
%
%
%
%
%

In this paper we focus on 4d $\mathcal{N}=1$ $\mathrm{SU}(N)$ with an antisymmetric  and four fundamental flavors.
When $N=2n$ this model is supposed to have a $\mathrm{D}_6 \times \mathrm{U}(1)^2$ global symmetry enhancement by opportunely flipping some chiral ring operators \cite{Razamat:2017wsk,Razamat:2018gbu}. 
This symmetry enhancement follows from the self-duality proposed in \cite{Csaki:1997cu,Spiridonov:2009za}.
The situation is very similar to the case of the $\mathrm{E}_7 \times \mathrm{U}(1)$ enhancement proven for $\mathrm{USp}(2n)$ with an antisymmetric and eight fundamentals \cite{Razamat:2017hda} (see \cite{Dimofte:2012pd} for the $n=1$ case).

However, differently from the case of $\mathrm{E}_7 $, where a strong argument corroborating the self duality of \cite{Csaki:1996eu} follows from the matching of the superconformal index \cite{rains2005transformationselliptichypergometricintegrals,Spiridonov:2008zr}, the situation for the $\mathrm{D}_6$ case is different. Indeed in this case these is no proof of the self duality neither from the index nor from other field theoretical arguments as tensor deconfinement, while in the  $\mathrm{E}_7 $ case such a proof was obtained in \cite{Bajeot:2022lah} by using the technique of sequential deconfinement pioneered in \cite{Pasquetti:2019uop}.
More broadly we refer the reader to \cite{Pasquetti:2019uop,Benvenuti:2020wpc,Etxebarria:2021lmq,Benvenuti:2021nwt,Bottini:2022vpy,Bajeot:2022lah,Bajeot:2022wmu,Amariti:2022wae,Amariti:2023wts,Amariti:2024sde,Jiang:2024ifv,Amariti:2024gco,Benvenuti:2024glr,Hwang:2024hhy,Amariti:2025jvi,Amariti:2025lem,Jia:2025koz} for recent applications of tensor deconfinement to prove IR dualities in various dimensions, elaborating on the seminal results of \cite{Berkooz:1995km,Luty:1996cg,Pouliot:1995me}.

Motivated but this last open question, here we start our analysis by providing such a proof of the self duality for the case of $\mathrm{SU}(4)$, where we show, through a rather involved series of deconfinements and dualities,
how to map all the self-dual phases one with each other. 

In the second part of the paper we consider the former model at generic $N$ and vanishing superpotential, and then we turn on various baryonic dangerously irrelevant superpotentials \footnote{See \cite{Kutasov:1995ss} for an extended discussion on such operators.}.
Deformations of this type have dramatic consequences in the IR dynamics of the model, on the chiral ring and on the vacuum structure. Furthermore the former self-duality 
(and the relative global symmetry enhancement) is generically broken by these types of deformations and new types of dualities emerge. Indeed, the baryonic dangerously irrelevant operator on one side breaks the multiple duality, keeping only a reduced amount of dual phases, while on the other side the RG flow triggered in the surviving dual phases is generically accompanied by an Higgsing of the dual gauge group. We will see that such an Higgsing will break the special unitary dual phase to a symplectic one, giving rise to a duality between $\mathrm{SU}(N)$ and $\mathrm{USp}(2M)$ gauge theories, where the value of $M$ depends  on the baryonic deformation. In any case the symplectic gauge theory is a flipped version of $\mathrm{USp}(2M)$ with an antisymmetric and eight fundamentals.

These dualities survive also upon circle compactification, where effective monopole superpotentials are generated, following the prescription of \cite{Aharony:2013dha}. It is also possible to remove the effective superpotentials through real mass deformations, obtaining “pure" 3d $\mathrm{SU}/\mathrm{USp}$ dualities. 

In the case of $\mathrm{SU}(2n)$ we find a pure 3d  duality that, upon a second  real mass flow, gives rise to a confining duality, previously discussed in the literature \cite{Nii:2019ebv}. 
The electric side of this duality corresponds to $\mathrm{SU}(2n)$ with an antisymmetric flavor and four fundamentals and it was claimed to 
not have a 4d parent \cite{Amariti:2024gco}. This is because there is no 4d confining gauge theory that gives rise to such 3d confining duality. Here we have shown that the 4d parent of this duality is indeed the $\mathrm{SU}/\mathrm{USp}$ duality obtained by a dangerously irrelevant baryonic deformation.

The effective duality can be also manipulated at the level of the squashed three sphere partition function by freezing some mass parameters and then applying the duplication formula for the hyperbolic Gamma functions. This operation has been already used in the literature \cite{Dolan:2008qi,Spiridonov:2010qv,Benini:2011mf,Amariti:2022wae,
Amariti:2024gco}  and it  “transforms" the one loop determinants of an antisymmetric or of a conjugate antisymmetric into the one loop determinant of a symmetric or a conjugate symmetric. On the other hand, the dual gauge group becomes an orthogonal one, of even or odd rank. The constraints on the mass parameters are modified as well. These new constraints can be interpreted at field theory level as new dangerously irrelevant baryonic deformations that trigger the new $\mathrm{SU}/\mathrm{SO}$ dualities in presence of linear monopole superpotentials. Also in this case we can trigger real mass flow, removing the monopole superpotential and recovering confining dualities already proposed in \cite{Amariti:2024gco}.

This paper is organized as follows. 
In Section \ref{Sec:proofself} we derive the self-duality for $4d$ $\mathrm{SU}(4)$ with an antisymmetric flavor and 4 fundamental flavors from Seiberg-like dualities. 
In Section \ref{sec:4d} we study baryonic-like deformations for $\mathrm{SU}(N)$ gauge group. We discover various dualities between the deformed theories and $\mathrm{USp}$ gauge theories which are supported by matching the index and via deconfinement sequences.
In Section \ref{sec:3d} we perform the circle reduction of these dualities to $3d$.
In Section \ref{sec:duplication} we discuss $3d$ dualities involving symmetric tensor matter or orthogonal gauge groups. 
At the level of the $S^3$ partition function these 
 dualities can be argued for by the duplication formula and we further provide an independent derivation via deconfinement techniques.
In Section \ref{sec:conc} we summarize our results and discuss various future directions.

%
%
%
%
%
\section{Proving the self-duality for $\mathrm{SU}(4)$}
\label{Sec:proofself}
%
%
%
%
%

In this section we provide a derivation of the self-duality discussed in \cite{Csaki:1997cu} for a 4d $\mathcal{N}=1$
$\mathrm{SU}(4)$ gauge theory with four pairs of fundamentals $Q$ and antifundamentals $\tilde Q$ and two antisymmetric 
tensors $A_{1,2}$. The antisymmetric representation is self-conjugate and in this case we can collect the two antisymmetrics into a single one, denoted as $A$, in the fundamental representation of a $\mathrm{U}(2)$ global symmetry.

The derivation consists of showing that the self duality under investigation is consequence of other elementary dualities,
essentially Seiberg \cite{Seiberg:1994pq} and Intriligator-Pouliot \cite{Intriligator:1995ne} duality.

The self-duality found in \cite{Csaki:1997cu} distinguishes three cases: there are three self-dual $\mathrm{SU}(4)$ gauge theories, with the same charged matter content and extra baryonic or mesonic singlets.
The self-dualities involve two mesons, $M_0 = Q \tilde Q$ and $M_2 = Q A^2 \tilde Q$ and/or two baryons
$B = A Q^2$ and $\tilde B = A \tilde Q^2$.
The three possibilities are distinguished by the superpotentials
\begin{itemize}
\item $W_B = M_0 q a^2 \tilde q + M_2 q \tilde q + B q^2 a + \tilde B \tilde q^2 a$,
\item $W_C = M_0 q a^2 \tilde q + M_2 q \tilde q $,
\item $W_D =B q^2 a + \tilde B \tilde q^2 a$,
\end{itemize}
where $a$ is the dual antisymmetric and $q$ and $\tilde q$ are the dual fundamentals.

\subsection{Derivation of $W_D$}

\begin{figure}
\begin{center}
  \includegraphics[width=15.5cm]{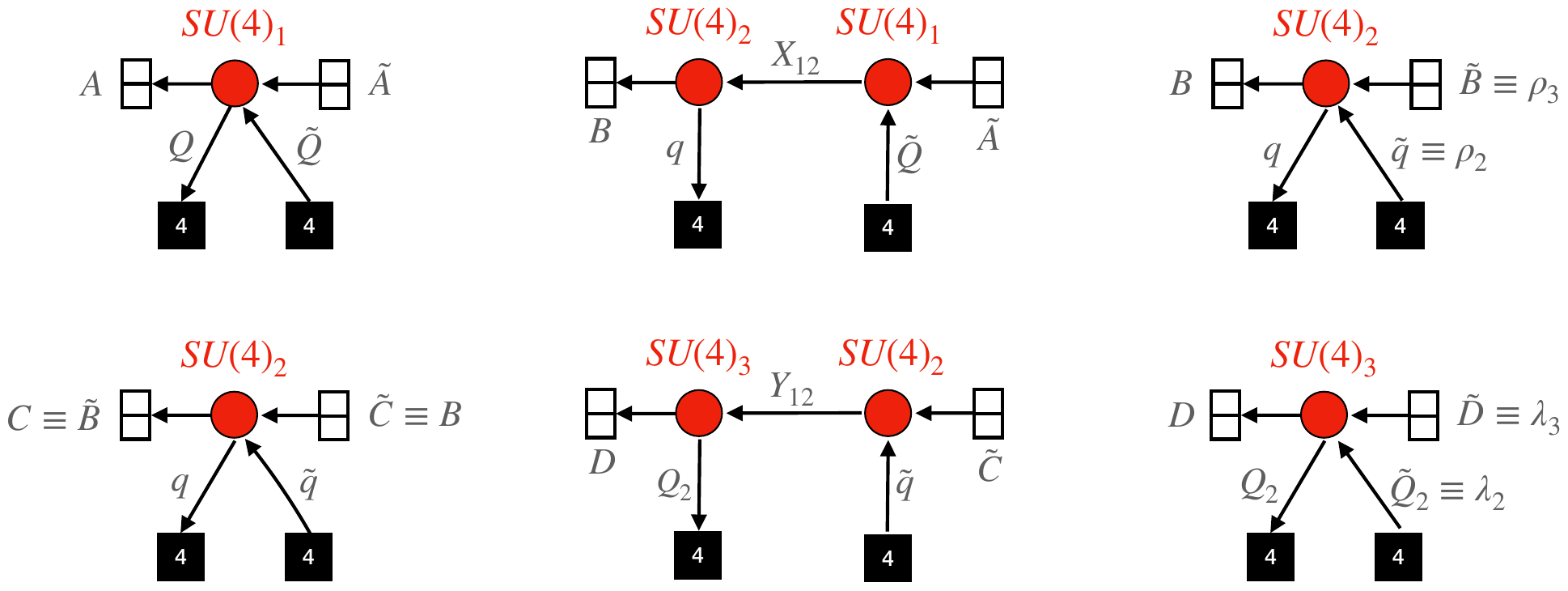}
  \end{center}
  \caption{In this figure we have plot the various steps of tensor deconfinements and ordinary dualities used to  derive the $\mathrm{SU}(4)$ (self-)dual model with superpotential $W_D$.}
    \label{figD}
\end{figure}

In this case we start considering the first quiver in Figure \ref{figD}, by distinguishing the conjugation of the two antisymmetric 
tensors, even if such distinction is immanent. Then we deconfine the antisymmetric using the confining duality for $\mathrm{SU}(4)$
with an antisymmetric, four fundamentals and four antifundamentals originally studied in \cite{Pouliot:1995me}. We obtain the second quiver in Figure \ref{figD}.
In order to understand the superpotential let us first describe the confinement of this second quiver leading to the first one in Figure \ref{figD}. The $\mathrm{SU}(4)_2$ gauge invariant operators are
$\varphi_1 = \mathrm{Pf} B$, $ \varphi_2 = q X_{12}$, $\varphi_3 = B X_{12}^2$, $ \varphi_4 = B q^2$, $\varphi_5 = q^4$ and $\varphi_6 = X_{12} ^4$.
In the case of vanishing superpotential for the second quiver, the original quiver would have superpotential
\begin{equation}
W=\varphi_1 \varphi_2^4 +\varphi_2^2  \varphi_3 \varphi_4 +\varphi_3^2  \varphi_5+ \varphi_1 \varphi_5 \varphi_6+\varphi_4^2  \varphi_6.
\end{equation}
On the other hand we want to have $W=0$ for the superpotential of the original theory, and we want to keep massless the fields $\varphi_2$ and $\varphi_3$, corresponding to the massless fields $Q$ and $A$ of the original model.
These two requirement are satisfied by the superpotential 
\begin{equation}
\label{flippato}
W=\alpha_1 \mathrm{Pf} B + \alpha_4  B q^2 + \alpha_5q^4 + \alpha_6 X_{12} ^4,
\end{equation}
in the deconfined quiver.
Then, we confine the $\mathrm{SU}(4)_1$ gauge nodes defining the singlets
$\rho_1 = \mathrm{Pf} \tilde A$, $ \rho_2 = \tilde Q X_{12} $, $ \rho_3 =\tilde A X_{12}^2 $, $ \rho_4 = \tilde A \tilde Q^2 $, 
$ \rho_5=\tilde Q^4 $ and $ \rho_6=X_{12} ^4$.
We obtain the third quiver in Figure \ref{figD} with superpotential
\begin{equation}
W=\rho_1 \tilde q^4 +\tilde q^2  \tilde B \rho_4 + \mathrm{Pf} \tilde B  \rho_5+ \rho_1 \rho_5 \rho_6+\rho_4^2  \rho_6
+\alpha_1 \mathrm{Pf} B + \alpha_4  B q^2 + \alpha_5q^4 + \alpha_6 \rho_6.
\end{equation}

This phase can be rearranged in a more symmetric way by integrating out the massive fields $\alpha_6 $ and $\rho_6$
and by redefining the singlets as
$\alpha_1 = \gamma$, 
$ \alpha_4= \beta$,
$\alpha_5= \eta$,
$\rho_1= \tilde \eta$,
$\rho_4= \tilde \beta$ and
$\rho_5 = \tilde \gamma$.
We also rename the two antisymmetric tensor $B$ and $\tilde B$ using $\tilde C$ and $C$ respectively, where we explicitly consider the conjugated representations. This trick is useful in the following deconfinements.
Summarizing, the quiver at this stage is represented by the fourth one in Figure \ref{figD}.
The superpotential for this phase is 
\begin{equation}
\label{almost}
W=  \eta  q^4 +\tilde \eta \tilde q^4 +\beta  q^2   \tilde C + \tilde \beta \tilde q^2  C +  \gamma \mathrm{Pf}  \tilde C+ \tilde \gamma \mathrm{Pf} C.
\end{equation}
We then proceed as above, by deconfining the antisymmetric $C$ using a $\mathrm{SU}(4)_3$ gauge node. The superpotential for the deconfined phase is 
\begin{equation}
W=  \eta  ( Q_2 Y_{12})^4 +\tilde \eta \tilde q^4 +\beta  ( Q_2 Y_{12})^2   \tilde C 
+ \tilde \beta \tilde q^2  D Y_{12}^2 +  \gamma \mathrm{Pf}  \tilde C+ \tilde \gamma (D Y_{12}^2)^2
+ \psi_4  D Q_2^2 +  \psi_6 Y_{12} ^4,
\end{equation}
where the last two terms include the flippers $\psi_{4,6}$  (the other two flippers appearing in (\ref{flippato}) do not appear here due to the presence of the interactions  $  \eta  q^4 $ and $\tilde \gamma \mathrm{Pf} C$ in (\ref{almost})).
Then we confine $\mathrm{SU}(4)_2$ an define the gauge invariant combinations
$\lambda_1 = \mathrm{Pf} \tilde C$, 
$\lambda_2 = \tilde q Y_{12}$, 
$\lambda_3 =\tilde C Y_{12}^2$, 
$\lambda_4 = \tilde C \tilde q^2$, 
$\lambda_5=\tilde q^4$ and 
$\lambda_6=Y_{12} ^4$.
We obtain the last quiver in Figure \ref{figD} with superpotential
\begin{eqnarray}
W&=&\lambda_1 \tilde Q_2^4 +\tilde Q_2^2  \tilde D \lambda_4 +\mathrm{Pf} \tilde D  \lambda_5+ \lambda_1 \lambda_5 \lambda_6+\lambda_4^2  \lambda_6 + \eta   Q_2^4 \lambda_6 
\nonumber \\
&+&
\tilde \eta \lambda_5 +\beta   Q_2^2 \tilde D
+ \tilde \beta  D \tilde Q_2^2 +  \gamma \lambda_1+ \tilde \gamma D^2 \lambda_6+\psi_4  D Q_2^2 + \psi_6 \lambda_6.
\end{eqnarray}
Integrating out the massive fields and defining the $\mathrm{SU}(2)$ doublets
$\mathbf{A} = \{D,\tilde D\}$,
$\mathbf{B} = \{\psi_4,\beta \}$ and
$\tilde{\mathbf{B}}= \{\tilde \beta,\lambda_4 \}$
we arrive to the final superpotential 
$W =
\mathbf{B} \mathbf{A} Q_2^2+  \tilde{\mathbf{B}} \mathbf{A} \tilde Q_2^2
$
that corresponds to $W_D$.

\subsection{Derivation of $W_B$}

\begin{figure}
\begin{center}
  \includegraphics[width=10cm]{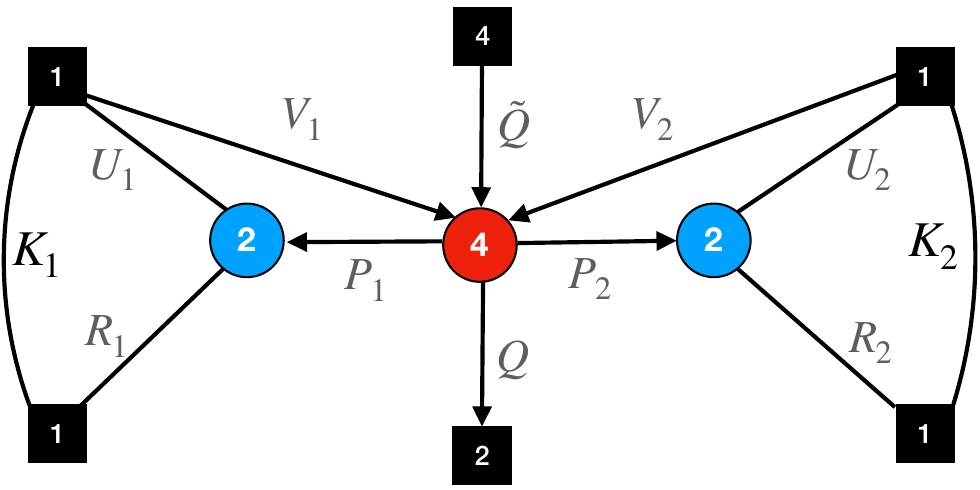}
  \end{center}
  \caption{Quiver obtained by deconfining the two antisymmetric of $\mathrm{SU}(4)$ (red node) in terms of two $\mathrm{SU}(2)$ gauge groups (blue nodes).}
    \label{figB}
\end{figure}

In this case we start by deconfining the two antisymmetric using two symplectic gauge groups.
We consider two antisymmetric tensors with the same conjugation and consistently we break 
one of the $\mathrm{SU}(4)$ flavor global symmetries to $\mathrm{SU}(2) \times \mathrm{U}(1)^2$. This breaking is 
visible in the quiver of Figure \ref{figB}, where the deconfined phase is represented.
The superpotential for this phase is 
\begin{equation}
W=U_1K_1R_1 + U_2 K_2 R_2 +U_1  V_1 P_1 +U_2  V_2 P_2.
\end{equation}
Then, we dualize the $\mathrm{SU}(4)$ gauge node, obtaining a $\mathrm{SU}(2)$ gauge group with superpotential
\begin{eqnarray}
W&=&
 \Phi_1 v_1 \tilde q+
 \Phi_2  v_2 \tilde q+
 \Phi_3 v_1 p_2+
\Phi_4 v_2 p_1+
\Phi_5  q p_1+
 \Phi_6 q p_2+
\Phi_7 q \tilde q
\nonumber \\
&+&
\Phi_8 v_2 p_2+
\Phi_9  v_1 p_1+
U_1K_1R_1 + U_2 K_2 R_2 +U_1   \Phi_9 +U_2  \Phi_8,
\end{eqnarray}
where the mesons of this phase are denoted as $\Phi_i$ and they corresponds to the $\mathrm{SU}(4)$ gauge invariant combinations
\begin{equation}
\vec \Phi = \{ Q V_1,Q V_2, P_2 V_1,P_1 V_2, P_1 \tilde Q,P_2 \tilde Q, Q \tilde Q,P_2V_2,P_1 V_1   \},
\end{equation}
while the  antifundamentals $\tilde q, p_1,p_2$ are dual to the fundamentals $Q,P_1,P_2$ and the 
fundamentals $v_1,v_2,q$ are dual to the antifundamentals $V_1,V_2,\tilde Q$.
Integrating out the massive fields we arrive at the superpotential
\begin{equation}
W=\Phi_1 v_1 \tilde q+ \Phi_2 v_2 \tilde q+\Phi_3 v_1 p_2+ \Phi_4  v_2 p_1+ \Phi_5  q p_1+\Phi_6  q p_2+ \Phi_7  q \tilde q+K_2 R_2  v_2 p_2+K_1 R_1  v_1 p_1.
\end{equation}

Then, we dualize the two $\mathrm{SU}(2)$ nodes treating them as $\mathrm{USp}(2)$ gauge nodes, sequentially.
Even if both the dual gauge groups are again $\mathrm{USp}(2)$, a rather intricate structure of singlets arises.
After the first duality the superpotential is
\begin{eqnarray}
W&=&  M_1  \tilde \Phi_5^2+
 M_2  \tilde p_1^2+
 M_3  \tilde p_1 \tilde \Phi_5+
 M_4  \tilde R_1 \tilde \Phi_5+
 M_5  \tilde \Phi_4 \tilde \Phi_5+ M_6  \tilde \Phi_4 \tilde R_1+ M_7  \tilde p_1 \tilde R_1+ M_8  \tilde p_1 \tilde \Phi_4
\nonumber\\
&+&\Phi_1 v_1 \tilde q+ \Phi_2 v_2 \tilde q+\Phi_3 v_1 p_2+ { M_8}  v_2 + q { M_3} +\Phi_6  q p_2+ \Phi_7  q \tilde q+K_2 R_2  v_2 p_2+K_1   v_1 M_7.
\nonumber\\
\end{eqnarray}
The mesons of this phase are denoted as $M_i$, and they correspond to the $\mathrm{USp}(2)$ gauge invariant combinations
\begin{equation}
\vec M = \{ \Phi_5^2,P_1^2,\Phi_5 P_1, \Phi_5R_1,\Phi_4\Phi_5,\Phi_4  R_1, P_1  R_1,P_1  \Phi_4
 \} ,
\end{equation}
while the  fundamentals $ \tilde p_1, \tilde \Phi_5,\tilde \Phi_4, \tilde R_1$ are dual to the fundamentals $P_1,  \Phi_5,\Phi_4,  R_1$.
Integrating out the massive fields we arrive at the superpotential
\begin{eqnarray}
W&=&\Phi_1 v_1 \tilde q+\Phi_3 v_1 p_2+ (\Phi_2  \tilde q+K_2 R_2  p_2) \tilde p_1 \tilde \Phi_4+ \tilde p_1 \tilde \Phi_5 ( \Phi_6   p_2+ \Phi_7   \tilde q) +K_1   v_1  M_7
\nonumber\\
&+&M_1 \tilde \Phi_5^2+
M_2 \tilde p_1^2+
 M_4  \tilde R_1 \tilde \Phi_5+
 M_5 \tilde \Phi_4 \tilde \Phi_5+ M_6 \tilde \Phi_4 \tilde R_1+ M_7  \tilde p_1 \tilde R_1.
\end{eqnarray}

After the second duality on the other $\mathrm{USp}(2)$ gauge group the superpotential becomes
\begin{eqnarray}
W&=& 
 N_1  \tilde \Phi_6^2+
 N_2  \tilde p_2^2+
 N_3  \tilde p_2 \tilde \Phi_6+
 N_4  \tilde R_2 \tilde \Phi_6+
 N_5  \tilde \Phi_3 \tilde \Phi_6+ N_6  \tilde \Phi_3 \tilde R_2+ N_7  \tilde p_2 \tilde R_2+ N_8  \tilde p_2\tilde \Phi_3
\nonumber\\
&+&
\Phi_1 v_1 \tilde q+v_1 { N_8}+ (\Phi_2  \tilde q+K_2 { N_7}) \tilde p_1 \tilde \Phi_4+ \tilde p_1 \tilde \Phi_5 ( { N_3}+ \Phi_7   \tilde q) +K_1   v_1  M_7
\nonumber\\
&+&
 M_1 \tilde \Phi_5^2+
M_2 \tilde p_1^2+
 M_4  \tilde R_1 \tilde \Phi_5+
 M_5 \tilde \Phi_4 \tilde \Phi_5+ M_6 \tilde \Phi_4 \tilde R_1+ M_7  \tilde p_1 \tilde R_1.
\end{eqnarray}
The mesons of this phase are denoted as $N_i$, and they correspond to the $\mathrm{USp}(2)$ gauge invariant combinations
\begin{equation}
\vec N= \{ \Phi_6^2,P_2^2,\Phi_6 P_2, \Phi_6R_2,\Phi_3\Phi_6,\Phi_3  R_2, P_2  R_2,P_2  \Phi_3
 \},
\end{equation}
while the  fundamentals $ \tilde p_2, \tilde \Phi_6,\tilde \Phi_3, \tilde R_2$ are dual to the fundamentals $P_2,  \Phi_6,\Phi_3,  R_2$.
Integrating out the massive fields we arrive at the superpotential
\begin{eqnarray}
W&=& N_1 \tilde \Phi_6^2+
 N_2 \tilde p_2^2+
N_4  \tilde R_2 \tilde \Phi_6+
 N_5 \tilde \Phi_3 \tilde \Phi_6+N_6 \tilde \Phi_3 \tilde R_2+ N_7 \tilde p_2 \tilde R_2
\nonumber\\
&+& M_1 \tilde \Phi_5^2+
M_2 \tilde p_1^2+
 M_4  \tilde R_1 \tilde \Phi_5+
 M_5 \tilde \Phi_4 \tilde \Phi_5+ M_6 \tilde \Phi_4 \tilde R_1+ M_7  \tilde p_1 \tilde R_1
\\
&+& N_3 (\tilde p_2 \tilde \Phi_6  +\tilde p_1 \tilde \Phi_5)  + \Phi_2  \tilde \Phi_4  \tilde q\tilde p_1 +  \tilde \Phi_5 \Phi_7   \tilde q \tilde p_1 + \Phi_1 \tilde q \tilde p_2 \tilde \Phi_3 + K_1 M_7 \tilde p_2 \tilde \Phi_3+K_2 N_7 \tilde p_1 \tilde \Phi_4. \nonumber
\end{eqnarray}
 
 At this point of the discussion we dualize the  $\mathrm{SU}(2)$ gauge node. In order to apply the rule of Seiberg duality on this node we need to specify the conjugation of the fundamental representations. 
 We choose the fields $\tilde p_1,\tilde p_2,\tilde q$ as antifundamentals and the fields $N_3,N_7,M_7$ as fundamentals.
 The dual gauge group becomes $\mathrm{SU}(4)$ and the superpotential for this phase is 
\begin{eqnarray}
\label{vomito}
W&=& M_1 \tilde \Phi_5^2+
M_2  \theta^2 \pi_2^2 +
 M_4  \tilde R_1 \tilde \Phi_5+
 M_5 \tilde \Phi_4 \tilde \Phi_5+ M_6 \tilde \Phi_4 \tilde R_1+ L_6\tilde R_1
\nonumber\\
&+& N_1 \tilde \Phi_6^2+
 N_2 \theta^2 \pi_1^2 +
N_4  \tilde R_2 \tilde \Phi_6+
 N_5 \tilde \Phi_3 \tilde \Phi_6+N_6 \tilde \Phi_3 \tilde R_2+ L_7 \tilde R_2
\nonumber\\
&+& L_9 \tilde \Phi_6  + L_8 \tilde \Phi_5 + (\Phi_2  \tilde \Phi_4  + \tilde \Phi_5 \Phi_7  ) \theta \pi_1 \pi_2^2  + \Phi_1 \tilde \Phi_3  \theta \pi_2 \pi_1^2  + K_1 L_2\tilde \Phi_3+K_2 L_3 \tilde \Phi_4 
\nonumber\\
&+& 
L_1 \tilde N_3 \theta + L_2 \tilde M_7 \pi_2 + L_3 \tilde N_7 \pi_1 + L_4 \tilde M_7 \theta
+ L_5 \tilde N_7 \theta 
\nonumber\\
&+&  L_6 \tilde M_7 \pi_1 + L_7 \tilde N_7 \pi_2 + L_8 \tilde N_3 \pi_1 + L_9 \tilde N_3 \pi_2.
\end{eqnarray}
The mesons of this phase are denoted as $L_i$, and they correspond to the $\mathrm{SU}(2)$ gauge invariant combinations
\begin{equation}
\vec N= \{ 
N_3 \tilde q , M_7 \tilde p_2 , N_7 \tilde p_1, M_7 \tilde q, N_7 \tilde q, M_7 \tilde p_1,N_7 p_2, N_3 \tilde p_1 ,N_3 \tilde p_2
 \},
\end{equation}
while the  antifundamentals $\theta , \pi_1,\pi_2$ are dual to the fundamentals $\tilde q,\tilde p_1,\tilde p_2$ and the 
fundamentals $\tilde N_3, \tilde N_7, \tilde M_7$ are dual to the antifundamentals $N_3,N_7,M_7$.
Observe that in the superpotential (\ref{vomito}) we have mapped explicitly the baryonic deformations, 
using the map
\begin{equation}
\tilde p_2^2 \rightarrow \theta^2 \pi_1^2, 
\quad
\tilde p_1^2 \rightarrow \theta^2 \pi_2^2, 
\quad
\tilde q \tilde p_1 \rightarrow \theta \pi_1 \pi_2^2, 
\quad
\tilde q \tilde p_2 \rightarrow \theta \pi_2 \pi_1^2.
\end{equation}
Integrating out the massive fields we arrive at the superpotential
\begin{eqnarray}
W &=& M_2  \theta^2 \pi_2^2 + N_2 \theta^2 \pi_1^2 +\Phi_2  \tilde \Phi_4 \theta \pi_1 \pi_2^2  + \Phi_1 \tilde \Phi_3 \theta \pi_2 \pi_1^2  + K_1 L_2\tilde \Phi_3+K_2 L_3 \tilde \Phi_4 
\nonumber\\
&+& L_1 \tilde N_3 \theta+L_2 \tilde M_7 \pi_2+L_3 \tilde N_7 \pi_1+L_4 \tilde M_7 \theta+L_5 \tilde N_7 \theta
\nonumber\\
&+& \tilde N_3 \pi_1( M_1 \tilde N_3 \pi_1+ M_4  \tilde M_7 \pi_1 + M_5 \tilde \Phi_4)
\nonumber\\
&+& \tilde N_3 \pi_2(N_1\tilde N_3 \pi_2+ N_4  \tilde N_7 \pi_2 + N_5 \tilde \Phi_3+\Phi_7   \theta \pi_2 \pi_1^2)
\nonumber\\
&+& \tilde M_7 \pi_1(M_4  \tilde N_3 \pi_1 + M_6 \tilde \Phi_4 )
+
\tilde N_7 \pi_2(N_4 \tilde N_3 \pi_2+
N_6 \tilde \Phi_3).
\end{eqnarray}
The last step consists of confining the two $\mathrm{USp}(2)$ gauge nodes. The two steps can be done simultaneously
and we obtain the superpotential 
\begin{eqnarray}
W &=&\tilde M_7 q_1 M_6 + M_1 \tilde N_3^2 A_1+ \tilde N_3 M_4  \tilde M_7 A_1+ \tilde N_3 M_5 q_1
+N_2 \theta^2 A_1+ \tilde N_7 w_1
\nonumber\\
&+&\tilde N_3  \Phi_7   \theta A_1 A_2+\Phi_2  \theta A_2 q_1 + \Phi_1 \theta  q_2 A_1+L_1 \tilde N_3 \theta+L_4 \tilde M_7 \theta+L_5 \tilde N_7 \theta
\nonumber\\
&+&\tilde N_7 q_2 
N_6 +N_1\tilde N_3^2 A_2+ \tilde N_3 N_4  \tilde N_7 A_2 + \tilde N_3  N_5 q_2+M_2  \theta^2 A_2+\tilde M_7 w_2
\nonumber\\
&+&
\mathrm{Pf} 
\left(
\begin{array}{ccc}
A_1 &w_1& q_1\\
\cdot &0& s_1\\
\cdot&\cdot&0
\end{array}
\right)+
\mathrm{Pf} 
\left(
\begin{array}{ccc}
A_2 &w_2& q_2\\
\cdot&0& s_2\\
\cdot&\cdot&0
\end{array}
\right),
\end{eqnarray}
where
\begin{eqnarray}
&&
A_1 = \pi_1^2,\quad
w_1 = \pi_1 L_3,\quad
q_1 = \pi_1 \tilde \Phi_4,\quad
s_1 = L_3 \tilde \Phi_4\nonumber \\
&&
A_2 = \pi_2^2,\quad
w_2 = \pi_2 L_2,\quad
q_2 = \pi_2 \tilde \Phi_3,\quad
s_2 = L_2 \tilde \Phi_3.
\end{eqnarray}
Integrating out the massive fields we arrive at the final superpotential
\begin{eqnarray}
W&=&
\tilde N_3 (M_5 q_1+N_5 q_2 + \theta L_1)+
A_1 A_2 (\tilde N_3  \Phi_7   \theta +  q_1 N_4 \tilde N_3  + q_2 M_4  \tilde N_3 )
+
\tilde N_3^2 (M_1 A_1+M_2 A_2)\nonumber \\
&+&
  \theta^2(N_2 A_1+M_2 A_2) + \theta  q_1 (L_5 A_1 + \Phi_2  A_2) +  \theta  q_2 (\Phi_1 A_1 + L_4 A_2)+q_1  q_2 (A_1
N_6+A_2 M_6). \nonumber \\
\end{eqnarray}
By collecting the fields in the $\mathrm{SU}(4)^2 \times \mathrm{SU}(2)$ flavor invariant combinations
\begin{eqnarray}
\{ \theta,q_1,q_2 \} &\rightarrow& \tilde Q \nonumber \\
\tilde N_3  &\rightarrow&  Q \nonumber \\
\{ M_1,M_2 \}  &\rightarrow&  \tilde{\mathbf{B}} \nonumber \\
\{ A_1,A_2 \}  &\rightarrow& \mathbf{A} \nonumber \\
\{ \{ N_2,L_5,\Phi_1,N_6\} ,\{M_2,\Phi_2,L_4,M_6 \} \}  &\rightarrow&  \mathbf{B} \nonumber \\
\{ L_1,M_5,N_5 \}  &\rightarrow& \mathbf{M}_2 \nonumber \\
\{ \Phi_7,M_4,N_4 \}  &\rightarrow& \mathbf{M}_0,
\end{eqnarray}
we arrive at the final form of the superpotential 
\begin{equation}
W =  \tilde{\mathbf{B}} Q^2 \mathbf{A} +\mathbf{B} \tilde Q^2 \mathbf{A}+ \mathbf{M}_2 Q \tilde Q + \mathbf{M}_0 Q A_1 A_2 \tilde Q,
\end{equation}
that corresponds to $W_B$.
We conclude observing that the derivation $W_B$ and $W_D$ concludes the derivation of the self duality, because $W_C$ can be obtained by combining the two derivations above.

%
%
%
%
%
\section{4d dualities}
\label{sec:4d}
%
%
%
%
%
In this section we focus on 4d $\mathcal{N}=1$ $\mathrm{SU}(N)$ gauge theories with an antisymmetric and four fundamental flavors. In the following we denote the antisymmetric as $A$, its conjugate as $\tilde A$, the fundamentals are denoted as $Q$ and the antifundamental as $\tilde Q$.
We have added to such models a dangerously irrelevant superpotential, for $N=2n$ it is proportional  to 
$A^{n-k} Q^{2k}$ with $k=0,1,2$ and for $N=2n+1$ it is proportional  to $A^{n-k} Q^{2k+1}$ with $k=0,1$.
In each of these five cases we have found that the deformed  model is dual to a $\mathrm{USp}(2m)$ gauge theory with an antisymmetric, eight fundamentals, in addition to some flippers depending on  the electric deformation.
Schematically we find the following dualities, modulo singlets discussed below:
\begin{equation}
\begin{tikzpicture}
\node at (-1,0) (SU) {$\begin{array}{c}\mathrm{SU}(2n) \\ \text{ with } A,\widetilde{A} \text{ and } 4 \square, 4 \overline{\square}\end{array}$};
\node at (9,2) (W4) {$\mathrm{USp}(2n)$ with $X$ and 8 fund};
\node at (9,0) (W2) {$\mathrm{USp}(2n-2)$ with $X$ and 8 fund};
\node at (9,-2) (W0) {$\mathrm{USp}(2n-4)$ with $X$ and 8 fund};
\draw[<->] (SU) -- node[midway,above,yshift=7pt] {$\delta\mathcal{W} = \widetilde{A}^{n-2} \widetilde{Q}^4$} (W4.west);
\draw[<->] (SU) -- node[midway,above] {$\delta\mathcal{W} = \widetilde{A}^{n-1} \widetilde{Q}^2$} (W2.west);
\draw[<->] (SU) -- node[midway,below,yshift=-7pt] {$\delta\mathcal{W} = \mathrm{Pf}(\widetilde{A})$} (W0.west);
\end{tikzpicture}
\end{equation}
and:
\begin{equation}
\begin{tikzpicture}
\node at (-1,0) (SU) {$\begin{array}{c}\mathrm{SU}(2n+1) \\ \text{ with } A,\widetilde{A} \text{ and } 4 \square, 4 \overline{\square}\end{array}$};
\node at (9,1) (W3) {$\mathrm{USp}(2n)$ with $X$ and 8 fund};
\node at (9,-1) (W1) {$\mathrm{USp}(2n-2)$ with $X$ and 8 fund};
\draw[<->] (SU) -- node[midway,above,yshift=7pt] {$\delta\mathcal{W} = \widetilde{A}^{n-1} \widetilde{Q}^3$} (W3.west);
\draw[<->] (SU) -- node[midway,below,yshift=-5pt] {$\delta\mathcal{W} = \widetilde{A}^{n} \widetilde{Q}$} (W1.west);
\end{tikzpicture}
\end{equation}
where $X$ is the antisymmetric of $\mathrm{USp}(2m)$.
In this section we are planning to study each single case in detail, showing how to obtain the dual description using tensor deconfinement and Higgsing the dual gauge group when necessary. We will provide in this way the matching of the superconformal indices and in addition we study the existence of unitary dualities in the conformal window.

\subsection{$\mathrm{SU}(2n)$}
\label{subsec2n}

Here we consider the case $N=2n$. There are three possible superpotential deformations. The first deformation is 
\begin{equation}
\label{Wdef1SU2n}
W = \tilde{A}^{n-2} \tilde Q^4,
\end{equation}
 the second deformation is 
\begin{equation}
\label{Wdef2SU2n}
W = \tilde{A}^{n-1} \tilde Q_3 \tilde Q_4,
\end{equation}
where the $\mathrm{SU}(4)$ flavor symmetry is explicitly broken by the deformation. 
The  third superpotential deformation is
\begin{equation}
\label{Wdef3SU2n}
W = \mathrm{Pf} \tilde A.
\end{equation}
In the following we will study the effect of each of these deformations in the IR behavior of the model.
Before distinguishing the three cases we can keep a common analysis by deconfining the antisymmetric in terms of another auxiliary $\mathrm{SU}(2n)$ gauge group, with an antisymmetric.

\begin{figure}
\begin{center}
  \includegraphics[width=15.5cm]{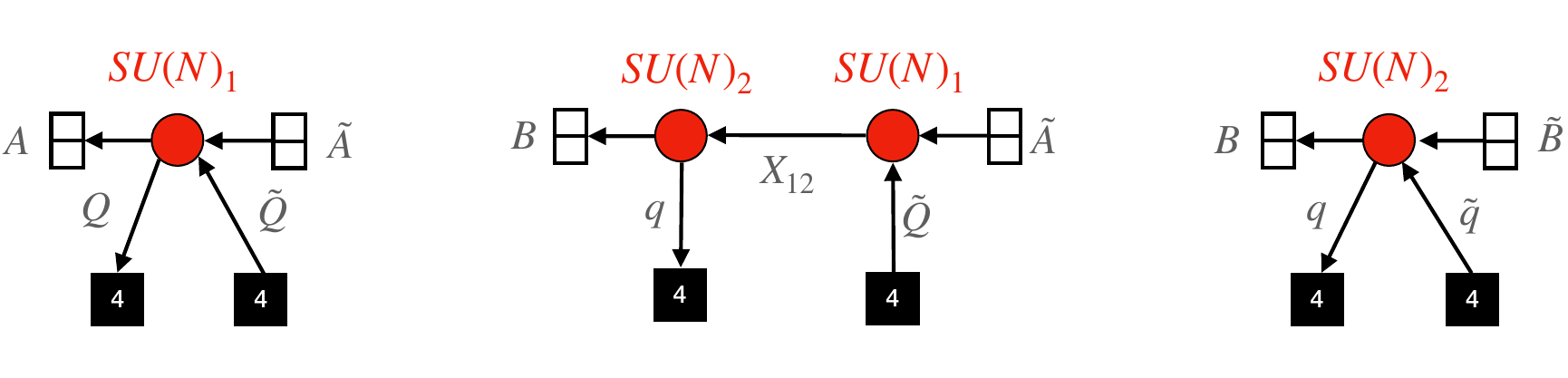}
  \end{center}
  \caption{First deconfinement sequence for $\mathrm{SU}(N)_1$ gauge theory with 4 fundamental flavors, 1 antisymmetric flavor and vanishing superpotential. The case $N=2n$ has been studied in subsection \ref{subsec2n} while the case of $N=2n+1$ has been studied in subsection \ref{subsec2np1}.}
    \label{figdec1su}
\end{figure}

Here we deconfine the antisymmetric $ A$ and the fundamentals $Q$, by trading them with an $\mathrm{SU}(2n)_2$ gauge node, 
with a new antisymmetric $B$, an $\mathrm{SU}(2n)_1 \times \mathrm{SU}(2n)_2$  bifundamental $X_{12}$ and four $\mathrm{SU}(2n)_2$ fundamentals 
$q$. The charged field content of this deconfined phase is depicted in the second quiver in Figure \ref{figdec1su} with $N=2n$.
The original fields $A$ and $Q$ are mapped to the combinations $B X_{12}^2$ and $q X_{12}$ respectively.
Starting with vanishing superpotential there are also new singlets $\alpha_{1,2,3,4}$ in the dual phase, interacting with the charged fields through a superpotential 
\begin{equation}
\label{decflipp1}
W=\alpha_1 \mathrm{Pf} B + \alpha_2  B^{n-1} q^2 + \alpha_3 B^{n-2}q^4 + \alpha_4 X_{12} ^{2n}.
\end{equation}
At this level we did not turn on any superpotential deformation, because it can be done later, such that the discussion here will apply also in the analysis below, where the deformations (\ref{Wdef1SU2n}), (\ref{Wdef2SU2n}) and (\ref{Wdef3SU2n}) will be separately considered.

Then we observe that the original $\mathrm{SU}(2n)_1$ gauge node is s-confining, and the confined degrees of freedom correspond to 
four  $\mathrm{SU}(2n)_2$ antifundamentals
$\tilde q = \tilde Q X_{12}$, an $\mathrm{SU}(2n)_2$ conjugate antisymmetric 
$\tilde B =\tilde A X_{12}^2$ and the $\mathrm{SU}(2n)_2$  singlets 
$\rho_1 = \mathrm{Pf} \tilde A$, $\rho_2 = \tilde A^{n-1} \tilde Q^2$, $\rho_3=\tilde A^{n-2} \tilde Q^4$ and 
$\rho_4=X_{12} ^{2n}$.

The charged field content of the  $\mathrm{SU}(2n)_2$ theory is represented in the third quiver in Figure \ref{figdec1su} and 
the superpotential is
\begin{equation}
\label{thirdspot}
W=\rho_1 \tilde q^4 \tilde B^{n-2}+\tilde q^2  \tilde B^{n-1} \rho_2 +\rho_3 \tilde B^n  + \rho_1 \rho_3 \rho_4+\rho_2^2  \rho_4+\alpha_1 B^n + \alpha_2  B^{n-1} q^2 + \alpha_3 B^{n-2} q^4 + \alpha_4 \rho_4,
\end{equation}
where $\rho_2$ and $\alpha_2$ are in the conjugate antisymmetric representation of $\mathrm{SU}(4)_L$ and $\mathrm{SU}(4)_R$, respectively. The fields $\rho_i$ and $\alpha_i$ therefore amount to 18 total singlet fields.

At this point of the discussion we can introduce the electric  deformation given by the superpotential (\ref{Wdef1SU2n}), (\ref{Wdef2SU2n}) and (\ref{Wdef3SU2n}) respectively.
The effects of such deformations are summarized below

\begin{enumerate}
\item The superpotential deformation \eqref{Wdef1SU2n}
gives rise to the linear term $\rho_3$ in \eqref{thirdspot}. The superpotential, after integrating out the massive fields becomes
\begin{equation}
\label{thirdspotrho3}
W=\rho_1 \tilde q^4 \tilde B^{n-2}+\tilde q^2  \tilde B^{n-1} \rho_2 +\rho_3 \tilde B^n  +\alpha_1 B^n + \alpha_2  B^{n-1} q^2 + \alpha_3 B^{n-2} q^4+\rho_3.
\end{equation}
The non-trivial $F$-term for the field $\rho_3$ gives
\begin{equation}
F_{\rho_3}  =\tilde B^n +1 = 0,
\end{equation}
where the equation is solved if $\tilde B$ acquire a non-zero vev, breaking $\mathrm{SU}(2n)$ to $\mathrm{USp}(2n)$.
\item The deformation (\ref{Wdef2SU2n}) breaks the $\mathrm{SU}(4)_L$  flavor symmetry, and it corresponds to breaking the $\mathrm{SU}(4)_L$ antisymmetric  $\rho_2$
into two singlets $\Gamma \equiv \rho_{2}^{(12)}$ and $\Omega \equiv \rho_{2}^{(34)}$ and an $\mathrm{SU}(2)^2$ bifundamental $\Psi_{ab} \equiv \rho_{2}^{(a,b+2)}$ with $a,b=1,2$.
 Analogously the antifundamentals $\tilde q$ are split into two antifundamentals $\tilde u_{1,2} \equiv \tilde q_{1,2}$ and two antifundamentals $\tilde v_{1,2} \equiv \tilde q_{3,4}$.
 The dual superpotential becomes
\begin{eqnarray}
\label{splitspotthird}
W&=&\rho_1 \tilde u^2 \tilde v^2 \tilde B^{n-2}+
\tilde u^2  \tilde B^{n-1} \Gamma+\tilde v^2  \tilde B^{n-1} \Omega + \tilde u \tilde v \tilde B^{n-1} \Psi
+ \rho_3 \tilde B^n  + \rho_1 \rho_3 \rho_4 \nonumber \\
&+&
(\Omega \Gamma +  \Psi^2 )\rho_4+\alpha_1 B^n + \alpha_2  B^{n-1} q^2 + \alpha_3 B^{n-2} q^4 + \alpha_4 \rho_4 +\Omega.
\end{eqnarray}

At this point of the discussion we can integrate out the massive fields $\alpha_4$ and $\rho_4$  and we are left with the non-trivial $F$-term for the field $\Omega$
\begin{equation}
F_\Omega  = \tilde v^2 \tilde B^{n-1} +1 = 0,
\end{equation}
where the equation is solved only if the fields $\tilde v$ and $\tilde B$ both acquire a non-zero vev.
Reintroducing color indices (lower indices) for $\mathrm{SU}(2n)_2$ we have:
\begin{equation}
\epsilon^{i_1,\dots, i_{2n}}
 \tilde B_{i_1, i_2} \dots \tilde{B}_{i_{2n-1}, i_{2n}} 
 \tilde v^{(1)}_{i_{2n-1}} \tilde v^{(2)}_{i_{2n}}
 +1 = 0.
\end{equation}
Without loss of generality we can take the vev to be aligned as:
\begin{equation}
\begin{array}{l}
\langle \tilde v^{(1)}_{{2n-1}} \rangle \neq 0,
\\
\langle \tilde v^{(2)}_{{2n}} \rangle \neq 0,
\\
\langle \tilde{B}_{2j-1, 2j} \rangle \neq 0.
\end{array}
\end{equation}
The vevs for $\tilde{v}^{(1,2)}$ Higgs $\mathrm{SU}(2n)_2$ to $\mathrm{SU}(2n-2)$, and the vevs for $\tilde{B}$ further Higgs it to $\mathrm{USp}(2n-2)$. 
\item The superpotential deformation \eqref{Wdef3SU2n} gives rise to the linear term $\rho_1$ in \eqref{thirdspot}. After integrating out the massive degrees of freedom we are left with
\begin{equation}
  W=\rho_1 \tilde q^4 \tilde B^{n-2}+\tilde q^2  \tilde B^{n-1} \rho_2 +\rho_3 \tilde B^n  +\alpha_1 B^n + \alpha_2  B^{n-1} q^2 + \alpha_3 B^{n-2} q^4+\rho_1.
\end{equation}
The $F$-term for the field $\rho_1$ gives
\begin{equation}
  F_{\rho_1} = \tilde{q}^4 \tilde{B}^{n-2} + 1 = 0,
\end{equation}
which implies the following non-zero vev for the fields
\begin{equation}
\begin{array}{l}
\langle \tilde q^{(i)} \rangle \neq 0 \quad \qquad i = 1,\dots, 4,
\\
\langle \tilde{B}_{2j-1, 2j} \rangle \neq 0.
\end{array}
\end{equation}
Such a vev for $\tilde{q}$ Higgs $\mathrm{SU}(2n)$ down to $\mathrm{SU}(2n-4)$ and then further down to $\mathrm{USp}(2n-4)$ because of the vev of $\tilde{B}$.
\end{enumerate}

At the level of the superconformal index the integral associated to the 
$\mathrm{SU}(2n)_1$ gauge theory is 
\begin{equation}
\frac{(p;p)_{\infty}^{2n-1}(q;q)_{\infty}^{2n-1}}{(2n)!}\int\displaylimits_{\mathbb{T}^{2n-1}} \prod_{i=1}^{2n-1} \frac{dz_i}{2\pi i z_i} 
\prod_{i=1}^{2n}  \prod_{a=1}^{4} \Gamma_e(z_i t_{a},z_i^{-1} s_{a})
\prod_{i<j}\frac{\Gamma_e\big(U_A z_i z_j,U_{\tilde A} z_i^{-1}z_j^{-1}\big)}{\Gamma_e((z_i/z_j)^{\pm 1})}
\end{equation}
with the $\mathrm{SU}(2n)$ constraint 
$
z_{2n}= \prod_{i=1}^{2n-1} z_i^{-1}
$. 
The fugacities are constrained by the balancing condition $
(U_A U_{\tilde A})^{2n-2} \prod_{a=1}^4 t_a s_a = (pq)^2$, corresponding to the
requirement on the axial anomaly, or equivalently of the anomaly freedom of the $\mathrm{U}(1)_R$ R-symmetry. There is a further constraint enforced by the superpotential deformation
\begin{enumerate}
\item (\ref{Wdef1SU2n}) $\rightarrow$  $U_{\tilde A}^{n-2} s_1 s_2 s_3 s_4 = pq$,
\item (\ref{Wdef2SU2n}) $\rightarrow$  $U_{\tilde A}^{n-1} s_3 s_4 = pq$,
\item (\ref{Wdef3SU2n}) $\rightarrow$  $U_{\tilde A}^{n} = pq$.
\end{enumerate}
Then the deconfined quiver has index
\begin{eqnarray}
\label{indexdec}
&&
\frac{(p;p)_{\infty}^{4n-2}(q;q)_{\infty}^{4n-2}}{((2n)!)^2}
\Gamma_e(pq/U_B^n)\prod_{a<b} \Gamma_e(pq m_a^{-1} m_b^{-1}U_B^{1-n})\Gamma_e(pq U_B^{2-n} /\prod_a m_a)\Gamma_e(pqV^{-2n})
 \\
&&
\int_{\mathbb{T}_{w,z}^{2n-1}} \prod_{i=1}^{2n-1} \frac{dz_i}{2\pi i z_i} 
 \frac{dw_i}{2\pi i w_i} 
\prod_{i=1}^{2n}  \prod_{a=1}^{4} \Gamma_e(w_i m_{a},z_i^{-1} s_{a})
\prod_{i,j} \Gamma_e(z_i w_j^{-1} V)
\prod_{i<j}\frac{\Gamma_e\big(U_B w_i w_j,U_{\tilde A} z_i^{-1}z_j^{-1}\big)}{\Gamma_e((z_i/z_j)^{\pm 1})\Gamma_e((w_i/w_j)^{\pm 1})}
\nonumber
\end{eqnarray}
where $U_B = U_A V^{-2}  $ and $m_a=t_a/V$
and the two balancing conditions are $U_B^{2n-2} \prod m_a V^{2n} = pq = U_{\tilde A}^{2n-2} \prod s_a V^{2n}$.
In addition, the constraint from (\ref{Wdef1SU2n}), (\ref{Wdef2SU2n}) or (\ref{Wdef3SU2n})  remains as before.
Observe that the first four elliptic Gamma functions in the first line of (\ref{indexdec}) 
represents the contributions of the singlets $\alpha_{1,2,3,4}$ respectively.

Then the index associated to the $\mathrm{SU}(2n)_2$ theory has index 
\begin{align}
   \mathcal{I}=&\frac{\left(p;p\right)_\infty^{2n-1}\left(q;q\right)_\infty^{2n-1}}{(2n)!} \Gamma_e\big(U_A^{n-2}\prod_{a=1}^{4}t_a; U_{\tilde{A}}^{n-2}\prod_{a=1}^{4}s_a;U_A^n;U_{\tilde{A}}^n\big) \!\! \prod_{1\leq a < b \leq 4} \!\!\!\! \Gamma_e\left({U_A^{n-1}t_a t_b} ; U_{\tilde{A}}^{n-1}s_a s_b\right) \nonumber \\
   &\int\displaylimits_{\mathbb{T}^{2n-1}} \!\!\!\prod_{i = 1}^{2n-1}\frac{\mathrm{d} \omega_i}{2\pi \mathrm{i} \omega_i}\prod_{i=1}^{2n}\prod_{a=1}^{4} \Gamma_e\left(\omega_i m_a;\omega_i^{-1} n_a\right) \!\!\!\! \prod_{1\leq i<j \leq 2n} \!\!\!\!\!\frac{\Gamma_e \left(\omega_i \omega_j U_B\right) \Gamma_e \left(\omega_i^{-1} \omega_j^{-1} U_{\tilde{B}} \right)}{\Gamma_e \left(\omega_i / \omega_j\right) \Gamma_e\left( \omega_j /\omega_i \right)},
   \label{eq:ind_andrea_pf}
\end{align}
where $n_a=s_a V$ and $U_{\tilde B}=U_{\tilde A} V^2$.
The cancellation of gauge anomalies imposes  constraint $(U_{B} U_{\Bt})^{2n-2}  \prod_{a=1}^{4} m_a n_a = (pq)^2$, while 
the superpotential deformations impose:
\begin{enumerate}
\item (\ref{Wdef1SU2n}) $\rightarrow$  $U_{\Bt}^{n}  = 1$,
\item (\ref{Wdef2SU2n}) $\rightarrow$  $U_{\Bt}^{n-1} n_1 n_2 = 1$,
\item (\ref{Wdef3SU2n}) $\rightarrow$  $U_{\Bt}^{n}  n_1 n_2 n_3 n_4 = 1$.
\end{enumerate}
The contour integral involved in  \eqref{eq:ind_andrea_pf} is pinched when these constraints are satisfied, and the integral can be (partially) resolved.
We refer the reader to \cite{Gaiotto:2012xa,Spiridonov:2014cxa} for further details on the pinching of the SCI and to \cite{Comi:2022aqo,Bajeot:2023gyl,Amariti:2024sde} for similar applications.
Below, we analyze the pinching in the presence of the three deformations separately.

%
%
%
%
\subsubsection{Dual Higgsing and pole pinching}
\label{sec:pinchev}

Here we reproduce the dual Higgsing at the level of the superconformal index, separating the analysis for the  three superpotential deformations  (\ref{Wdef1SU2n}),  (\ref{Wdef2SU2n})  and (\ref{Wdef3SU2n}).
In this way we find three different dualities between the original $\mathrm{SU}(2n)$ model equipped with one of these superpotential deformations and a $\mathrm{USp}(2m)$ gauge theory, with $m=n$, $m=n-1$ and $m=n-2$ respectively, an antisymmetric, eight fundamentals and a flipped superpotential. 

\begin{itemize}
\item \underline{The case of  $W= \tilde A^{n-2} \tilde Q^4$} \\

The superpotential deformation \eqref{Wdef1SU2n} imposes the constraint
\begin{equation}
    U_{\tilde{A}}^{n-2}\prod_{a=1}^{4}s_a = pq.
\end{equation}
Such constraint cannot be straightforwardly imposed at the level of the index \eqref{eq:ind_andrea_pf} as it is a singular limit signalling the presence of a Higgsing, and it must be treated carefully. We define
\begin{equation}
    U_{\tilde{A}}^{n-2}\prod_{a=1}^{4}s_a \coloneqq pq e^\varepsilon, \quad U_{\tilde{B}}\coloneqq e^{-\varepsilon/n},
\end{equation}
such that the balancing conditions are satisfied. The effect of the superpotential deformation  \eqref{Wdef1SU2n}  can now be studied by considering the limit $\varepsilon \to 0$ of the index.
We consider the following combination of Gamma functions:
\begin{equation}
    \prod_{1\leq i<j \leq 2n}\Gamma_e\left(\omega_i^{-1}\omega_j^{-1} U_{\tilde{B}}\right).
\end{equation}
They define the family of poles
\begin{align}
    \omega_i \omega_j = U_{\tilde{B}}\, p^{k}q^{l}, \quad 1\leq i<j\leq 2n, \quad k,l \geq 0.
\end{align}
Let us focus on the poles with $k,l = 0$ and consider the family of poles defined by the $n$ pairings of $2n$ elements 
\begin{equation}
    \omega_{i_1}= \omega_{i_2}^{-1} U_{\tilde{B}}, \quad \dots \quad \omega_{i_{2n-1}} = \omega_{i_{2n}}^{-1} U_{\tilde{B}}.
\end{equation}
Without loss of generality, for any fixed pairing we can always relabel the integration variables and consider only the single ordered pairing
\begin{equation}
       \omega_{1} = \omega_{2}^{-1} U_{\tilde{B}}, \quad\dots \quad \omega_{2n-1} = \omega_{2n}^{-1} U_{\tilde{B}},
\end{equation}
together with the appropriate degeneracy factor $\frac{(2n)!}{2^n n!}$, as each pairing of poles will contribute equally to the index.
Enforcing the $\mathrm{SU}(2n)$ constraint $\prod_{i=1}^{2n} \omega_i = 1$, the holonomies need also to satisfy
\begin{equation}
    \omega_{2n-1}\omega_{2n} = U_{\tilde{B}}^{1-n}, \implies \Gamma_e\left( \omega_{2n-1}^{-1}\omega_{2n}^{-1} U_{\tilde{B}}\right) = \Gamma_e\left(U_{\tilde{B}}^{n}\right),
\end{equation}
pinching the integration contour as $U_{\tilde{B}}\xrightarrow{\varepsilon \to 0} 1$. 
The pairing of $2n$ variables into $n$ pairs, together with the $\mathrm{SU}(2n)$ constraint, allows for a partial evaluation of $n-1$ integrals out of the $2n-1$ ones.

After relabeling $y_i = \tfrac{\sqrt{U_{\tilde{B}}}}{\omega_{2i-1}} = \tfrac{\omega_{2i}}{\sqrt{U_{\tilde{B}}}}, \; i = 1,\dots,n $ the various charged fields contribute as:
\begin{align}
    q &\to \prod_{a=1}^{4}\prod_{i=1}^{n} \Gamma_e \left(y_i^{\pm 1} \sqrt{U_{\tilde{B}}}m_a\right) \nonumber \\
    \tilde{q} &\to \prod_{a=1}^{4}\prod_{i=1}^{n} \Gamma_e \left(y_i^{\pm 1} \frac{n_a}{\sqrt{U_{\tilde{B}}}}\right)\nonumber \\
    B & \to \Gamma_e\left(U_{\tilde{B}}U_B\right)^n \prod_{i<j}^{n} \Gamma_e \left(y_i^{\pm 1} y_j^{\pm 1}  U_{\tilde{B}}U_B\right) \nonumber \\
    \tilde{B} &\to \Gamma_e\left(U_{\tilde{B}}^n\right) \prod_{i<j}^{n} \Gamma_e \left(y_i^{\pm 1} y_j^{\pm 1}\right) \nonumber \\
    A & \to  \prod_{i<j}^{n-1} \Gamma_e \left(y_i^{\pm 1} y_j^{\pm 1} \right)^2  \prod_{i = 1}^{n-1}\Gamma_e \left(y_i^{\pm 2} \right),\nonumber \\
\end{align}
By noticing that for any fixed non-zero $\varepsilon$
\begin{equation}
    \Gamma_e\left(U_{\tilde{A}}^{n-2}\prod_{a=1}^{4}s_a\right) \Gamma_e \left(U_{\tilde{B}}^n\right) = \Gamma_e (pq e^\varepsilon) \Gamma_e (e^{-\varepsilon}) = 1,
\end{equation}
the $\varepsilon\to0$ limit of the index \eqref{eq:ind_andrea_pf} is regular and well-defined, and we obtain
\begin{align}
\label{findapolo2}
\mathcal{I} = & \frac{\left(p;p\right)_\infty^{n}\left(q;q\right)_\infty^{n}}{2^n n!} \Gamma_e\big(U_A^{n-2}\prod_{a=1}^{4}t_a; U_A^n;U_{\tilde{A}}^n \big) \!\! \prod_{1\leq a < b \leq 4} \!\!\!\! \Gamma_e\left({U_A^{n-1}t_a t_b} ; pq \,U_{\tilde{A}} s_a^{-1} s_b^{-1}\right)\! \Gamma_e\left(U_A U_{\tilde{A}}\right)^n \nonumber \\
& \int\displaylimits_{\mathbb{T}^{n}} \prod_{i = 1}^{n}\frac{\mathrm{d} y_i}{2\pi \mathrm{i} y_i}
\prod_{i=1}^{n}\prod_{a=1}^{4} \Gamma_e\left(y_i^{\pm 1} t_a U_{\tilde{A}}^{1/2};y_i^{\pm 1} s_a U_{\tilde{A}}^{-1/2}\right) 
\frac{\prod\displaylimits_{1\leq i<j \leq n}\Gamma_e \left(y_i^{\pm 1} y_j^{\pm 1} U_A U_{\tilde{A}}\right)}{\prod\displaylimits_{1\leq i<j \leq n}\Gamma_e \left(y_i^{\pm 1} y_j^{\pm 1}\right) \prod\displaylimits_{i=1}^{n} \Gamma_e\left( y_i^{\pm 2} \right)}
\end{align}
after accounting for the $\frac{(2n)!}{2^n n!}$ degeneracy of the sequence of pinching poles, and employing the dictionary for the fugacities and the balancing conditions.
The result is compatible with a $\mathrm{USp}(2n)$ gauge theory with 8 fundamentals with fugacities $\vec{u}$ and a totally antisymmetric with fugacity $U_X$ with:
\begin{equation}
U_X = U_A U_{\tilde{A}},
\qquad
\vec{u} = \left(
t_a \sqrt{U_{\tilde{A}}}, \frac{s_a}{ \sqrt{U_{\tilde{A}}}}
\right).
\end{equation}

\item \underline{The case of  $W= \tilde A^{n-1} \tilde Q_3 \tilde Q_4$} \\

The superpotential deformation \eqref{Wdef2SU2n} imposes the constraint
\begin{equation}
U_{\tilde A}^{n-1}\prod_{a=1}^{2}s_a = p q.
\end{equation}
We define 
\begin{equation}
  U_{\tilde A}^{n-1}\prod_{a=1}^{2} s_a \coloneqq p q e^\varepsilon, \quad U_{\tilde B} \coloneqq  U_{\tilde B}^{n-1} n_1 n_2 = e^{-\varepsilon}.
\end{equation}
The contour integral involved in \eqref{eq:ind_andrea_pf} is pinched as $\varepsilon \to 0$ when these constraints are satisfied, and the integral can be partially resolved.
To see this we consider the following combination of Gamma functions appearing in the integrand:
\begin{equation}	
\label{eq:gammas_poles_Q2_1}
\prod_{1\leq i < j\leq 2n-2} \Gamma_e \left(\omega_i^{-1}\omega_{j}^{-1}U_{\tilde{B}} \right)
\end{equation}
which have poles for the following values of the gauge fugacities $\omega_i$
\begin{align}
    \omega_i \omega_j = U_{\tilde{B}}\, p^{k}q^{l}, \quad 1\leq i<j\leq 2n-2, \quad k,l \geq 0.
\end{align}
Let us focus on the poles with $k,l = 0$ and consider the following sequence of poles
\begin{equation}
    \omega_{1}\omega_{2} = U_{\tilde{B}}, \quad \dots \quad \omega_{2n-3}\omega_{2n-2}= U_{\tilde{B}}.
\end{equation}
Consider also 
\begin{equation}
  \prod_{a = 1}^{2}\Gamma_e \left( \omega_{2n-2+a}^{-1} n_a \right) \implies \omega_{2n-2+a}= n_a.
\end{equation}
Enforcing the $\mathrm{SU}\left(2n\right)$ constraint, the contour gets pinched as $\varepsilon \to 0$ and 
\begin{equation}
  \Gamma_e \left(\omega_{2n}^{-1}n_2\right) = \Gamma_e \left( e^{-\varepsilon} \right).
\end{equation}

Such sequence of poles allows us to perform $n$ out of the $2n-1$ integrations. Accounting for all the possible equivalent ways of constructing the family of poles we also obtain a degeneracy factor $\frac{2n!}{2^{n-1}(n-1)!}$. The resulting integral corresponds to the superconformal index of $\mathrm{USp}(2n-2)$ with 8 fundamentals and an antisymmetric. We find it convenient to write the resulting integral in terms of the following gauge fugacities:
\begin{equation}
y_i \equiv \frac{\omega_{2i}}{\sqrt{U_{\Bt}}} = \frac{\sqrt{U_{\Bt}}} {\omega_{2i-1}}, \qquad i=1, \dots, n-1.
\end{equation}
Then the contributions of the various charged fields reduce to:
\begin{align}
&q \to \prod_{a=1}^{4} \prod_{i=1}^{n-1}  \Ge[y_i^{\pm} \sqrt{U_{\Bt}} m_a] \prod_{b=1}^2 \Ge[m_a n_b] \nonumber \\
&\tilde{q} \to \prod_{i=1}^{n-1} \Ge[y_i^{\pm}
  \frac{n_{1,2}}{\sqrt{U_{\Bt}}}] \Ge[y_i^{\pm} \frac{n_{3,4}}{\sqrt{U_{\Bt}}}]
  \Ge[\left(\frac{n_1}{n_2}\right)^{\pm} ]
  \Ge[\frac{n_{3,4}}{n_{1,2}} ] \nonumber \\
&B \to \prod_{i<j}^{n-1}\Ge[y_i^{\pm} y_j^{\pm} U_B U_{\Bt}]
  \prod_{i=1}^{n-1} \prod_{a=1}^2 \Gamma_e\left(y_i^{\pm} U_B \sqrt{U_{\Bt}} n_a\right) \Gamma_e\left(U_B n_1 n_2\right)\\
&\Bt \to 
	\prod_{i<j}^{n-1}\Ge[y_i^{\pm} y_j^{\pm} ]
	\prod_{i=1}^{n-1} \prod_{a=1}^2 \Ge[y_i^{\pm} \frac{ \sqrt{U_{\Bt}} } { n_a}] \Ge[\frac{U_{\tilde{B}}}{n_1 n_2}] \nonumber \\ 
& A \to \prod_{i<j = 1}^{n-1}\Ge[y_i^{\pm} y_j^{\pm} ] ^{-1}
  \prod_{i=1}^{n-1}\prod_{a=1}^{2} \Gamma_e\left(  y_i{\pm} 
	\left( \frac{n_a}{ \sqrt{U_{\Bt}}} \right)^{\pm}
	\right)^{-1} \Gamma_e \left(\left(\frac{n_1}{n_2}\right)^{\pm}\right)
\end{align}

plus singlets described below. 
Furthermore, from the leftover contributions of $\tilde B$ and $A_\mu$  we read the contributions corresponding to the vector multiplet of a $\mathrm{USp}(2n-2)$ gauge group with an antisymmetric $X$ and 8 fundamentals $u_i$ with:
\begin{equation}
	U_{X} = U_A U_{\At} 
	,\quad
	\vec{u} = \left(
		\sqrt{U_{\At}} \vec{t}\,;  \frac{s_{3,4}}{\sqrt{U_A}},  \;U_A \sqrt{U_{\At}} s_{1,2}		\right)
\end{equation}
where we translated the mass parameters in terms of the original ones.

Observe that there are also extra singlets arising from the first line of \eqref{eq:ind_andrea_pf}.
Some of such singlets cancel with the contributions of the singlets leftover from the contributions of the charged fields.
Explicitly we have
\begin{eqnarray}
\Gamma_e(U_{\tilde A}^{n-1} s_{1,2} s_{3,4}) \Gamma_e\left(\frac{n_{3,4}}{n_{1,2}} \right)&=& 1 \nonumber \\
\Gamma_e(U_{\tilde A}^{n-2} \prod_{a=1}^{4} s_a)\Gamma_e \left(\frac{U_{\tilde B}}{n_1 n_2}\right)&=& 1 
\end{eqnarray}
Furthermore, we have $\Gamma_e(U_B n_1 n_2) = \Gamma_e(U_A s_1 s_2)$.

The final integral becomes 
\begin{eqnarray}
\label{findapolo}
&&
\Gamma_e(U_A^n,U_{\tilde A}^n, U_A^{n-2} \prod_{i=1}^4 t_a, U_{\tilde A}^{n-1} s_1 s_2)
\prod_{a<b} \Gamma_e (t_a t_b U_A^{n-1})
\prod_{a=1}^4 \prod_{r=1}^2 \Gamma_e(s_rt_a)
 \Gamma_e(U_A s_1 s_2)
 \nonumber \\
&&
 \Gamma_e (U_A U_{\tilde A})^{n-1}
\frac{(p;p)_{\infty}^{n-1}(q;q)_{\infty}^{n-1}}{(n-1)! 2^{n-1}}
\int\displaylimits_{\mathbb{T}^{n-1}} \prod_{\ell=1}^{n-1} 
\frac{\mathrm{d} y_\ell}{2 \pi i y_\ell}
\frac{\prod_{\ell<k}\Gamma_e (U_A U_{\tilde A} y_\ell^{\pm 1} y_k^{\pm 1} )}
{
\prod_{\ell < k}
\Gamma_e 
(y_\ell^{\pm 1} y_k^{\pm 1} )
\prod_{\ell=1}^{n-1} \Gamma_e(y_\ell^{\pm 2})
}
\nonumber \\
&&
\prod_{\ell=1}^{n-1} 
\prod_{a=1}^2
\Gamma_e (y_\ell^{\pm 1}  s_a U_A \sqrt{U_{\tilde A}})
\prod_{a=3}^4 \Gamma_e\big(y_\ell^{\pm 1} \frac{s_a}{\sqrt {U_{\tilde A}}}
\big)
\prod_{a=1}^4
\Gamma_e (y_\ell^{\pm 1} \sqrt{U_{\tilde A}} t_a ).
\end{eqnarray}
\item \underline{The case of $W = \mathrm{Pf}\tilde{A}$} \\

The superpotential deformation  imposes the further constraint
\begin{equation}
    U_{\tilde{A}}^{n} = pq.
\end{equation}
By defining
\begin{equation}
    U_{\tilde{B}}^{n-2}\prod_{a=1}^{4}n_a \coloneqq  e^{-\varepsilon}, \quad U_{\tilde{A}}\coloneqq pq e^{\varepsilon/n},
\end{equation}
such that the balancing conditions  are satisfied, we can consider the limit $\varepsilon \to 0$ of the index, which implements the Higgsing of the theory due to the superpotential deformation \eqref{Wdef3SU2n}.
We consider the following sequence of Gamma functions:
\begin{equation}
    \prod_{1\leq i<j \leq 2n-4}\Gamma_e\left(\omega_i^{-1}\omega_j^{-1} U_{\tilde{B}}\right).
\end{equation}
They define the family of poles
\begin{align}
    \omega_i \omega_j = U_{\tilde{B}}\, p^{k}q^{l}, \quad 1\leq i<j\leq 2n-4, \quad k,l \geq 0.
\end{align}
Let us focus on the poles with $k,l = 0$.
Considering the family of poles defined by
\begin{equation}
    \omega_{1}= \omega_{2}^{-1} U_{\tilde{B}}, \quad \dots \quad \omega_{2n-5} = \omega_{2n-4}^{-1} U_{\tilde{B}}.
\end{equation}
Consider also
\begin{equation}
    \prod_{a=1}^{4}\Gamma_e\left(\omega_{2n-4+a}^{-1}n_a\right) \implies  \omega_{2n-4+a} = n_a.
\end{equation}

Enforcing the $\mathrm{SU}(2n)$ constraint $\prod_{i=1}^{2n} \omega_i = 1$, the holonomies also satisfy
\begin{equation}
    \omega_{2n-3}\omega_{2n-2}\omega_{2n-1}\omega_{2n} = U_{\tilde{B}}^{2-n}, \implies \Gamma_e\left( \omega_{2n}^{-1}n_4\right) = \Gamma_e\left(e^{-\varepsilon}\right),
\end{equation}
pinching the integration contour as $U_{\tilde{B}}^{n-2}\prod_{a=1}^{4}n_a \xrightarrow{\varepsilon \to 0} 1$.

Such sequence of poles allows for an evaluation of $n+1$ integrals out of the $2n-1$, when the superpotential deformation is implemented, reproducing the Higgsing of the $\mathrm{SU}(2n)$ gauge group down to $\mathrm{USp}(2n-4)$. 

Accounting for all the possible equivalent ways of constructing the same family of poles, we get a $\frac{2n!}{2^{n-2}(n-2)!4!}$ degeneracy factor arising from the pairings of $2n$ terms in $2n-4$ pairs and an extra $4!$ contributions from the permutations of $n_a, \, a=1,\dots,4$, leaving us with a total degeneracy factor of $\frac{2n!}{2^{n-2}(n-2)!}$, reconstructing the Weyl of $\mathrm{USp}(2n-4)$.

After relabeling $y_{i} \equiv \frac{\omega_{2i}}{\sqrt{U_{\tilde{B}}}}, \; i = 1,\dots,n-2 $ the various charged fields contribute as 
\begin{align}
    q & \to \prod_{a=1}^{4}\prod_{i=1}^{n} \Gamma_e \left(y_i^{\pm 1}\sqrt{U_{\tilde{B}}} m_a\right) \prod_{a,b}^{4}\Gamma_e \left(m_a n_b\right) \nonumber \\
    \tilde{q} & \to \Gamma_e \left(e^{-\varepsilon}\right) \prod_{a=1}^{4}\prod_{i=1}^{n} \Gamma_e \left(\frac{y_i^{\pm 1}}{\sqrt{U_{\tilde{B}}}} n_a\right) \prod_{a \neq b}\Gamma_e \left(n_a / n_b\right) \nonumber \\
    B & \to \Gamma_e\left(U_B U_{\tilde{B}}\right)^{n-2} \prod_{i<j}^{n-2} \Gamma_e \left(y_i^{\pm 1} y_j^{\pm 1}  U_B U_{\tilde{B}}\right) \prod_{a<b}^{4} \Gamma_e \left(n_a n_b U_B \right) \prod_{i=1}^{n-2}\prod_{a=1}^{4} \Gamma_e \left(y_i^{\pm 1} n_a\sqrt{U_{\tilde{B}}}U_B \right) \nonumber \\
    \tilde{B} & \to \prod_{i<j}^{n-2} \Gamma_e \left(y_i^{\pm 1} y_j^{\pm 1}\right) \prod_{i=1}^{n-2}\prod_{a=1}^{4} \Gamma_e \left(y_i^{\pm 1} n_a^{-1}\sqrt{U_{\tilde{B}}}\right) \prod_{a<b}^{4} \Gamma_e \left( n_a^{-1} n_b^{-1}U_{\tilde{B}}\right) \nonumber \\
    A & \to  \prod_{i<j}^{n-2} \Gamma_e \left(y_i^{\pm 1} y_j^{\pm 1} \right)^2  \prod_{i = 1}^{n-2}\Gamma_e \left(y_i^{\pm 2} \right) \prod_{a\neq b}^{4} \Gamma_e \left(n_a/ n_b \right) \prod_{i=1}^{n-2}\prod_{a=1}^{4} \Gamma_e \left(y_i^{\pm 1} \left(\frac{n_a}{\sqrt{U_{\tilde{B}}}}\right)^{\pm1}\right). \nonumber \\
\end{align}
By noticing that for any fixed non-zero $\varepsilon$
\begin{equation}
    \Gamma_e\left(U_{\tilde{B}}^{n-2}\prod_{a=1}^{4}n_a\right) \Gamma_e \left(U_{\tilde{A}}^n\right) = \Gamma_e (e^{-\varepsilon}) \Gamma_e (pq e^{\varepsilon}) = 1,
\end{equation}
we can take the $\varepsilon\to0$ limit of the index \eqref{eq:ind_andrea_pf} and we obtain, after many simplifications and employing the balancing conditions together with the extra superpotential deformation,
\begin{align}
&\mathcal{I} = \frac{\left(p;p\right)_\infty^{n-2}\left(q;q\right)_\infty^{n-2}}{2^{n-2} (n-2)!} \Gamma_e\left(U_A U_{\tilde{A}}\right)^{n-2} \Gamma_e\big(U_A^n \big) \Gamma_e\big(U_A^{n-2}\prod_{a=1}^{4}t_a; U_{\tilde{A}}^{n-2}\prod_{a=1}^{4}s_a) \prod_{a,b=1}^{4}\!\Gamma_e \left(s_a t_b\right)\nonumber \\
&\prod_{a < b}^{4} \Gamma_e\left({U_A^{n-1}t_a t_b};U_A s_a s_b\right) \int\displaylimits_{\mathbb{T}^{n-2}} \!\!\!\prod_{i = 1}^{n-2}\frac{\mathrm{d} y_i}{2\pi \mathrm{i} y_i}\prod_{i=1}^{n-2}\prod_{a=1}^{4} \Gamma_e\left(y_i^{\pm 1} \sqrt{U_{\tilde{A}}} t_a;y_i^{\pm 1} \sqrt{U_{\tilde{A}}} U_A s_a\right) \nonumber \\
& \!\frac{ \prod_{i<j}^{n-2}\Gamma_e \left(y_i^{\pm 1} y_j^{\pm 1} U_A U_{\tilde{A}} \right)}{\prod_{i<j}^{n-2}\Gamma_e \left(y_i^{\pm 1} y_j^{\pm1}\right) \prod_{i=1}^{n-2}\Gamma_e\left( y_i^{\pm 2} \right)},
\label{deffinalPfA}  
\end{align}
which defines a $\mathrm{USp}(2n-4)$ gauge theory with one antisymmetric and 8 fundamentals.
\end{itemize}
%
%
%
%
%
\subsubsection{An alternative deconfinement}
%
%
%
%
%
Here we study the duality just obtained by a different deconfinement of the antisymmetric tensors $\tilde A$.
The three cases deserve a different analysis.
\begin{itemize}
\item We start by considering the first deformation (\ref{Wdef1SU2n}).

The deconfined theory in this case corresponds to the quiver in Figure \ref{figdec2def2} with superpotential
\begin{equation}
W = C \tilde R^2.
\end{equation}
In this case we have used a $\mathrm{USp}(2n)$ gauge group in order to deconfine the antisymmetric $\tilde A$, that corresponds in the deconfined  model to the combination $\tilde A = \tilde P^2$. Furthermore,  the antifundamentals $\tilde Q$ correspond in the deconfined quiver to the combinations $\tilde Q = \tilde P \tilde R $.
The $\mathrm{SU}(4)$ antisymmetric singlet $C$ is crucial in order to reproduce the superpotential deformation
 (\ref{Wdef1SU2n}) of the original theory. Indeed if we confine the $\mathrm{USp}(2n)$ node we obtain the original $\mathrm{SU}(2n)$ model with superpotential 
\begin{equation}
W =  \mathrm{Pf} \tilde A \,\mathrm{Pf} \,\mathcal{C} + \tilde A^{n-1} \mathcal{C} \tilde Q^2 + \tilde A^{n-2} \tilde Q^4+  C \mathcal{C},
\end{equation}
with $\mathcal{C} = \tilde R^2$
This superpotential coincides with (\ref{Wdef1SU2n}) after integrating out the massive fields $\mathcal{C} $ and $C$. Moreover by solving these $F$-terms we find that the singlet $C$ in the deconfined phase coincides with the operator $\tilde A^{n-1} \tilde Q^2$.

\begin{figure}
\begin{center}
  \includegraphics[width=8cm]{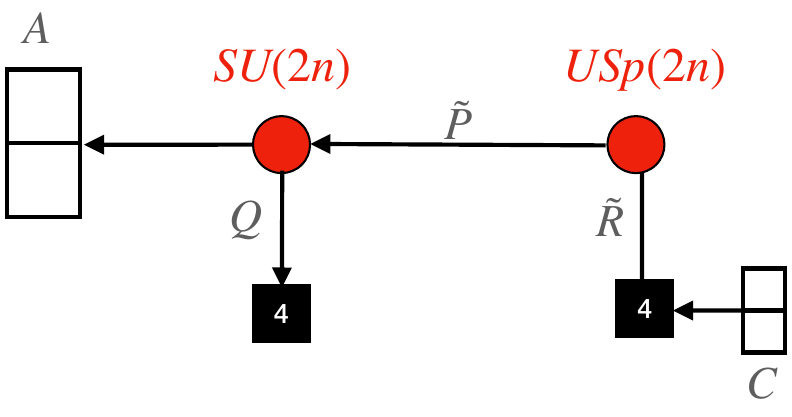}
  \end{center}
  \caption{Quiver gauge theory obtained by deconfining the conjugate antisymmetric 
  $\tilde A$  for the $\mathrm{SU}(2n)$ gauge theory with superpotential deformation (\ref{Wdef1SU2n}).}
    \label{figdec2def2}
\end{figure}

The next step consists of observing that the $\mathrm{SU}(2n)$ gauge theory is confining, indeed it has $2n$ antifundamentals, four fundamentals and an antisymmetric. 
The gauge invariant degrees of freedom are
\begin{equation}
\eta_1=\mathrm{Pf} \, A
, \quad
\eta_2=A^{n-1}  Q^2
, \quad
 \eta_3=A^{n-2}  Q^4
 , \quad
 \eta_4=\tilde P^{2n}
,\quad
R =\tilde P Q 
, \quad
 B_{As}=A \tilde P^2,
 \end{equation}
where $B_{As}$ is an (reducible\footnote{From now on we will omit to mention that the $\mathrm{USp}(2m)$ antisymmetric tensors considered in this paper are always reducible.}) antisymmetric , $R$ are fundamentals  and the other combinations are singlets of $\mathrm{USp}(2n)$.

The superpotential of this model is 
\begin{eqnarray}
W=\eta_1 R^4 B_{As}^{n-2}+R^2  B_{As}^{n-1} \eta_2 + \eta_3 \mathrm{Pf}\, B_{As} + \eta_1 \eta_3 \eta_4+\eta_2^2  \eta_4+\tilde R^2 C.
 \end{eqnarray}

In order to complete the analysis we can also map the chiral ring operators 
of the electric theory and the ones of the magnetic dual. We found the following mapping

\begin{equation}
\begin{array}{c|c|c}
\mathrm{SU}(2n) & \mathrm{SU}(2n) \times \mathrm{USp}(2n) & \mathrm{USp}(2n) \\
\hline
 Q_\alpha  (A \tilde A)^k \tilde Q_\beta
 &
 Q_\alpha (A \tilde P^2)^k  \tilde P \tilde R_{\beta} 
 &
 B_{as}^k R_\alpha \tilde R_\beta
\\
\tilde A (A \tilde A)^j  Q_{[\alpha} Q_{\beta]}
&
\tilde P^2 (A \tilde P^2)^j  Q_{[\alpha} Q_{\beta]} 
&
B_{as}^{j}   R_{[\alpha}    R_{\beta]} 
\\
A (A \tilde A)^j \tilde Q_{[\alpha} \tilde Q_{\beta]}
&
A (A \tilde P^2)^j  \tilde P \tilde R_{[\alpha} \tilde P \tilde R_{\beta]} 
&
B_{as}^{j+1} \tilde R_{[\alpha} \tilde   R_{\beta]} 
\\
 (A \tilde A)^m
 &
 (A \tilde P^2)^m
 &
 B_{as}^m
  \\
\mathrm{Pf} A
&
\mathrm{Pf} A
&
\eta_1
\\
\mathrm{Pf}  \tilde A 
&
 \tilde P^{2n}
 &
 \eta_4
 \\
\tilde  A^{n-1} \tilde  Q_{[\alpha} \tilde Q_{\beta]}
&
C
&
C
\\
 A^{n-1} Q_{[\alpha} Q_{\beta]}
 &
 A^{n-1} Q_{[\alpha} Q_{\beta]}
 &
  \eta_2
  \\
  A^{n-2} Q_1Q_2Q_3Q_4
&
A^{n-2} Q_1Q_2Q_3Q_4
&
\eta_3
\end{array}
\label{Eq:GISU2n1}
\end{equation}
with $k=0,\dots,n-1$, $j=0,\dots,n-2$  and $m=1,\dots,n-1$.

At the level of the superconformal index, starting from the index of the original model, the deconfined quiver is given by

\begin{eqnarray}
&&
\frac{(p;p)_{\infty}^{3n-1}(q;q)_{\infty}^{3n-1}}{(2n)!n! 2^n}
\prod_{a<b} \Gamma_e(p q s_a^{-1}s_b^{-1} U_{\tilde A} )
\int_{\mathbb{T}^{3n-1}} \prod_{i=1}^{2n-1} \frac{dz_i}{2\pi i z_i} 
\prod_{\ell=1}^{n} 
\frac{dw_\ell}{2 \pi i w_\ell}
\prod_{i=1}^{2n}  \prod_{a=1}^{4} \Gamma_e(z_i t_{a})
\nonumber \\
&&
\prod_{i<j}\frac{\Gamma_e\big(U_A z_i z_j\big)}{\Gamma_e((z_i/z_j)^{\pm 1})}
\frac{\prod_{i=1}^{2n} \prod_{\ell=1}^{n} \Gamma_e\big(z_i^{-1} w_\ell^{\pm 1} \sqrt {U_{\tilde A}}\big)
\prod_{a=1}^4 \prod_{\ell=1}^{n} \Gamma_e\big(w_\ell^{\pm 1} s_a /\sqrt {U_{\tilde A}}
\big)}
{
\prod_{\ell < k}
\Gamma_e 
(w_\ell^{\pm 1} w_k^{\pm 1} )
\prod_{\ell=1}^n \Gamma_e(w_\ell^{\pm 2})
}.
 \nonumber \\
\end{eqnarray}
Then, confining the $\mathrm{SU}(2n)$ node we arrive at the expected result  (\ref{findapolo2}).

\item Next, we consider the second deformation (\ref{Wdef2SU2n}).

The deconfined theory in this case corresponds to the quiver in Figure \ref{figdec2def1} with superpotential
\begin{equation}
W = \sigma R_3 R_4
\end{equation}
In this case we have used a $\mathrm{USp}(2n-2)$ gauge group in order to deconfine the antisymmetric $\tilde A$, that corresponds in the deconfined  model to the combination $\tilde A = \tilde D^2$. Furthermore, the antifundamentals $\tilde{Q}_{3,4}$ correspond in the deconfined quiver to the combinations $\tilde Q_{3,4} = \tilde D R_{3,4}$.
The singlet $\sigma$ is crucial in order to reproduce the superpotential deformation
(\ref{Wdef2SU2n}) of the original theory. Indeed, if we confine the $\mathrm{USp}(2n-2)$ node we obtain the original $\mathrm{SU}(2n)$ model with superpotential 
\begin{equation}
W = \varphi \mathrm{Pf} \tilde A + \tilde A^{n-1} \tilde Q_3 \tilde Q_4 + \sigma \varphi.
\end{equation}
This superpotential coincides with (\ref{Wdef2SU2n}) after integrating out the massive fields $\sigma$ and $\varphi$. Moreover, by solving these $F$-terms we find that the singlet $\sigma$ in the deconfined phase coincides with the operator $\mathrm{Pf} \tilde A$.

The next step consists of observing that the $\mathrm{SU}(2n)$ gauge theory is confining. Indeed, it has $2n$ antifundamentals, four fundamentals and an antisymmetric. Actually its ``global" $\mathrm{SU}(2n)$ flavor symmetry in this case is broken to $\mathrm{USp}(2n-2)\times \mathrm{SU}(2)$, by the gauging and its gauge invariant degrees of freedom are
\begin{eqnarray}
&&
\eta_1=\mathrm{Pf} \, A
, \quad
\eta_2=A^{n-1}  Q^2
, \quad
 \eta_3=A^{n-2}  Q^4
 , \quad
 \eta_4=\tilde D^{2n-2}\tilde Q_1 \tilde Q_2
  \\
&&
\phi_{\alpha} =\tilde D Q_{\alpha}, \quad \psi_{\alpha;1,2}=Q_\alpha \tilde Q_{1,2} 
, \quad
 B_{As}=A \tilde D^2
 , \quad
 B_s=A \tilde Q_1 \tilde Q_2
 , \quad
 B_{V_{1,2}}=A \tilde Q_{1,2} \tilde D, \nonumber
 \end{eqnarray}
where $B_{As}$ is an antisymmetric, $B_V$ and $\Phi$ are fundamentals  and the other combinations are singlets of $\mathrm{USp}(2n-2)$.

The superpotential of this model is 
\begin{eqnarray}
\label{425}
W &=& \eta_1 \phi^4 (B_{As}^{n-3} B_s + B_{As}^{n-4} B_V^2)
+\eta_2 ( B_{As}^{n-1}\psi^2+ B_{As}^{n-2} B_s \phi^2+ B_{As}^{n-2} B_V \phi \psi
\nonumber \\
&+&
B_{As}^{n-3} B_V^2\phi^2) + \eta_3 (B_{As}^{n-1} B_s + B_{As}^{n-2} B_V^2 ) + \eta_1 \eta_3 \eta_4+\eta_2^2  \eta_4+R_3 R_4 \sigma.
 \end{eqnarray}

In order to complete the analysis we can also map the chiral ring operators 
of the electric theory and the ones of the magnetic dual. We found the following mapping
\begin{equation}
\begin{array}{c|c|c}
\mathrm{SU}(2n) & \mathrm{SU}(2n) \times \mathrm{USp}(2n-2) & \mathrm{USp}(2n-2)\\
\hline
 Q_\alpha  \tilde Q_{1,2} & Q_\alpha   \tilde Q_{1,2} &
 \psi_{\alpha;1,2}   \\
  Q_\alpha  (A \tilde A)^{m} \tilde Q_{1,2} & Q_\alpha  (A  \tilde D^2)^{m} \tilde Q_{1,2} &
 \phi_{\alpha} B_{As}^{m-1} B_{V_{1,2}}  \\
 Q_\alpha  (A \tilde A)^j \tilde Q_{3,4} &
Q_\alpha  (A  \tilde D^2)^j \tilde  D R_{3,4}&
\phi_{\alpha} B_{As}^j R_{3,4}\\
 \tilde A (A \tilde A)^j  Q_{[\alpha} Q_{\beta]} & \tilde D^2 (A \tilde D^2)^j  Q_{[\alpha} Q_{\beta]}&
 B_{As}^j  \phi_{[\alpha} \phi_{\beta]} \\
 A (A \tilde A)^j \tilde Q_{1,2} \tilde Q_{3,4} & A (A \tilde D^2)^j \tilde Q_{1,2} \tilde D R_{3,4} &  
B_{V_{1,2}} B_{As}^j R_{3,4}
 \\
 A \tilde Q_{1} \tilde Q_2
 &
 A  \tilde Q_{1} \tilde Q_2&
 B_s
 \\
 A (A \tilde A)^{j'} \tilde Q_{1} \tilde Q_2
 &
 A (A \tilde D^2)^{j'} \tilde Q_{1} \tilde Q_2&
 B_{As}^{j'-1} B_{V_1} B_{V_2}
  \\
A (A \tilde A)^{\ell} \tilde Q_{3} \tilde Q_4
 &
A (A \tilde D^2)^{\ell} \tilde D R_{3} \tilde D R_4
 &
 B_{As}^{\ell+1} R_3 R_4
 \\
 (A \tilde A)^m
 &
 (A \tilde D^2)^m
 &
 B_{As}^m
 \\
 \mathrm{Pf} A & \mathrm{Pf} A &\eta_1\\
 \mathrm{Pf} \tilde A & \sigma&\sigma\\
 A^{n-1} Q_{[\alpha} Q_{\beta]}
 &
 A^{n-1} Q_{[\alpha} Q_{\beta]} &\eta_2
 \\
 \tilde  A^{n-1} \tilde  Q_1 \tilde Q_2
&
\tilde  D^{2n-2} \tilde  Q_1 \tilde Q_2&\eta_4 \\
 A^{n-2} Q^4& A^{n-2} Q^4&\eta_3\\
\end{array}
\label{Eq:GISU2n2}
\end{equation}
Where 
$m=1,\dots,n-1$, $\ell=0,\dots,n-3$, $j=0,\dots,n-2$ and  $j'=1,\dots,n-2$. 
At the level of the superconformal index, starting from the index of the original model, the deconfined quiver is given by 
\begin{eqnarray}
&&
\frac{(p;p)_{\infty}^{3n-2}(q;q)_{\infty}^{3n-2}}{(2n)!(n-1)! 2^{n-1}}
 \Gamma_e(U_{\tilde A}^n)
 \int_{\mathbb{T}^{2n-1}} \prod_{i=1}^{2n-1} \frac{dz_i}{2\pi i z_i} 
  \prod_{\ell=1}^{n-1} 
\frac{dw_\ell}{2 \pi i w_\ell}
\nonumber \\
&&
\prod_{i=1}^{2n} \Big(
\prod_{a=1}^{4} \Gamma_e(z_i t_{a})
\prod_{b=1}^{2}
\Gamma_e(z_i^{-1} s_{b})\Big)
\frac{\prod_{i=1}^{2n} \prod_{\ell=1}^{n-1} \Gamma_e\big(z_i^{-1} w_\ell^{\pm 1} \sqrt {U_{\tilde A}}
\big)}
{
\prod_{\ell < k}
\Gamma_e 
(w_\ell^{\pm 1} w_k^{\pm 1} )
}
\nonumber \\
&&
\prod_{i<j}\frac{\Gamma_e\big(U_A z_i z_j\big)}{\Gamma_e((z_i/z_j)^{\pm 1})}
\frac{\prod_{i=1}^{2n} \prod_{\ell=1}^{n-1} \Gamma_e\big(z_i^{-1} w_\ell^{\pm 1} \sqrt {U_{\tilde A}}
\big)}
{
\prod_{\ell=1}^n \Gamma_e(w_\ell^{\pm 2})
}.
\end{eqnarray}
Then, confining the $\mathrm{SU}(2n)$ node we arrive at \eqref{findapolo}.

\begin{figure}
\begin{center}
  \includegraphics[width=8cm]{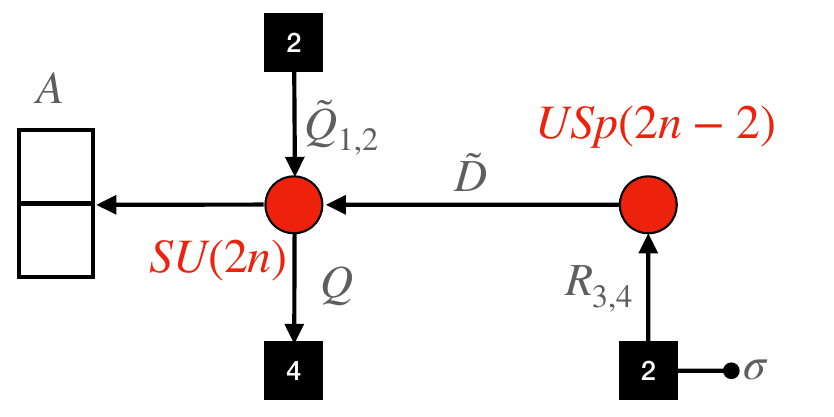}
  \end{center}
  \caption{Quiver gauge theory obtained by deconfining the conjugate antisymmetric 
  $\tilde A$  for the $\mathrm{SU}(2n)$ gauge theory with superpotential deformation (\ref{Wdef2SU2n}).}
    \label{figdec2def1}
\end{figure}

\item We conclude with the third deformation (\ref{Wdef3SU2n}).

The deconfined theory in this case corresponds to the quiver in Figure \ref{figusp2nm4} with superpotential
\begin{equation}
W =0.
\end{equation}
In this case we have used an $\mathrm{USp}(2n-4)$ gauge group in order to deconfine the antisymmetric $\tilde A$, that corresponds in the deconfined  model to the combination $\tilde A = \tilde P^2$.

The next step is to observe that the $\mathrm{SU}(2n)$ gauge theory is confining; indeed, it contains 2n antifundamentals, four fundamentals, and one antisymmetric. Actually its ``global" $\mathrm{SU}(2n)$ flavor symmetry in this case is broken to $\mathrm{USp}(2n-4)\times \mathrm{SU}(4)$, by the gauging and its gauge invariant degrees of freedom are
\begin{eqnarray}
&&
\eta_1=\mathrm{Pf} \, A
, \quad
\eta_2=A^{n-1}  Q^2
, \quad
 \eta_3=A^{n-2}  Q^4
 , \quad
 \eta_4=\tilde P^{2n-4}\tilde Q^4  \\
&&
M_{ab} = \tilde Q_a Q_b , \quad
R_{Q_a} = \tilde P Q_a , \quad
 B_{As}=A \tilde P^2
 , \quad
 B_S=A \tilde Q^2
 , \quad
 B_{V_{a}}=A \tilde  P \tilde Q_a, 
\nonumber
 \end{eqnarray}
where $B_{As}$ is a $\mathrm{USp}(2n-4)$ reducible antisymmetric, $B_V$ and $R_Q$ are $4+4$ $\mathrm{USp}(2n-4)$ fundamentals  and the other combinations are singlets of $\mathrm{USp}(2n-4)$.

The superpotential of this model is 
\begin{eqnarray}
\label{spotpf}
W &=& \eta_1 (R_Q^4 B^{n-4}B_S^2+R_Q^3 M B_V B^{n-4}B_S+R_Q^2 M^2 B^{n-4}B_V^2
+R_Q^2 M^2 B^{n-3}B_S
\nonumber \\
&+&
M^3 R_Q B_V B_{As}^{n-3}+M^4 B_{As}^{n-2})
+
\eta_2(M R_Q B_{AS}^{n-4} B_V^3 + M R_Q B_{AS}^{n-4} B_V B_S^2
\nonumber \\
&+&
M^2 B_{AS}^{n-3} B_V^2 + M^2 B_{AS}^{n-2}  B_S+R_Q^2 B_{As}^{n-3} B_S^2)
\nonumber \\
&+&
\eta_3(B_{As}^{n-2} B_S^2+B_{As}^{n-3} B_S B_V^2+B_{As}^{n-4}  B_V^4)+\eta_1 \eta_3 \eta_4+\eta_2^2  \eta_4.
 \end{eqnarray}

In order to complete the analysis we can also map the chiral ring operators 
of the electric theory and the ones of the magnetic dual. We found the following mapping
\begin{equation}
\begin{array}{c|c}
\mathrm{SU}(2n)  & \mathrm{USp}(2n-4)\\
\hline
 Q \tilde Q & M  \\
 Q (A \tilde A)^{k} \tilde Q &  B_{As}^{k-1} B_VR_Q \\ 
 \tilde A (A \tilde A)^{j}  Q^2 & B_{As}^j R_Q^2\\
 A   \tilde Q^2  & B_S
 \\
 A (A \tilde A)^{k}   \tilde Q^2  & B_{As}^{k-1} B_V^2
 \\ 
 (A \tilde A)^m & B_{As}^m\\
\mathrm{Pf} A & \eta_1\\
A^{n-1} Q^2& \eta_2\\
A^{n-2} Q^4& \eta_3\\
\tilde A^{n-2} \tilde Q^4& \eta_4\\
\end{array}
\end{equation}
with  $m=1,\dots,n-2$, $k=1,\dots,n-2$, $j=0,\dots,n-3$.
The electric superpotential \eqref{Wdef3SU2n} sets to zero, in the chiral ring,  the  operators involving
 $(n-1)$ factors of the antisymmetric $\tilde A$ and this forces the constraints on the labels $k$, $j$ and $m$ above. Consistently, in the dual $\mathrm{USp}(2n-4)$ theory the highest power of the antisymmetric $B_{As}$ contracted with two fundamentals is indeed $n-3$.

At the level of the superconformal index, starting from the index of the original model we have checked that by following the steps explained at field theory level, i.e. deconfining the conjugate antisymmetric and confining the 
$\mathrm{SU}(2n)$ we have recovered the result (\ref{deffinalPfA}) obtained above. We omit the details of the derivation leaving them to the interested reader.

\begin{figure}
\begin{center}
  \includegraphics[width=6cm]{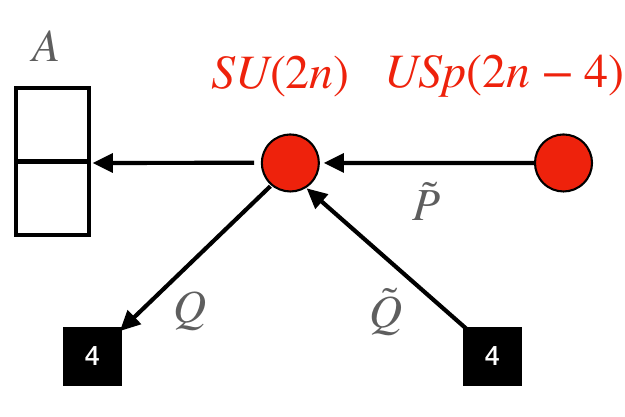}
  \end{center}
  \caption{Quiver gauge theory obtained by deconfining the conjugate antisymmetric 
  $\tilde A$  for the $\mathrm{SU}(2n)$ gauge theory with superpotential deformation (\ref{Wdef3SU2n}).}
    \label{figusp2nm4}
\end{figure}

\end{itemize}

\subsubsection{Phase structure of the dualities}

We have considered so far dualities between SQFTs with $\mathrm{SU}(2n)$ and $\mathrm{USp}(2m)$ gauge groups, in presence of matter fields in the fundamental and in the  antisymmetric representation, but we have not discussed the phase structure of such dualities.

The electric descriptions correspond to special unitary UV-free gauge theories that, in absence of the superpotential deformations \eqref{Wdef1SU2n},\eqref{Wdef2SU2n} and \eqref{Wdef3SU2n}, flow to a superconformal fixed point if some gauge invariant operators, hitting the bound of unitarity, are removed from  the chiral ring.  Let us review such removal as originally discussed in \cite{Kutasov:2003iy,Barnes:2004jj}.

Consider a gauge invariant operator $\mathcal{O}$ in the chiral ring. After performing the a-maximization procedure \cite{Intriligator:2003jj}, if we have $\Delta_{\mathcal{O}}<1$ the presence of such an operator in the spectrum is not consistent with the existence of an interacting fixed point. For this reason the operator needs to be removed from the chiral ring. 
In order to remove the operator we need to modify the UV description by adding two gauge singlets, say $L$ and $M$ respectively and considering the superpotential interaction
\begin{equation}
\label{UVmod}
W_{UV}= L(\mathcal{O} + \epsilon M)
\end{equation}
with a small UV coupling $\epsilon$.
Such a modified UV picture does not modify the IR fixed point if $\Delta_{\mathcal{O} }>1$. Indeed if we consider the UV theory with 
$\epsilon=0$ then $\Delta_M=1$ exactly.
If there is a fixed point with $\Delta_{\mathcal{O} }>1$ then (\ref{UVmod}) fixes $R_L<4/3$.  Using the fact that $R_M=2/3$ the second term in (\ref{UVmod}) is relevant and it can be integrated out.
The $F$ terms $F_{L,M}$ then impose $W_{UV}=0$ and this is the original description, that indeed does not require to add any extra singlets.
On the other hand, the modified UV picture becomes crucial in the case of 
$\Delta_{\mathcal{O} }<1$, because in such a case the second coupling in 
(\ref{UVmod})  becomes irrelevant, i.e. the fields $M$ are free and decoupled at the fixed point.
The surviving superpotential term sets the operator  $\mathcal{O}$, hitting the bound of unitarity, to zero in the chiral ring.  
The field $L$ is commonly denoted as a flipper in the literature (see \emph{e.g.} \cite{Benvenuti:2017lle}).
Observe that  flippers can be added in general also outside the conformal window, but here we have reviewed their role in taking care of accidental symmetries.

Coming back to the electric descriptions at hand we have observed that 
$\mathrm{SU}(2n)$ with an antisymmetric flavor and four fundamental flavors is conformal if we remove from the chiral ring the operators Pf $A$,  Pf $\tilde A$, $Q^2 (A \tilde A)^{0,\dots, j_{max}} \tilde A $,  $\tilde Q^2 (A \tilde A)^{0,\dots, j_{max}}  A $, $Q (A \tilde A)^{0,\dots, j_{max}-1}  \tilde Q$ and $(A \tilde A)^{0,\dots,k_{max}}$ where $j_{max}$ and $k_{max}$ depend on $n$. At large $n$ for example we have $j_{max} \sim 0.341 n$ and $k_{max} \sim 0.658 n$.

At such fixed point the superpotential deformations \eqref{Wdef1SU2n} and \eqref{Wdef2SU2n} are relevant, and they can trigger a flow to an IR fixed point, providing the fact that if further gauge invariant operators hit the bound of unitarity they need to be removed.
On the other hand, such a flow cannot be triggered by \eqref{Wdef3SU2n}, because this operator has been already removed from the chiral ring in order to reach the fixed point.

Nevertheless, the flows triggered by the relevant deformations \eqref{Wdef1SU2n} and \eqref{Wdef2SU2n} do not necessarily lead to  an IR interacting CFT, and this possibility needs to be checked explicitly for any gauge rank $n$ for both the deformations.
Alternatively, one can study the flow from the UV 
asymptotically-free theory directly by adding the (potentially dangerously) irrelevant  deformations \eqref{Wdef1SU2n}, \eqref{Wdef2SU2n} and \eqref{Wdef3SU2n}. In the first two cases the argument given above suggest that such deformations are actually dangerously irrelevant, while more work is necessary for the third case.

In general the analysis for the three deformations requires to determine the exact superconformal R-charges and central charges  through an a-maximization process, taking care of the possible accidental symmetries as well.

A careful study of the phase diagrams for the different theory as $n$ varies is then required, analyzing for which $n$ an operator in the chiral ring of the theory hits the unitary bound and becomes free. Unfortunately the rich matter content of the theories did not allow for such a detailed analysis for generic gauge rank $n$, so we relied on a case-by-case study, leaving for the future a more detailed comprehension of the phase diagrams.

 We flip in an iterative way the specific gauge invariant combinations hitting the unitarity bound with lowest R-charges. For many cases this is sufficient to prove the existence of a conformal window, but in some low-rank case  or for the deformation \eqref{Wdef3SU2n} we found obstructions to the existence of an interacting fixed point at the end of the flow triggered by such 
 deformations.

Let us finally comment upon the existence of a conformal window for the three different deformations.

\begin{itemize}
\item In the case of the deformation \eqref{Wdef1SU2n} we performed the analysis for $2 \leq n \leq 20 $ and found that for each $n$ is possible to flip part of the towers in \eqref{Eq:GISU2n2} to reach an interacting fixed point. The number of the flippers increases with $n$, 
but we did not recognize a generic pattern.
\item  In the case of the deformation \eqref{Wdef2SU2n} we carried out the analysis for $n \leq 11$ and we found that for $6\leq n \leq 11$ a conformal window is present after having flipped some part of the different towers in \eqref{Eq:GISU2n1}. For $2\leq n \leq 5$ the iterative procedure does not give rise to an interacting CFT.
\item  In the case of the deformation \eqref{Wdef3SU2n} for $3\leq n \leq 20$ the iterative procedure does not give rise to an interacting CFT.
This is consistent with the fact that the operator Pf $\tilde{A}$  is  removed from the chiral ring in order to reach a CFT starting with  $W=0$, signalling  that the deformation cannot become relevant at such fixed point.
\end{itemize} 

\subsection{$\mathrm{SU}(2n+1)$}
\label{subsec2np1}
Here we consider the case $N=2n+1$. There are three possible superpotential deformations. The first deformation is 
\begin{equation}
\label{Wdef1SU2np1}
W = \tilde{A}^{n-1} \tilde Q_2 \tilde Q_3\tilde Q_4,
\end{equation}
 the second deformation is 
\begin{equation}
\label{Wdef2SU2np1}
W = \tilde{A}^{n} \tilde Q_4,
\end{equation}
where in both cases the $\mathrm{SU}(4)$ flavor symmetry is explicitly broken by the deformation. 

In the following we will study the effect of each of these deformations in the IR behavior of the model.
Before distinguishing the two cases we can keep a common analysis by deconfining the antisymmetric in terms of another auxiliary $\mathrm{SU}(2n+1)$ gauge group, with an antisymmetric.

Here we deconfine the antisymmetric $ A$ and the fundamentals $Q$, by trading them with a $\mathrm{SU}(2n+1)_2$ gauge node, 
with a new antisymmetric $B$, an $\mathrm{SU}(2n+1)_1 \times \mathrm{SU}(2n+1)_2$  bifundamental $X_{12}$ and four $\mathrm{SU}(2n+1)_2$ fundamentals 
$q$. The charged field content of this deconfined phase appears in the second quiver in Figure \ref{figdec1su}, with $N= 2n+1$.
The original fields $A$ and $Q$ are mapped to the combinations $B X_{12}^2$ and $q X_{12}$ respectively.
Starting with vanishing superpotential there are also new singlets $\alpha_{1,2,3}$ in the dual phase, interacting with the charged fields through a superpotential 
\begin{equation}
\label{decflipp12np1}
W=\alpha_1 B^n q + \alpha_2  B^{n-1} q^3 + \alpha_3 X_{12} ^{2n+1}.
\end{equation}
At this level we did not turn on any superpotential deformation in the electric picture, because it can be done later, such that the discussion here will apply also in the analysis below, where the dangerously irrelevant deformations (\ref{Wdef1SU2np1}) and (\ref{Wdef2SU2np1})   will be considered.

Then we observe that the original $\mathrm{SU}(2n+1)_1$ gauge node is s-confining, and the confined degrees of freedom correspond to 
four  $\mathrm{SU}(2n+1)_2$ antifundamentals
$\tilde q = \tilde Q X_{12}$, an $\mathrm{SU}(2n+1)_2$ conjugate antisymmetric 
$\tilde B =\tilde A X_{12}^2$ and the $\mathrm{SU}(2n+1)_2$  singlets 
$\rho_1 = A^n Q$, $\rho_2 = \tilde A^{n-1} \tilde Q^3$ and 
$\rho_3=X_{12} ^{2n+1}$.

The charged field content of the  $\mathrm{SU}(2n+1)_2$ theory is represented in a quiver analog to the third one in Figure \ref{figdec1su} with $2n \rightarrow 2n+1$ and 
the superpotential, after integrating out the massive fields, is
\begin{equation}
\label{thirdspot2np1}
W=\rho_1 \tilde q^3 \tilde B^{n-1}+\rho_2 \tilde q \tilde B^n+\alpha_1 B^n q + \alpha_2  B^{n-1} q^3.
\end{equation}

At this point of the discussion we can introduce the electric  deformation given by the superpotential (\ref{Wdef1SU2np1}) and (\ref{Wdef2SU2np1}) respectively.
The first deformation is mapped to a linear superpotential $\rho_2$ in the $\mathrm{SU}(2n+1)_2$ theory and gives rise to the $F$-term
$\tilde q_1 \tilde B^{n}\neq 0$, while the second deformation is a linear term $\rho_1$ and gives the $F$-term $\tilde q_1 \tilde q_2 \tilde q_3 \tilde B^{n-1}\neq 0$.
In the first case the gauge group is broken to $\mathrm{USp}(2n)$ while in the second case it becomes $\mathrm{USp}(2n-2)$.
The index associated to the $\mathrm{SU}(2n+1)_2$ theory is given by:

\begin{align}
   \mathcal{I}=&\frac{\left(p;p\right)_\infty^{2n}\left(q;q\right)_\infty^{2n}}{(2n+1)!} \prod_{a=1}^{4}\Gamma_e \big( U_A^{n} \,t_a; U_{\tilde{A}}^{n} \,s_a;U_A^{n-1} \prod_{b\neq a}t_b;U_{\tilde{A}}^{n-1} \prod_{b\neq a}s_b \big)
   \nonumber \\
   &\int\displaylimits_{\mathbb{T}_{2n}} \! \prod_{i = 1}^{2n}\frac{\mathrm{d} \omega_i}{2\pi \mathrm{i} \omega_i}\prod_{i=1}^{2n+1}\prod_{a=1}^{4} \Gamma_e\left(\omega_i m_a;\omega_i^{-1} n_a\right) \!\! \prod_{ i<j}^{2n+1} \frac{\Gamma_e \left(\omega_i \omega_j U_B\right) \Gamma_e \left(\omega_i^{-1} \omega_j^{-1} U_{\tilde{B}} \right)}{\Gamma_e \left(\omega_i / \omega_j\right) \Gamma_e\left( \omega_j /\omega_i \right)},
   \label{eq:ind_suodd}
\end{align}
with $U_{\tilde{B}} = U_{\tilde{A}}v^2$, $U_B = U_A v^{-2}$, $n_a = s_a v$ and $m_a = t_a v^{-1}$, together with the balancing conditions arising from the cancellations of the gauge anomalies 
\begin{align}
   &\left(U_A  U_{\tilde{A}}\right)^{2n-1} \prod_{a=1}^{4} s_a t_a = (pq)^2, \nonumber \\
   &U_{\tilde{A}}^{2n-1} v^{2n+1} \prod_{a = 1}^{4} s_a  = pq, \nonumber \\
   & U_{B}^{2n-1}v^{2n+1} \prod_{a = 1}^{4} m_a  = pq. 
   \label{eq:bc_and_odd}
\end{align}

Turning on the two deformations imposes the following constraints:
\begin{enumerate}
\item (\ref{Wdef1SU2np1}) $\rightarrow$  $U_{\Bt}^{n} n_1 = 1$
\item (\ref{Wdef2SU2np1}) $\rightarrow$  $U_{\Bt}^{n-1} n_1 n_2 n_3 = 1$

\end{enumerate}
The contour integral involved in  \eqref{eq:ind_suodd} is pinched when these constraints are satisfied, and the integral can be partially resolved.
Below we analyze the pinching in the presence of the two deformations separately.

%
%
%
%
%
\subsubsection{Deconfinement and pole pinching}
%
%
%
%
%
Here we reproduce the dual Higgsing at the level of the superconformal index, separating the analysis for the two superpotential deformations  (\ref{Wdef1SU2np1}) and  (\ref{Wdef2SU2np1}).
In this way we find two different dualities between the original $\mathrm{SU}(2n+1)$ model equipped with one of these superpotential deformation and an $\mathrm{USp}(2m)$ gauge theory, with $m=n$ and $m=n-1$ respectively, an antisymmetric, eight fundamentals and a flipped superpotential. 

\begin{itemize}
\item  \underline{The case of  $W = \tilde{A}^{n-1}\tilde{Q}_2\tilde{Q}_3\tilde{Q}_4$}\\

The superpotential deformation \eqref{Wdef1SU2np1} imposes the further constraint
\begin{equation}
    U_{\tilde{A}}^{n-1}\,s_2 s_3 s_4 = pq.
\end{equation}
Similarly to the analysis of subsection \ref{sec:pinchev}, we define
\begin{equation}
    U_{\tilde{A}}^{n-1} s_2 s_3 s_4 \coloneqq pq e^\varepsilon, \quad U_{\tilde{B}}^{n} \, n_1 \coloneqq e^{-\varepsilon},
\end{equation}
such that the balancing conditions \eqref{eq:bc_and_odd} are satisfied. The effect of the superpotential deformation \eqref{Wdef1SU2np1} can now be studied by considering the limit $\varepsilon \to 0$ of the index.
We consider the following sequence of Gamma functions
\begin{equation}
    \prod_{i<j}^{2n}\Gamma_e\left(\omega_i^{-1}\omega_j^{-1} U_{\tilde{B}}\right).
\end{equation}
and focus on the family of poles
\begin{align}
    \omega_{i} \omega_{i+1} = U_{\tilde{B}}\, p^{k}q^{l},\quad i=1,\dots, 2n, \quad k,l \geq 0.
\end{align}
Consider also 
\begin{equation}
    \Gamma_e\left(\omega_{2n+1}^{-1} n_1\right) \implies \omega_{2n+1} = n_1. 
\end{equation}

Enforcing the $\mathrm{SU}(2n+1)$ constraint, the holonomies also satisfy
\begin{equation}
    \omega_{2n+1} = n_1 e^\varepsilon, \implies \Gamma_e\left(\omega_{2n+1}^{-1}n_1\right) = \Gamma_e\left(e^{-\varepsilon}\right),
\end{equation}
pinching the integration contour as the superpotential deformation is implemented by sending $\varepsilon \to 0$. 

Such sequence of poles allows for a partial evaluation of $n$ integrals out of the $2n$ ones, with a degeneracy factor $\frac{(2n+1)!}{2^{n} n!}$ from all the equivalent arrangements of the $2n+1$ variables, reproducing the expected Weyl group of $\mathrm{USp}(2n)$ gauge group.

After relabeling $y_i = \tfrac{\omega_{2i}}{\sqrt{U_{\tilde{B}}}} = \tfrac{\sqrt{U_{\tilde{B}}}}{\omega_{2i-1}}, \; i = 1,\dots,n $ the various charged fields contribute as:
\begin{align}
    q &\to \prod_{a=1}^{4}\prod_{i=1}^{n} \Gamma_e \left(y_i^{\pm 1} \sqrt{U_{\tilde{B}}}m_a\right) \prod_{a=1}^{4} \Gamma_e \left(m_a n_1\right) 
    \nonumber \\
    \tilde{q} &\to \Gamma_e \left(e^{-\varepsilon}\right) \prod_{a=1}^{4}\prod_{i=1}^{n} \Gamma_e \left(y_i^{\pm 1} \frac{n_a}{\sqrt{U_{\tilde{B}}}}\right) \prod_{a=2}^{4}\Gamma_e \left(\frac{n_a}{n_1}\right) 
    \nonumber \\
    B &\to \Gamma_e \left(U_{\tilde{B}}U_B\right)^{n} 
    \prod_{i<j}^{n} \Gamma_e \left(y_i^{\pm 1} y_j^{\pm 1}  U_{\tilde{B}}U_B \right) 
    \prod_{i=1}^{n}\Gamma_e \left(y_i^{\pm 1} n_1  \sqrt{U_{\tilde{B}}}U_B \right)
    \nonumber \\
    \tilde{B} &\to \prod_{i<j}^{n} \Gamma_e \left(y_i^{\pm 1} y_j^{\pm 1}\right) \prod_{i=1}^{n} \Gamma_e\left(y_i^{\pm 1} \frac{\sqrt{U_{\tilde{B}}}}{n_1}\right)
    \nonumber \\
    A &\to  \prod_{i<j}^{n} \Gamma_e \left(y_i^{\pm 1} y_j^{\pm 1} \right)^2  \prod_{i = 1}^{n}\Gamma_e \left(y_i^{\pm 2} \right)\prod_{i = 1}^{n}\Gamma_e \left(y_i^{\pm 1} \left(\frac{\sqrt{U_{\tilde{B}}}}{n_1}\right)^{\pm 1} \right).
    \nonumber \\
\end{align}
Combining all these contributions and simplifying some singlets after enforcing the balancing conditions
\begin{align}
    \prod_{a=2}^{4}\Gamma_e\left(U_{\tilde{A}}^{n-1}\prod_{b\neq 1,a}s_b s_1 \right) =&  
    \Gamma_e\left( pq \frac{n_1}{n_a}\right),
\end{align} 
we get
\begin{align}
\label{cubicidefndex2np1}
\mathcal{I}=&\frac{\left(p;p\right)_\infty^{n}\left(q;q\right)_\infty^{n}}{2^{n} n!} \prod_{a=1}^{4}\Gamma_e \left( U_A^{n} \,t_a ; U_{\tilde{A}}^{n} \,s_a;  U_A^{n-1} \prod_{b\neq a}^{3}t_b\right) \prod_{a=1}^{4} \Gamma_e \left(t_a s_1 \right) \nonumber \\
& \Gamma_e \left(U_A U_{\tilde{A}}\right)^{n} \int_{\mathbb{T}^{n}} \prod_{i=1}^{n}\frac{\mathrm{d}y_i}{2 \pi \mathrm{i} y_i }\frac{\prod\displaylimits_{1\leq i<j \leq n}\Gamma_e \left(y_i^{\pm 1} y_j^{\pm 1} U_A U_{\tilde{A}}\right)}{\prod\displaylimits_{1\leq i<j \leq n}\Gamma_e \left(y_i^{\pm 1} y_j^{\pm 1}\right) \prod\displaylimits_{i=1}^{n} \Gamma_e\left( y_i^{\pm 2} \right)} 
\nonumber \\
& \prod_{b=1}^{4} \Gamma_e\left( y_i^{\pm 1}t_b \sqrt{U_{\tilde{A}}}\right) \prod_{i=1}^{n} \Gamma_e \left(y_i^{\pm 1} s_1 \sqrt{U_{\tilde{A}}}U_A \right) \prod_{a=2}^{4} \Gamma_e \left(y_i^{\pm 1} \frac{s_a} {\sqrt{U_{\tilde{A}}}}\right).
\end{align}

The result defines a $\mathrm{USp}(2n)$ gauge theory with 8 fundamentals with fugacities $\vec{u}$ and a totally antisymmetric with fugacity $U_X$ with
\begin{equation}
U_X = U_A U_{\tilde{A}},
\qquad
\vec{u} = \left(
t_b \sqrt{U_{\tilde{A}}},\, s_1 \sqrt{U_{\tilde{A}}}U_A,\, \frac{s_a}{\sqrt{U_{\tilde{A}}}}
\right).
\end{equation}
\item \underline{The case of  $W = \tilde{A}^{n}\tilde{Q}_4$ } \\
The superpotential deformation \eqref{Wdef2SU2np1} imposes the further constraint
\begin{equation}
    U_{\tilde{A}}^{n}\,s_4 = pq.
\end{equation}
We define
\begin{equation}
    U_{\tilde{A}}^{n}s_4 \coloneqq pq e^\varepsilon, \quad U_{\tilde{B}}^{n-1}\prod_{a=1}^{3}n_a \coloneqq e^{-\varepsilon},
\end{equation}
We consider the following sequence of Gamma functions
\begin{equation}
    \prod_{i<j}^{2n-2}\Gamma_e\left(\omega_i^{-1}\omega_j^{-1} U_{\tilde{B}}\right).
\end{equation}
and focus on the family of poles
\begin{align}
    \omega_{i} \omega_{i+1} = U_{\tilde{B}}\, p^{k}q^{l},\quad i=1,\dots, 2n-2, \quad k,l \geq 0.
\end{align}
Consider also 
\begin{equation}
    \Gamma_e\left(\omega_{2n-2+a}^{-1} n_a\right) \implies \omega_{2n-2+a} = n_a \quad a=1,2,3. 
\end{equation}

Enforcing the $\mathrm{SU}(2n+1)$ constraint, the holonomies also satisfy
\begin{equation}
    \omega_{2n+1} = n_3 e^\varepsilon, \implies \Gamma_e\left(\omega_{2n+1}^{-1}n_3\right) = \Gamma_e\left(e^{-\varepsilon}\right),
\end{equation}
pinching the integration contour as the superpotential deformation is implemented by sending $\varepsilon \to 0$. 

Such sequence of poles allows for a partial evaluation of $n+1$ integrals out of the $2n$ ones reproducing a $\mathrm{USp}(2n-2)$ gauge group, after accounting for the usual $\frac{(2n+1)!}{2^{n-1}(n-1)!}$ degeneracy factor.

After relabeling $y_i = \tfrac{\omega_{2i}}{\sqrt{U_{\tilde{B}}}} = \tfrac{\sqrt{U_{\tilde{B}}}}{\omega_{2i-1}}, \; i = 1,\dots,n-1 $ the various charged fields contribute as:
\begin{align}
    q & \to \prod_{a=1}^{4}\prod_{i=1}^{n-1} \Gamma_e \left(y_i^{\pm 1} \sqrt{U_{\tilde{B}}}m_a\right) \prod_{a=1}^{4}\prod_{b=1}^{3} \Gamma_e \left(m_a n_b\right) 
    \nonumber \\
    \tilde{q} & \to \prod_{a=1}^{4}\prod_{i=1}^{n-1} \Gamma_e \left(y_i^{\pm 1} \frac{n_a}{\sqrt{U_{\tilde{B}}}}\right) \Gamma_e \left(e^{-\varepsilon}\right) \prod_{a\neq b}^{3} \Gamma_e \left(\frac{n_a}{n_b}\right) \prod_{a=1}^{3}\Gamma_e \left(\frac{n_4}{n_a}\right) 
    \nonumber \\
    B & \to \Gamma_e \left(U_{\tilde{B}}U_B\right)^{n-1} \prod_{i<j}^{n-1} \Gamma_e \left(y_i^{\pm 1} y_j^{\pm 1}  U_{\tilde{B}}U_B \right)\prod_{a=1}^{3} \prod_{i=1}^{n-1}\Gamma_e \left(y_i^{\pm 1} n_a  \sqrt{U_{\tilde{B}}}U_B \right) \prod_{a<b}^{3} \Gamma_e\left(n_a n_b U_B\right)
    \nonumber \\
    \tilde{B} &\to \prod_{i<j}^{n-1} \Gamma_e \left(y_i^{\pm 1} y_j^{\pm 1}\right) \prod_{a=1}^{3}\prod_{i=1}^{n-1} \Gamma_e\left(y_i^{\pm 1} \frac{\sqrt{U_{\tilde{B}}}}{n_a}\right) \prod_{a<b}^{3}\Gamma_e\left(\frac{U_{\tilde{B}}}{n_a n_b}\right)
    \nonumber \\
    A & \to  \prod_{i<j}^{n-1} \Gamma_e \left(y_i^{\pm 1} y_j^{\pm 1} \right)^2  \prod_{i = 1}^{n-1}\Gamma_e \left(y_i^{\pm 2} \right)\prod_{a \neq b}^{3}\Gamma_e \left(\frac{n_a}{n_b}\right) \prod_{a = 1}^{3}\prod_{i = 1}^{n-1}\Gamma_e \left(y_i^{\pm 1} \left(\frac{\sqrt{U_{\tilde{B}}}}{n_a}\right)^{\pm 1} \right).
    \nonumber \\
\end{align}
Combining all these contributions and simplifying some singlets after enforcing the balancing conditions
\begin{align}
    \prod_{a=1}^{3}\Gamma_e\left(U_{\tilde{A}}^{n-1}\prod_{b\neq 4,a}s_b s_4 \right) =& \prod_{a<b}^{3}  
    \Gamma_e\left( pq \frac{n_a n_b}{U_{\tilde{B}}}\right), \nonumber \\
    \prod_{b=1}^{3}\Gamma_e\left(U_{\tilde{A}}^{n}\,s_b \right) =& \prod_{a = 1}^{3}\Gamma_e\left( pq\, \frac{n_a}{n_4}\right), \nonumber \\
\end{align} 
we get
\begin{align}
\label{linearidefndex2np1}
\mathcal{I}=&\frac{\left(p;p\right)_\infty^{n-1}\left(q;q\right)_\infty^{n-1}}{2^{n-1} (n-1)!} \prod_{a=1}^{4}\Gamma_e \left( U_A^{n} \,t_a ; U_A^{n-1} \prod_{b\neq a}^{3}t_b\right) \Gamma_e \left(U_{\tilde{A}}^{n-1} s_1 s_2 s_3 \right) \prod_{a<b}^{3} \Gamma_e(U_A s_a s_b) \nonumber \\
& \prod_{a=1}^{4} \prod_{b=1}^{3}\Gamma_e \left(t_a s_b \right) \Gamma_e \left(U_A U_{\tilde{A}}\right)^{n-1}
\int_{\mathbb{T}_{n-1}} \prod_{i=1}^{n-1}\frac{\mathrm{d}y_i}{2 \pi \mathrm{i} y_i }\frac{\prod\displaylimits_{1\leq i<j \leq n}\Gamma_e \left(y_i^{\pm 1} y_j^{\pm 1} U_A U_{\tilde{A}}\right)}{\prod\displaylimits_{1\leq i<j \leq n}\Gamma_e \left(y_i^{\pm 1} y_j^{\pm 1}\right) \prod\displaylimits_{i=1}^{n} \Gamma_e\left( y_i^{\pm 2} \right)} 
\nonumber \\
&\prod_{i=1}^{n-1} \prod_{a=1}^{3} \Gamma_e \left(y_i^{\pm 1} s_a \sqrt{U_{\tilde{A}}}U_A \right) \prod_{b=1}^{4} \Gamma_e\left( y_i^{\pm 1}t_b \sqrt{U_{\tilde{A}}}\right) \Gamma_e \left(y_i^{\pm 1} \frac{s_4} {\sqrt{U_{\tilde{A}}}}\right).
\end{align}

The result is compatible with a $\mathrm{USp}(2n-2)$ gauge theory with 8 fundamentals with fugacities $\vec{u}$ and a totally antisymmetric with fugacity $U_X$ with
\begin{equation}
U_X = U_A U_{\tilde{A}},
\qquad
\vec{u} = \left(
t_a \sqrt{U_{\tilde{A}}},\, s_b \sqrt{U_{\tilde{A}}}U_A,\, \frac{s_4}{\sqrt{U_{\tilde{A}}}}
\right).
\end{equation}

\end{itemize}

\subsubsection{An alternative deconfinement}

Analogously to the case of $\mathrm{SU}(2n)$ here we study the dualities just obtained by a different deconfinement of the antisymmetric tensors $\tilde A$.
Again the two cases deserve a different analysis.
\begin{itemize}
\item We start by considering the first deformation (\ref{Wdef1SU2np1}).

The deconfined theory in this case corresponds to the quiver in Figure \ref{fig:Alternative3} with superpotential
\begin{equation}
W = \epsilon^{abc } C_a \tilde R_b  \tilde R_c
\end{equation}
with $a,b,c=2,3,4$. 
In this case we have used a $\mathrm{USp}(2n)$ gauge group in order to deconfine the antisymmetric $\tilde A$, that corresponds in the deconfined  model to the combination $\tilde A = \tilde P^2$. Furthermore, the antifundamentals $\tilde Q$ correspond in the deconfined quiver to the combinations $\tilde Q_{234} = \tilde P \tilde R_{234} $.
The  $\mathrm{SU}(3)$ fundamental  $C$ is crucial in order to reproduce the superpotential deformation
 (\ref{Wdef1SU2np1}) of the original theory. Indeed, if we confine the $\mathrm{USp}(2n)$ node we obtain the original $\mathrm{SU}(2n+1)$ model with superpotential 
\begin{equation}
W =  \tilde A^{n} \sum_{i=2}^4 \mathcal{C}^{\overline i} \tilde Q_i + \tilde A^{n-1} \tilde Q_2  \tilde Q_3 \tilde Q_4+  C \mathcal{C} 
\end{equation}
with $\mathcal{C} = \tilde R^2$.
This superpotential coincides with (\ref{Wdef1SU2np1}) after integrating out the massive fields $\mathcal{C} $ and $C$. Moreover, by solving these $F$-terms we find that the singlet $C_a$ in the deconfined phase coincides with the operator $\tilde A^{n} \tilde Q_a$.

\begin{figure}
\begin{center}
  \includegraphics[width=13cm]{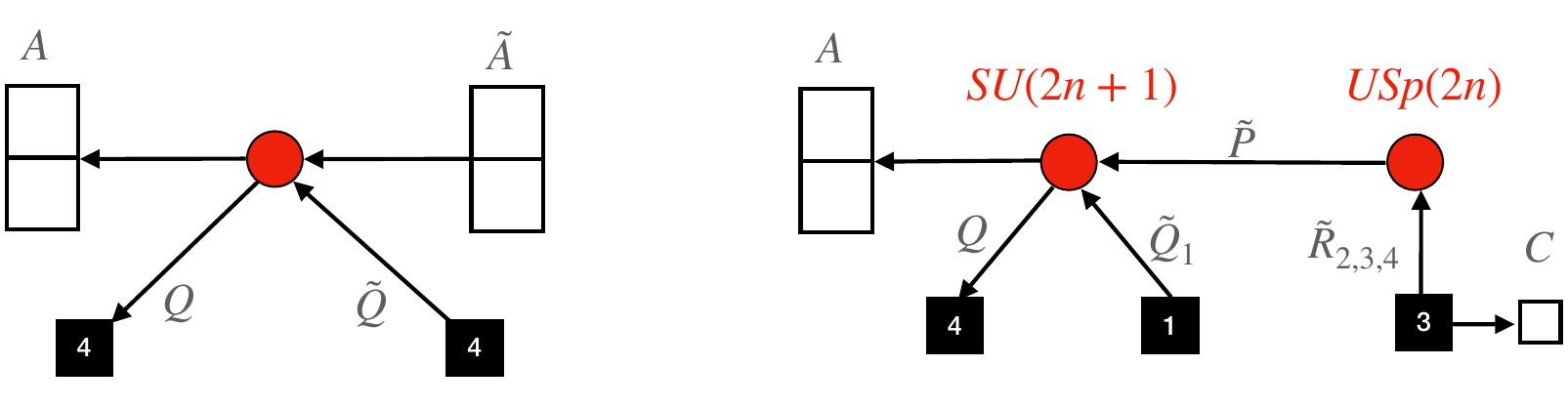}
  \end{center}
  \caption{Quiver gauge theory obtained by deconfining the conjugate antisymmetric 
  $\tilde A$  for the $\mathrm{SU}(2n+1)$ gauge theory with superpotential deformation (\ref{Wdef1SU2np1}).}
    \label{fig:Alternative3}
\end{figure}

The next step consists of observing that the $\mathrm{SU}(2n+1)$ gauge theory is confining, indeed it has $2n+1$ antifundamentals, four fundamentals and an antisymmetric. 
The gauge invariant degrees of freedom are
\begin{eqnarray}
&&
\eta_1=A^{n}  Q
, \quad
\eta_2=A^{n-1}  Q^3
 , \quad
 \eta_3=\tilde P^{2n} \tilde Q_1
,\quad
R_Q =\tilde P Q 
, \quad\nonumber \\
&&
 B_{As}=A \tilde P^2
 , \quad
  B_{V}=A \tilde P \tilde Q_1
 , \quad
M_1=\tilde Q_1 Q,
 \end{eqnarray}
where $B_{As}$ is an antisymmetric, $R_Q$ and $B_V$ are fundamentals  and the other combinations are singlets of $\mathrm{USp}(2n)$.

The superpotential of this model is 
\begin{equation}
W=\eta_1 (M_1 R_Q^2 B_{As}^{n-1}+R_Q^3 B_{As}^{n-2} B_V)+\eta_2 (M_1 B_{AS}^{n} + B_V R_Q B_{AS}^{n-1}) +\eta_1 \eta_2 \eta_3+ R^2 C.
 \end{equation}

In order to complete the analysis we can also map the chiral ring operators 
of the electric theory and the ones of the magnetic dual. We found the following mapping
\begin{equation}
\begin{array}{c|c}
\mathrm{SU}(2n+1) & \mathrm{USp}(2n) \\
\hline
Q \tilde Q_1       &    M_1\\
Q(A \tilde A)^{k'} \tilde Q_1       &    B_{As}^{k'-1} R_Q B_V \\
Q (A \tilde A)^k \tilde Q_{a}      &   B_{As}^k R_Q  \tilde R_{a}\\
\tilde A(A \tilde A)^k Q^2             &   B_{As}^k R_Q^2 \\
 A(A \tilde A)^k \tilde Q_1 \tilde Q_{a}  &     B_{As}^k B_V \tilde R_{a}\\
 A(A \tilde A)^\ell  \tilde Q_{[a} \tilde Q_{b]}   &     B_{As}^{\ell+1}   \tilde R_{[a} \tilde R_{b]} \\
(A \tilde A)^m     &     B_{As}^m\\
A^n Q                 &       \eta_1\\
\tilde A^n \tilde Q_1 &    \eta_3\\
\tilde A^n \tilde Q_a   &      C\\
A^{n-1} Q^3               &      \eta_2\\
 \end{array}
 \label{Eq:GISU2np11}
 \end{equation}
 with $k=0,\dots,n-1$, $k'=1,\dots,n-1$, $\ell=0,\dots,n-2$, $m=1,\dots,n$ and $a,b=2,\dots,4$.
 
 The operators $(\tilde A^{n-1} \tilde Q_1 \tilde Q_{[a} \tilde Q_{b]}  )$ are set to zero on the chiral ring by the F-term equations of $\tilde Q_{c}$, with $c \neq a,b,1$, due to the electric superpotential \eqref{Wdef1SU2np1}.

At the level of the superconformal index, starting from the index of the original model we have checked that by following the steps explained at field theory level, i.e. deconfining the conjugate antisymmetric and confining the 
$\mathrm{SU}(2n+1)$ we have recovered the result (\ref{cubicidefndex2np1}) obtained above. We omit the details of the derivation leaving them to the interested reader.

\item The second and last  deformation corresponds to (\ref{Wdef2SU2np1}).

\begin{figure}
\begin{center}
  \includegraphics[width=13cm]{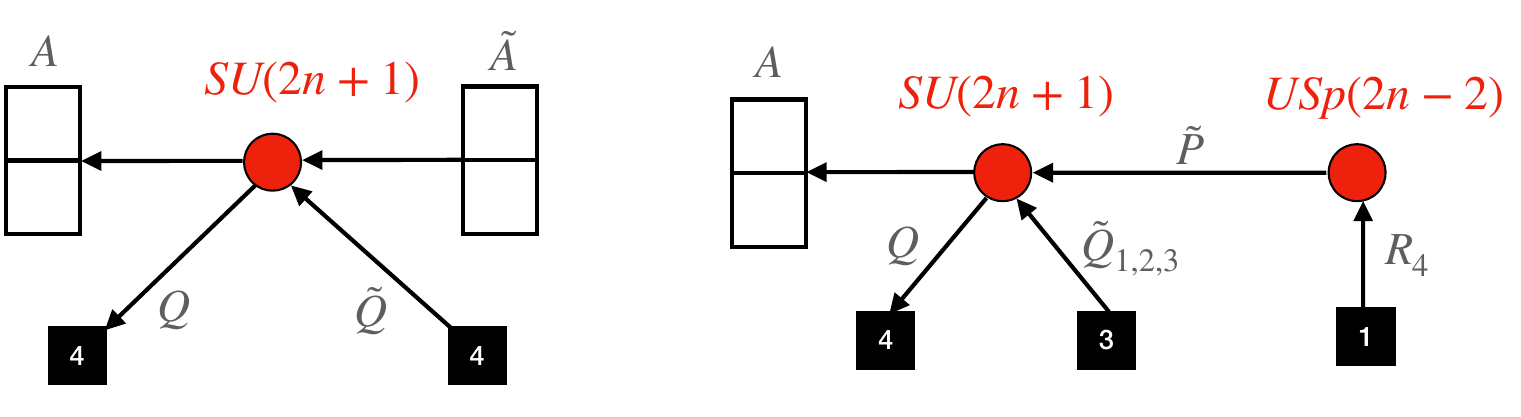}
  \end{center}
  \caption{Quiver gauge theory obtained by deconfining the conjugate antisymmetric 
  $\tilde A$  for the $\mathrm{SU}(2n+1)$ gauge theory with superpotential deformation (\ref{Wdef2SU2np1}).}
    \label{fig:Alternative4}
\end{figure}

The deconfined theory in this case corresponds to the quiver in Figure  \ref{fig:Alternative4} with superpotential
\begin{equation}
W = 0.
\end{equation}
In this case we have used a $\mathrm{USp}(2n-2)$ gauge group in order to deconfine the antisymmetric $\tilde A$, that corresponds in the deconfined model to the combination $\tilde A = \tilde P^2$. Furthermore, the antifundamental $\tilde Q_4$ corresponds in the deconfined quiver to the combinations $\tilde Q_{4} = \tilde P \tilde R_{4} $.

The next step consists of observing that the $\mathrm{SU}(2n+1)$ gauge theory is confining, indeed it has $2n+1$ antifundamentals, four fundamentals and an antisymmetric. 
The gauge invariant degrees of freedom are
\begin{eqnarray}
&&
\eta_1=A^{n}  Q
, \quad
\eta_2=A^{n-1}  Q^3
 , \quad
 \eta_3=\tilde P^{2n-2} \tilde Q_1\tilde Q_2\tilde Q_3
,\quad
R_Q =\tilde P Q 
, \quad\nonumber \\
&&
B_{S_{ab}} = A  \tilde Q_a \tilde Q_b
, \quad
 B_{As}=A \tilde P^2
 , \quad
  B_{V_a}=A \tilde P \tilde Q_a
 , \quad
M_a=\tilde Q_a Q
 \end{eqnarray}
 with $a,b=1,2,3$ and 
where $B_{As}$ is an antisymmetric, $R_Q$ and $B_V$ are fundamentals  and the other combinations are singlets of $\mathrm{USp}(2n-2)$.
The superpotential of this model is 
\begin{eqnarray}
W&=&\eta_1(M^2 R_Q B_V B_{As}^{n-3} B_{S}+M^3 B_{As}^{n-2} B_{S})\nonumber \\
&+&
\eta_2(B_{As}^{n-1} B_S M+R_Q B_V B_{As}^{n-2} B_S)+\eta_1 \eta_2 \eta_3.
 \end{eqnarray}
In order to complete the analysis we can also map the chiral ring operators 
of the electric theory and the ones of the magnetic dual. We found the following mapping
\begin{equation}
\begin{array}{c|c}
\mathrm{SU}(2n+1) & \mathrm{USp}(2n-2) \\
\hline
 Q\tilde Q_a  &  M_a  \\
Q(A \tilde A)^{k'} \tilde Q_a  &  B_{As}^{k'-1} R_Q  B_{V_a}  \\
Q(A \tilde A)^\ell \tilde Q_4  & B_{As}^\ell R_Q R_{4}\\
\tilde A(A \tilde A)^\ell Q^2 &  B_{As}^\ell R_Q^2 \\
 A(A \tilde A)^\ell \tilde Q_a\tilde Q_4   & B_{As}^\ell B_{V_a} R_{4}\\
 A\tilde Q_{[a}  \tilde Q_{b]}    &   B_{S_{ab}} \\
  A(A \tilde A)^{k'} \tilde Q_{[a}  \tilde Q_{b]}    &   B_{As}^{k'-1}B_{V_a} B_{V_b}\\
(A \tilde A)^m    &              B_{As}^m\\
A^n Q                  &             \eta_1\\
A^{n-1} Q^3             &           \eta_2\\
\tilde A^{n-1} \tilde Q_1 \tilde Q_2 \tilde Q_3          &      \eta_3\\
 \end{array}
 \label{Eq:GISU2np12}
 \end{equation}
 with $k=0,\dots,n-1$, $\ell=0,\dots,n-2$, $k'=1,\dots,n-1$, $m=1,\dots,n-1$ and $a,b=1,\dots,3$.
 The operators $(\tilde A^n \tilde Q_a)$ and $(\tilde A^{n-1} \tilde Q_{[b} \tilde Q_{c]} \tilde Q_4) $ are set to zero in the chiral ring by the F-terms of $\tilde{Q}_4$ and $\tilde{A}$, respectively, due to the electric superpotential $\mathcal{W}=\tilde{A}^{n} \tilde{Q}_4$. 

At the level of the superconformal index, starting from the index of the original model we have checked that by following the steps explained at field theory level, i.e. deconfining the conjugate antisymmetric and confining the 
$\mathrm{SU}(2n+1)$ we have recovered the result (\ref{linearidefndex2np1}) obtained above. We omit the details of the derivation leaving them to the interested reader.

\end{itemize}

\subsubsection{Phase structure of the dualities}
Also in this case the a-maximization procedure is in order. The general comments are the same as in the even case, so we refer the reader to that section for the general analysis.
\begin{itemize}
\item For the first deformation \eqref{Wdef1SU2np1} we performed the a-maximization for $1 \leq n \leq 10$ and we found that also here an interacting CFT can exist only when we flip parts of the operator towers in \eqref{Eq:GISU2np11}. As always the number of flippers increases with $n$.
\item We studied the second deformation \eqref{Wdef2SU2np1} for $2 \leq n \leq 10$ and we found that for $n \geq 6$ an interacting CFT can exist only when we flip parts of the operator towers in \eqref{Eq:GISU2np12}, while for $n = 5$ we found that 
a fixed point can exist  when all the operators in the chiral ring are flipped. Finally, for $n=2,3,4$,  it is not possible to have and interacting CFT.
\end{itemize}
%
%
%
%
%
\section{3d reduction}
\label{sec:3d}
%
%
%
%
%
In this section we study the reduction to three dimensions of the dualities found above in 4d. We follow the ARSW prescription \cite{Aharony:2013dha}, i.e. we first obtain an effective 
duality   on $S^1$. This duality has the same field content of the 4d parent, but in addiction there is a monopole superpotential (a KK monopole in such case) that enforces the same constraints on the global symmetries imposed by the anomalies in 4d.
Then we perform, when possible, a real mass flow, integrating out some of the matter fields and  removing the monopole superpotential. We focus only on cases that give origin to new 3d dualities\footnote{
Furthermore, we will not discuss the reduction of $\mathrm{SU}(2n)$ with $W =$ Pf $\tilde A$.}. Such models have  two types of singlets in the dual phases, i.e. mesons and electric monopoles. These last are ubiquitous in 3d dualities in absence of monopole superpotentials and CS deformations, and originate from the real mass flow discussed above, arising from massless combinations of mesons associated to massive charged fields in the electric phase.

\subsection{$\mathrm{SU}(2n)$ with $W=\tilde A^{n-2} \tilde Q^4$}

The effective reduced duality reflects in the matching of the $3d$ partition function on the squashed three-sphere
\footnote{Here and in the rest of the paper we adopt the notation spelled out in \cite{Amariti:2024gco} to identify the gauge and matter content of the squashed three partition functions of \cite{Hama:2011ea}.}
\begin{eqnarray}
\label{onS1defo1}
&&
Z_{\mathrm{SU}(2n)}^{(4;4;\cdot;1;1;\cdot;\cdot)} (\vec \mu;\vec \nu;\cdot;\tau_A;\tau_{\tilde A};\cdot;\cdot)
=
\Gamma_h\left((n-2)\tau_A +\sum_{a=1}^4 \mu_a\right)
\Gamma_h(\tau_A+\tau_{\tilde A})
\Gamma_h(n\tau_A,n\tau_{\tilde A})
\nonumber \\
&&
\prod_{a<b} \Gamma_h((n-1)\tau_A+\mu_a+\mu_b,2\omega+\tau_{\tilde A}-\nu_a-\nu_b)
Z_{\mathrm{USp}(2n)}^{(8;\cdot;1)} \left(\vec \nu-\frac{\tau_{\tilde A}}{2},\vec \mu+\frac{\tau_{\tilde A}}{2}; \tau_{\tilde A}+\tau_{ A } \right).
\nonumber \\
\end{eqnarray}
Observe that in the RHS we added the contribution of the singlet 
with  mass $\tau_A+\tau_{\tilde A}$, stressing that the antisymmetric in the argument of $Z_{\mathrm{USp}(2n)}$ is irreducible. The same comment applies to all the cases below.

The identity (\ref{onS1defo1}) is valid provided two constraints are satisfied by the mass parameters
\begin{equation}
\label{BCsuusp1}
(n-2)\tau_{\tilde A}+\sum_{a=1}^4\nu_a=2\omega
,\quad
(2n-2) (\tau_{\tilde A} +\tau_A ) + \sum_{a=1}^4 (\mu_a +\nu_a)= 4 \omega,
\end{equation}
and it corresponds to an effective duality with the same field content of the 4d model, interacting with the same superpotential (forcing the first constraint in (\ref{BCsuusp1})) in addition to the contribution of the KK monopole, that indeed enforces the second constraint in
(\ref{BCsuusp1}).

It is possible to remove the effects from the monopole superpotential by suitable real mass flows. 
For example, we can assign large and opposite masses to a pair of fundamentals, obtaining a $\mathrm{SU}(2n)$ theory with an antisymmetric flavor, two fundamentals and four antifundamentals, again with the superpotential $W = \tilde A^{n-2} \tilde Q^4$.

The $\mathrm{USp}(2n)$ dual theory in this case has six fundamentals and one antisymmetric in addition to various baryonic singlets. There are also two new types of singlet originating from the former baryons $A^{n-2} Q^4$ and 
$A^{n-1} Q_3 Q_4$.
We conclude observing that  this last duality can be proven directly using tensor deconfinement in 3d. 

\subsection{$\mathrm{SU}(2n)$ with $W=\tilde A^{n-1} \tilde Q^2$}
\label{defnmen1sus1}

The effective reduced duality reflects in the matching of the $3d$ partition function on the squashed three sphere
\begin{eqnarray}
\label{onS1sefnm1}
&&
Z_{\mathrm{SU}(2n)}^{(4;4;\cdot;1;1;\cdot;\cdot)} (\vec \mu;\vec \nu;\cdot;\tau_A;\tau_{\tilde A};\cdot;\cdot)
=
\Gh[n \tau_A,n \tau_{\tilde A},(n-2)\tau_A+\sum_{a=1}^{4} \mu_a,(n-1) \tau_{\tilde A}+\nu_1 +\nu_2] 
\nonumber \\
&&
\Gamma_h(\tau_A+\tau_{\tilde A})
\Gh[\tau_A+\nu_1+\nu_2]
\prod_{1\leq a<b\leq 4} \Gh[(n-1)\tau_A+\mu_a +\mu_b]
\prod_{a=1}^{4} \prod_{r=1,2} \Gh[\mu_a +\nu_r]
\nonumber \\
&&
Z_{\mathrm{USp}(2n-2)}^{(8;\cdot;1)} \left(\nu_{1,2}+\tau_A+\frac{\tau_{\tilde A}}{2},\nu_{3,4}-\frac{\tau_{\tilde A}}{2},\vec \mu +\frac{\tau_{\tilde A}}{2};\cdot; \tau_{\tilde A}+\tau_{ A } \right).
\end{eqnarray}
The identity is valid provided two constraints are satisfied by the mass parameters
\begin{equation}
\label{BAonS1sefnm1}
(2n-2) (\tau_{\tilde A} +\tau_A ) + \sum_{a=1}^4 (\mu_a +\nu_a)= 4 \omega,\quad
(n-1) \tau_{\tilde A} +\nu_3+\nu_4=2\omega,
\end{equation}
which descend from the 4d balancing conditions imposed by the cancellation of the axial anomaly and by the superpotential (\ref{Wdef2SU2n}) respectively.
Here, while the second constraint is still imposed by the superpotential deformation, the first constraint is imposed by a linear monopole deformation, corresponding to the KK monopole.

The effective duality discussed above can be further reduced to a pure 3d duality, by removing the linear monopole superpotential through a real mass flow.

Such real mass flow corresponds to assigning large opposite real masses to the antifundamentals $\tilde Q_1$ and $\tilde Q_2$.

On the dual side two $\mathrm{USp}(2n-2)$ fundamentals are integrated out as well. Furthermore, the singlets $\tilde A^{n-1}  \tilde Q_1 \tilde Q_2$ and $A \tilde Q_1 \tilde Q_2$ are massless, and they are left in the IR spectrum as (dressed) monopoles.

\begin{figure}
\begin{center}
  \includegraphics[width=15.5cm]{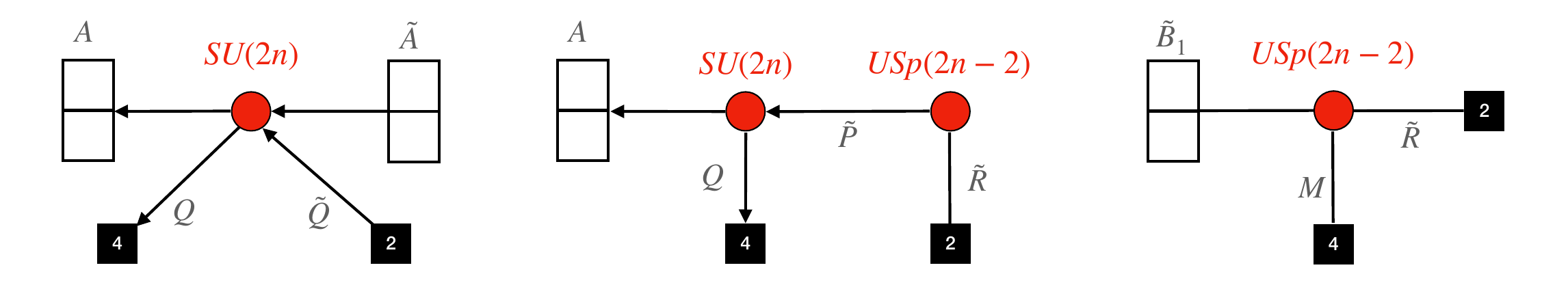}
  \end{center}
  \caption{Schematic description of the derivation of the SU/USp duality for the 3d 
  $SU(2n)$ model with an antisymmetric pair, four fundamentals and two antifundamentals, through tensor deconfinement and elementary dualities.}
    \label{fig:decASpure3d1}
\end{figure}

Alternatively, we can study the duality directly at 3d level by deconfining the antisymmetric tensors and then by dualizing the original $\mathrm{SU}(2n)$ gauge node. 
Such procedure can be schematically represented with the aim of the quiver description in Figure \ref{fig:decASpure3d1}.
The original gauge theory has superpotential $W = \tilde A^{n-1} \tilde Q^2$.
then we deconfine the antisymmetric $\tilde A$, by considering an $\mathrm{USp}(2n-2)$ gauge theory with superpotential $W = Y_{\mathrm{USp}(2n-2)} + \sigma \tilde R^2$.

In this case there is a monopole superpotential for the $\mathrm{USp}(2n-2)$ gauge group in the second quiver of Figure \ref{fig:decASpure3d1} and the flipper $\sigma$ corresponds to $ \mathrm{Pf} \tilde A$. 
This can be shown by re-confining the $\mathrm{USp}(2n-2)$ gauge node and obtaining an $\mathrm{SU}(2n)$ gauge theory with superpotential 
\begin{equation}
W = \sigma s + \tilde A^{n-1} \tilde Q^2 + s \mathrm{Pf} \tilde A.
\end{equation}
By integrating massive fields we obtain the original superpotential and in
addition the relation $\sigma = \mathrm{Pf} \tilde A$.
At the level of the three sphere partition function we observe that the singlet $\sigma$ has mass parameter
\begin{equation}
m_\sigma = 2\omega - m_{\tilde R_1}- m_{\tilde R_2},
\end{equation}
where
$m_{\tilde R_{1,2}} = \nu_{3,4}- \frac{\tau_{\tilde A}}{2}$.
In this case the superpotential imposes the balancing condition 
\begin{equation}
\nu_3 +\nu_4 +(n-1) \tau_{\tilde A} = 2 \omega,
\end{equation}
which implies $m_\sigma = n \tau_{\tilde A}$, that is consistent with the duality map $\sigma = \mathrm{Pf} \tilde A$.

The next step consists of confining the $\mathrm{SU}(2n)$ gauge theory, with four fundamentals, $2n-2$ antifundamentals, an antisymmetric and vanishing superpotential.
Such confining duality was originally proposed in \cite{Nii:2019ebv} and it can be obtained by dimensional reduction of the 4d confining duality  for  $\mathrm{SU}(2n)$ with 
four fundamentals, $2n-2$ antifundamentals, an antisymmetric and vanishing superpotential of \cite{Csaki:1996zb}. By following the ARSW prescription one first reduces on $S^1$, with a KK monopole superpotential and then assigns two opposite real masses to a pair of antifundamentals. In this way one finds a pure 3d confining duality with two Coulomb branch variables (dressed monopole operators) that originate from the massless baryonic variables that involve the two massive antifundamentals.
In this case such two dressed monopoles correspond to the combinations
\begin{equation}
Y_{A}^{dressed} = Y_{\mathrm{SU}(2n-2)}^{(bare)} A,
\quad
Y_{\tilde P^{2n-2}}^{dressed} = Y_{\mathrm{SU}(2n-2)}^{(bare)} \tilde P^{2n-2},
\end{equation}
where we slightly modified the label of the dressing of the second one with respect to the notation of \cite{Nii:2019ebv}.
On the other hand, the mesonic and baryonic combinations in the WZ dual description are
\begin{equation}
T=A^n,\quad  B_{n-1}=A^{n-1} q^2,\quad  B_{n-2}=A^{n-2} q^4,\quad M=Q\tilde P ,\quad \tilde B_1= A \tilde P^2.
\end{equation}
Confining the $\mathrm{SU}(2n)$ node we are left with the third quiver in  Figure \ref{fig:decASpure3d1} with superpotential\footnote{Observe that the term $Y_{A}^{dressed} \tilde B_1^{n-1} B_{n-2}$ was omitted in \cite{Nii:2019ebv}, but it can be proven to arise both from dimensional reduction and from pure 3d tensor deconfinement.}
\begin{equation}
\label{dadec}
W =
Y_{A}^{dressed} (\tilde B_1^{n-2} M^2 B_{n-1} + T \tilde B_1^{n-3} M^4
+\tilde B_{1}^{n-1}  B_{n-2} )
+
Y_{\tilde P^{2n-2}}^{dressed} (B_{n-1}^2 + TB_{n-2})
+
\sigma R^2.
\end{equation}

It is interesting to compare this superpotential with the one that we would obtain
from (\ref{425}) by applying the real mass flow.
Under the real mass flow the fields in (\ref{425}) become
\begin{eqnarray}
\label{flow}
&&
\eta_1 \rightarrow T,
\quad
\eta_2 \rightarrow B_{n-1},
\quad
\eta_3 \rightarrow B_{n-2},
\quad
\phi \rightarrow M,
\nonumber \\
&&\sigma \rightarrow \sigma, \quad
 R_{3,4} \rightarrow \tilde R_{1,2},
\quad
B_{As} \rightarrow \tilde B_{1},
\quad
\eta_4 \rightarrow Y_{\tilde P^{2n-2}}^{dressed},
\quad
 B_s \rightarrow Y_{A}^{dressed},
\end{eqnarray}
while the fields $B_V$ and $\psi$ are massive and disappear from the low energy spectrum.

The relevant superpotential terms from (\ref{425})  are then 
\begin{equation}
W =  B_s (\eta_1 \phi^4 B_{As}^{n-3} +B_{As}^{n-2} \phi^2\eta_2+\eta_3 B_{As}^{n-1} )+\eta_4(\eta_2^2+\eta_1 \eta_3) + \sigma R^2,
\end{equation}
that under (\ref{flow}) become

\begin{eqnarray}
\label{damasse}
W &=&  Y_{A}^{dressed}  (\tilde B_{1}^{n-2} M^2  B_{n-1} + T \tilde B_1^{n-3} M^4+
 \tilde B_{1}^{n-1}  B_{n-2}  ) \nonumber \\
&+&
Y_{\tilde P^{2n-2}}^{dressed} (B_{n-1}^2+T B_{n-2}) + \sigma \tilde R^2.
\end{eqnarray}

At this point we can perform another real mass flow on the two antifundamentals involved in the superpotential.
In this case the finite parts of the real masses of such fields must be carefully be 
assigned, because they are constrained before the flow by the superpotential 
term $\tilde A^{n-1} \tilde Q_3 \tilde Q_4$.
Such scaling is rather evident at the level of the partition function, where the masses are assigned as
\begin{equation}
\label{flow34}
\nu_{3}=\omega+ s -\frac{(n-1)}{2}\tau_{\tilde A} \quad\quad \text{and} 
\quad \quad
\nu_{4}=\omega- s -\frac{(n-1)}{2}\tau_{\tilde A}\, .
\end{equation}
In the electric theory the divergent term arises from  $\prod_{i=1}^{2n} \Gamma_h(\nu_{3,4}-z_i)$ with $\prod_{i=1}^{2n}
z_i=1$, which cancels with the  divergent term of the  magnetic theory  which arises from $\prod_{j=1}^{n-1}
\Gamma_h\left(\nu_{3,4}-\frac{1}{2} \tau_{\tilde A} \pm w_j\right)$.

At the level of the deconfinement discussed above we observe that in the second quiver of Figure \ref{fig:decASpure3d1}
the fields denoted as  $\tilde R$ are massive and they are integrated out. However the flipper $\sigma$ involved in the superpotential $W = \sigma \tilde R^2$ survives in the low energy spectrum with the same charge. It is indeed associated to the operator $\mathrm{Pf} \tilde A$ and it is consistent with the claim that in this phase an interaction $W= \sigma \, Y_{\mathrm{USp}(2n-2)} $ is expected (see {\bf Appendix B.1} of \cite{Amariti:2024gco}). 

At this point of the discussion we observe that flow (\ref{flow34}) applied to the third quiver depicted in Figure \ref{fig:decASpure3d1} gives rise to an $\mathrm{USp}(2n-2)$ gauge theory with an antisymmetric and four fundamentals with superpotential 
\begin{eqnarray}
\label{damasse2}
W &=&  Y_{A}^{dressed}  (\tilde B_{1}^{n-2} M^2  B_{n-1} + T \tilde B_1^{n-3} M^4+
\tilde B_{1}^{n-1}  B_{n-2}  ) \nonumber \\
&+&
Y_{\tilde P^{2n-2}}^{dressed} (B_{n-1}^2+T B_{n-2}) + \sigma B_{n-2} \hat Y_{\mathrm{USp}(2n-2)}^{(n-2)},
\end{eqnarray}
which coincides with the one studied in formula {\bf (B.2)} of \cite{Amariti:2024gco}, except the presence of the term $ Y_{A}^{dressed}   \tilde B_{1}^{n-1}  B_{n-2}  $  as discussed above.
This superpotential can be obtained from (\ref{damasse}) by studying the real mass flow, except the last term which involves the dressed monopole 
$ \hat Y_{\mathrm{USp}(2n-2)}^{(n-2)}= \hat Y_{\mathrm{USp}(2n-2)}^{(bare)} \tilde B_1^{n-2}$ that is claimed to be dynamically generated.

The difference with the dualities studied in the previous paragraphs is that in this case is that this model can be further confined to a WZ model. The details of such confinement have been discussed in details in  \cite{Amariti:2018wht,Benvenuti:2018bav}, and we refer the interested reader to these references for further details.

%
%
%
\subsection{$\mathrm{SU}(2n+1)$ with $W=\tilde A^{n} \tilde Q$}
%
%
%

The effective reduced duality reflects in the matching of the $3d$ partition function on the squashed three-sphere
\begin{eqnarray}
\label{id2np1firstonS1}&&
Z_{\mathrm{SU}(2n+1)}^{(4;4;\cdot;1;1;\cdot;\cdot)}
(\vec \mu;\vec \nu;\cdot;\tau_A;\tau_{\tilde A};\cdot;\cdot)=
\Gamma_h(\tau_{\tilde A}+\tau_ A)
\prod_{a=1}^4 \left( \Gamma_h\left(n \tau_A + \mu_a\right)
\prod_{b=1}^3 \Gamma_h\left(\mu_a+\nu_{b}\right)\right)
\nonumber \\
&&
\prod_{a<b<c} \! \Gamma_h\left((n-1) \tau_A + \mu_a+ \mu_b+ \mu_c \right) \!
\prod_{1\leq b < c \leq 3}  \!\Gamma_h(\tau_A +\nu_a +\nu_b)
\Gamma_h \bigg((n-1) \tau_{\tilde A}+\sum_{b=1}^3 \nu_b\bigg)
\nonumber \\
&&
Z_{\mathrm{USp}(2n-2)}^{(8;\cdot;1)}\left(\vec \mu+\frac{\tau_{\tilde A}}{2},
\nu_{1,2,3} +\tau_A +\frac{\tau_{\tilde A}}{2},\nu_{4}-\frac{\tau_{\tilde A}}{2}
;\cdot;
\tau_{\tilde A}+\tau_ A
\right). \nonumber \\
\end{eqnarray}
The identity is valid provided the following two constraints are satisfied by the mass parameters
\begin{equation}
\label{BC2np1firstonS1}
(2n-1)(\tau_A+\tau_{\tilde A}) + \sum_{a=1}^4 (\mu_a +\nu_a)= 4 \omega
\quad \& \quad
n\tau_{\tilde A} +\nu_4=2 \omega,
\end{equation}
which descend from the 4d balancing conditions imposed by the cancellation of the axial anomaly and by the superpotential $W=\tilde A^{n} \tilde Q_4$ respectively.
Here, while the second constraint is still imposed by the superpotential deformation, the first constraint is imposed by a linear monopole deformation, corresponding to the KK monopole.

Similarly to the cases studied above it is possible to remove the monopole superpotential by real mass flow. Various options are possible, either involving fundamentals with the same conjugation or with opposite conjugation.  The analysis is straightforward and we will not pursue it here leaving it to the interested reader.
The only comment which is in order is that it is not possible in this case to reach a confining duality for  $\mathrm{SU}(2n+1)$ with an antisymmetric flavor and four fundamentals. The reason is that if we remove the monopole superpotential by assigning two real masses to the antifundamentals that are not involved in the superpotential term $\tilde A^{n} \tilde Q_4$, then, the second real mass flow, involving also $\tilde Q_4$, is obstructed in the dual phase. The situation in this sense is different with respect to the one of $\mathrm{SU}(2n)$ with $W = \tilde A^{n-2} \tilde Q_3 \tilde Q_4$, where such a second flow was possible. The result is consistent with the fact that a 3d confining duality for $\mathrm{SU}(2n+1)$ with an antisymmetric flavor and four fundamentals has not been obtained in the literature.

%
%
%
\subsection{$\mathrm{SU}(2n+1)$ with $W=\tilde A^{n-1} \tilde Q^3$}
%
%
%
The effective reduced duality reflects in the matching of the $3d$ partition function on the squashed three-sphere
\begin{eqnarray}
\label{id2np1fsecondonS1}
&&
Z_{\mathrm{SU}(2n+1)}^{(4;4;\cdot;1;1;\cdot;\cdot)}
(\vec \mu;\vec \nu;\cdot;\tau_A;\tau_{\tilde A};\cdot;\cdot)=
\Gamma_h(\tau_{\tilde A}+\tau_ A,n \tau_{\tilde A}+\nu_1 )
\prod_{b=2}^4 \Gamma_h(n \tau_{\tilde A}+\nu_{b} )
\nonumber \\
\times &&
\prod_{a=1}^4 \Gamma_h\left(\mu_a+\nu_{1},n \tau_A + \mu_a\right)
\prod_{a<b<c} \!\!\!\! \Gamma_h\left((n-1) \tau_A + \mu_a+ \mu_b+ \mu_c \right)
 \\
\times &&
Z_{\mathrm{USp}(2n)}^{(8;\cdot;1)}\left(\vec \mu+\frac{\tau_{\tilde A}}{2},
\nu_1 +\tau_A +\frac{\tau_{\tilde A}}{2},\nu_{2,3,4}-\frac{\tau_{\tilde A}}{2}
;\cdot;
\tau_{\tilde A}+\tau_ A
\right).
\nonumber 
\end{eqnarray}
The identity is valid provided the following two constraints are satisfied by the mass parameters
\begin{equation}
\label{BC2np1secondonS1}
(2n-1)(\tau_A+\tau_{\tilde A}) + \sum_{a=1}^4 (\mu_a +\nu_a)= 4 \omega
\quad \& \quad
(n-1)\tau_{\tilde A} + \sum_{a=2}^4 \nu_a=2 \omega,
\end{equation}
which descend from the 4d balancing conditions imposed by the cancellation of the axial anomaly and by the superpotential $W=\tilde A^{n-1} \tilde Q^3$ respectively.
Here, while the second constraint is still imposed by the superpotential deformation, the first constraint is imposed by a linear monopole deformation, corresponding to the KK monopole.
Analogously to the cases discussed above it is possible to remove the monopole superpotential by real mass flows. 
In this case the only sensible option corresponds to assigning to opposite masses to a pair of fundamentals.
Again we leave the analysis the interested reader.

%
%
%
%
%
\section{Duplication formula}
\label{sec:duplication}
%
%
%
%
%

In this section we study the effective duality on $S^1$ derived in Section \ref{sec:3d} above, by operating with the duplication formula for the hyperbolic Gamma functions 
\begin{equation}
\label{duplication}
\Gamma_h(2z) 
= 
\Gamma_h \left(z \right)  
 \Gamma_h \left( z+\frac{\omega_1}{2} \right)   
 \Gamma_h \left( z+\frac{\omega_2}{2}  \right)  
 \Gamma_h \left( z+\omega \right).  
\end{equation}

Despite the fact that the formula does not have a clear physical interpretation in 3d (see \cite{Kim:2023qwh,Hayashi:2023boy,Kim:2024vci} for a 5d interpretation of a similar formula), it has been used in various papers in order to convert symplectic gauge groups into orthogonal ones and/or antisymmetric tensors into symmetric ones.
Here we are not willing to face the problem of the interpretation of the formula at physical level, but we explore the consequences of its application to the effective dualities obtained in Section \ref{sec:3d}.

Then, we proceed by \emph{freezing} the values of some of the mass parameters for the (anti)-fundamentals to opportune values, in order to allow the application of formula (\ref{duplication}).
Some of the mass parameters involved in the formula are proportional to $\omega_{1,2}$ and it is not clear what is the physical interpretation of such freezing in terms of the global symmetries. However, if we choose opportune values the final result on the integral associated to the squashed three-sphere partition function can be physically interpreted with a sensible gauge and field content and with sensible interactions. Once we find a sensible field content in the electric phase we apply the duality map and study the fate of the dual partition function upon the dual freezing and the application of the duplication formula. The procedure does not in principle guarantees a sensible gauge and field content on the dual side. However, restricting ourselves to the case where it is possible, we obtain a new integral identity, which we interpret as an evidence of a new duality.
In order to corroborate this last interpretation we then proceed by providing a proof of the new duality by tensor deconfinement along the lines of the discussion in the previous sections.

Before proceeding a comment is in order. One may wonder why we did not perform a similar discussion in the 4d cases studied above. The reason is that in such cases the duplication formula would have required to \emph{freeze} more than four fugacities for the fundamentals and/or the antifundamentals in order to provide a sensible physical result. In the models studied here such a large number of fugacities is not available and this forced us to concentrate on the 3d cases.
However, this observation has a physical interpretation for the effective dualities that we found. While we started by considering models with antisymmetric matter in presence of a linear KK monopole superpotential, the one that we obtain after the application of the duplication formula is not a  KK monopole superpotential, but it is another linear monopole. This signals the absence of anomaly-free 4d parent with the same field content of 3d effective models obtained from the application of the freezing and of the duplication formula.

We will distinguish two class of dualities. The first class regards dualities among $\mathrm{SU}(N)$ and $\mathrm{USp}(2M)$ gauge groups, while the second class regards dualities between $\mathrm{SU}(N)$ and $\mathrm{SO}(M)$ gauge groups.

In the first case the dualities are obtained by freezing (some of) the mass parameters for the fundamentals, while in the second case the dualities are obtained by freezing (some of) the mass parameters for the antifundamentals.

\subsection{SU/USp dualities}
\label{subsec:USp}

Here we propose new dualities by considering the ones derived in Section \ref{sec:3d}.
As discussed in the introduction of this section the proposal originates from the application of the duplication formula on the identities of  Section \ref{sec:3d}, after freezing some of the mass parameters for the fundamentals. 
We proceed by freezing the vector associated to the masses $\mu_a$ as
\begin{equation}
\label{freddo}
\vec \mu = \frac{\tau_S}{2}+\vec v,\quad \text{with} \quad \vec v=\left\{0,\frac{\omega_1}{2},\frac{\omega_2}{2}, \mu-\frac{\tau_S}{2} \right\},
\end{equation}
where with a slight abuse of notation we redefined the free parameter $\mu_4$ as $\mu$.
Furthermore, we redefined $\tau_A$ as $\tau_S$.

By applying the duplication formula after such freezing and redefinitions, the $\mathrm{SU}(N)$ integrands are modified by the substitution
\begin{equation}
\label{duplapp}
\prod_{1\leq i<j \leq N} \!\!\!\! \Gamma_h(\sigma_i + \sigma_j+\tau_A)\prod_{i=1}^N \prod_{a=1}^{4} \Gamma_h(\sigma_i+\mu_a )
\rightarrow \!\!\!\! \!\!\!\! 
\prod_{1\leq i\leq j \leq N} \!\!\!\! \!\!  \Gamma_h(\sigma_i + \sigma_j+\tau_S)
\prod_{i=1}^N \Gamma_h\left(\sigma_i+\mu,\omega\!-\!\sigma_i\!-\!\frac{\tau_S}{2} \right).
\end{equation}
Furthermore, the balancing conditions are modified accordingly.

The interpretation of formula (\ref{duplapp}) is that in the electric field content we have converted an $\mathrm{SU}(N)$ antisymmetric $A$ and four $\mathrm{SU}(N)$ fundamentals $Q$ into a $\mathrm{SU}(N)$ symmetric $S$, $\mathrm{SU}(N)$ fundamental $Q$ and one $\mathrm{SU}(N)$ antifundamental $\tilde Q_S$.
This last field does not have a free mass parameter, and it implies the presence of a superpotential interaction
\begin{equation}
W \subset S \tilde Q_S^2.
\end{equation}
Observe that for each model under investigation other superpotential terms, either involving the charged field or the monopoles, are allowed, as will see in the various examples below. 
In the following, we will study the fate of the effective dualities of section \ref{sec:3d} under the application of the freezing (\ref{freddo}) and of the duplication formula.

%
%
%
%
\subsubsection{$\mathrm{SU}(2n)$ with the deformation $W = \tilde A^{n-1} \tilde Q^2$}
%
%
%
%

Here we start our analysis with the $\mathrm{SU}(2n)/\mathrm{USp}(2n)$ duality obtained after deforming the electric theory by 
the superpotential (\ref{Wdef2SU2n}) and then reducing on $S^1$.
The starting point is then the identity (\ref{onS1sefnm1}) provided the validity of the balancing conditions (\ref{BAonS1sefnm1}).

We already discussed the consequences of the freezing on the LHS of the identity.
Furthermore, the balancing conditions (\ref{BAonS1sefnm1}) become
\begin{equation}
\label{balafterdup}
(2n-2)\tau_{\tilde A}+\left(2n-\frac{1}{2}\right) \tau_S+\mu+\sum_{a=1}^4\nu_a=3\omega
,\quad
(n-1)\tau_{\tilde A}+\nu_3+\nu_4=2\omega.
\end{equation}
It follows that the superpotential for the $\mathrm{SU}(2n)$ in the electric gauge theory is
\begin{equation}
\label{spot1S}
W = Y_{\mathrm{SU}(2n-2)}^{(bare)} + S \tilde Q_S^2 + \tilde A^{n-1} \tilde Q_3 \tilde Q_4,
\end{equation}
where the linear monopole superpotential forces the first constraint in (\ref{balafterdup})  and it is gauge invariant.

On the other hand, we can provide a candidate dual field theory by applying the freezing of the mass parameters in the partition function on the RHS of the identity (\ref{onS1sefnm1}).
In this case the integrand is modified by the substitution 
\begin{eqnarray}
\label{subso}
&&
\Gamma_h(\tau_{\tilde A}+\tau_A)^n
\prod_{1\leq i<j \leq n} \Gamma_h(\pm \sigma_i \pm \sigma_j+\tau_{\tilde A}+\tau_A)\prod_{i=1}^n \prod_{a=1}^{4} \Gamma_h\left(\pm \sigma_i+\mu_a +\frac{\tau_{\tilde A}}{2}\right)  
\nonumber \\
\rightarrow&&
\Gamma_h(\tau_{\tilde A}+\tau_A)^{-n}
\prod_{1\leq i\leq  j \leq n} \Gamma_h(\pm \sigma_i \pm \sigma_j+\tau_{\tilde A}+\tau_S)
 \\
&\times&
\prod_{i=1}^n \Gamma_h\left(\pm \sigma_i +\mu+\frac{1}{2}\tau_{\tilde A},\pm \sigma_i + \omega-\frac{\tau_S+\tau_{\tilde A}}{2}\right). \nonumber
\end{eqnarray}

In this case we have a $\mathrm{USp}(2n)$ gauge theory with an adjoint $X$, and six fundamentals. 
One fundamental, that we denote as $q_X$, has mass parameter given by the last term in the second line of (\ref{subso}) and it interacts with the adjoint $X$ through a superpotential term $W \subset q_X^2 X$. We denote as $\phi$
the other fundamentals read from the second line of (\ref{subso}). The other four fundamentals in the LHS of (\ref{onS1sefnm1}) are blind to the freezing, and we denote them as $\tilde R_{3,4}$ and $B_{V_{1,2}}$.

Furthermore, in this dual phase the hyperbolic gamma functions corresponding to gauge invariant operators of the electric phase acting as singlets in the dual phase are modified by the freezing accordingly.
After some massage the expression becomes 
\begin{eqnarray}
&&
\Gamma_h(n  \tau_{\tilde A},2n\tau_S,(2n-1)\tau_S+2\mu,(n-1) \tau_{\tilde A} +\nu_1 +\nu_2) \nonumber \\
&&
\prod_{r=1}^{2} \Gamma_h(\mu + \nu_r ,(n-1)\tau_{\tilde A}+\left(2n-1\right) \tau_S+\mu+\nu_r)
\prod_{r,s=1}^{2} \Gamma_h(\tau_S + \nu_r+\nu_s ).
\end{eqnarray}

The other terms are interpreted as follows:
\begin{itemize}
\item $ n \tau_{\tilde A}$: this is the electric operator $\tilde b=\mathrm{Pf} \tilde A$;
\item $ 2n \tau_{S}$: this is the electric operator $\Phi=\det S$;
\item $(2n-1)\tau_S+2\mu$: this is the electric operator $K= S^{2n-1} Q^2$;
\item $(n-1) \tau_{\tilde A} +\nu_1 +\nu_2$: this is the electric operator $L =\tilde A^{n-1} \tilde Q_1 \tilde Q_2$;
\item $\mu + \nu_r $: this is the electric operator $M_r= Q \tilde Q_r$;
\item $(n-1)\tau_{\tilde A}+\left(2n-1\right) \tau_S+\mu+\nu_r)$: this is the electric operator $J= S^{2n-1} \tilde A^{n-1} Q \tilde Q_r$;
\item $\tau_S + \nu_r+\nu_s$: this is the electric operator $H_{rs} = S \tilde Q_r \tilde Q_s$.
\end{itemize}

The identity obtained from the application of the duplication formula then relates this dual $\mathrm{USp}(2n)$  model with the $\mathrm{SU}(2n)$ gauge theory discussed above. The duality map is rather non trivial, as one can see from the singlets appearing in the RHS of the new identity. 
A rather complex superpotential compatible with the global symmetries is then expected. 
By looking at the charge structure we found that the following superpotential is allowed by the global symmetries
\begin{eqnarray}
W &=& b \tilde R^2+\Phi(K L^2 
+ 
\phi^2 H^2 X^{2n-3}  
+
 M^2 H X^{2n-2}+
L \phi \phi_{X}  \nonumber \\
&+&  L M  J+
B_V H X^{2n-3} M \phi)+
X \phi_{X}^2+
B_V \phi_{X} J
+
H J^2+ Y_{\mathrm{USp(2n-4)}}^{(bare)}
  \nonumber \\
&+&
K (H^2 X^{2n-2}+X^{2n-3} B_V^2 H  +X^{2n-4} B_V^4 ), 
\end{eqnarray}
where $Y_{\mathrm{USp(2n-4)}}^{(bare)}$ is the gauge invariant bare monopole of the  breaking   $\mathrm{USp}(2n-2) \rightarrow  \mathrm{USp}(2n-4) \times \mathrm{U}(1)$.
In presence of a $\mathrm{USp}(2n-2)$ adjoint in the dual phase the linear monopole superpotential forces the first  constraint in   \eqref{balafterdup}
in the dual phase. 

In order to corroborate the validity of this duality, proposed from the application of the duplication formula, we 
will show that it can be obtained by tensor deconfinement.
In this case we need to deconfine a symmetric tensor, through a confining duality involving an $\mathrm{SO}(N)$ gauge group.
Such duality was originally found in \cite{Benvenuti:2021nwt} and further studied in \cite{Amariti:2024gco}. We refer the reader to {\bf appendix D} of \cite{Amariti:2024gco} for the conventions adopted here.

\begin{figure}
\begin{center}
  \includegraphics[width=12.5cm]{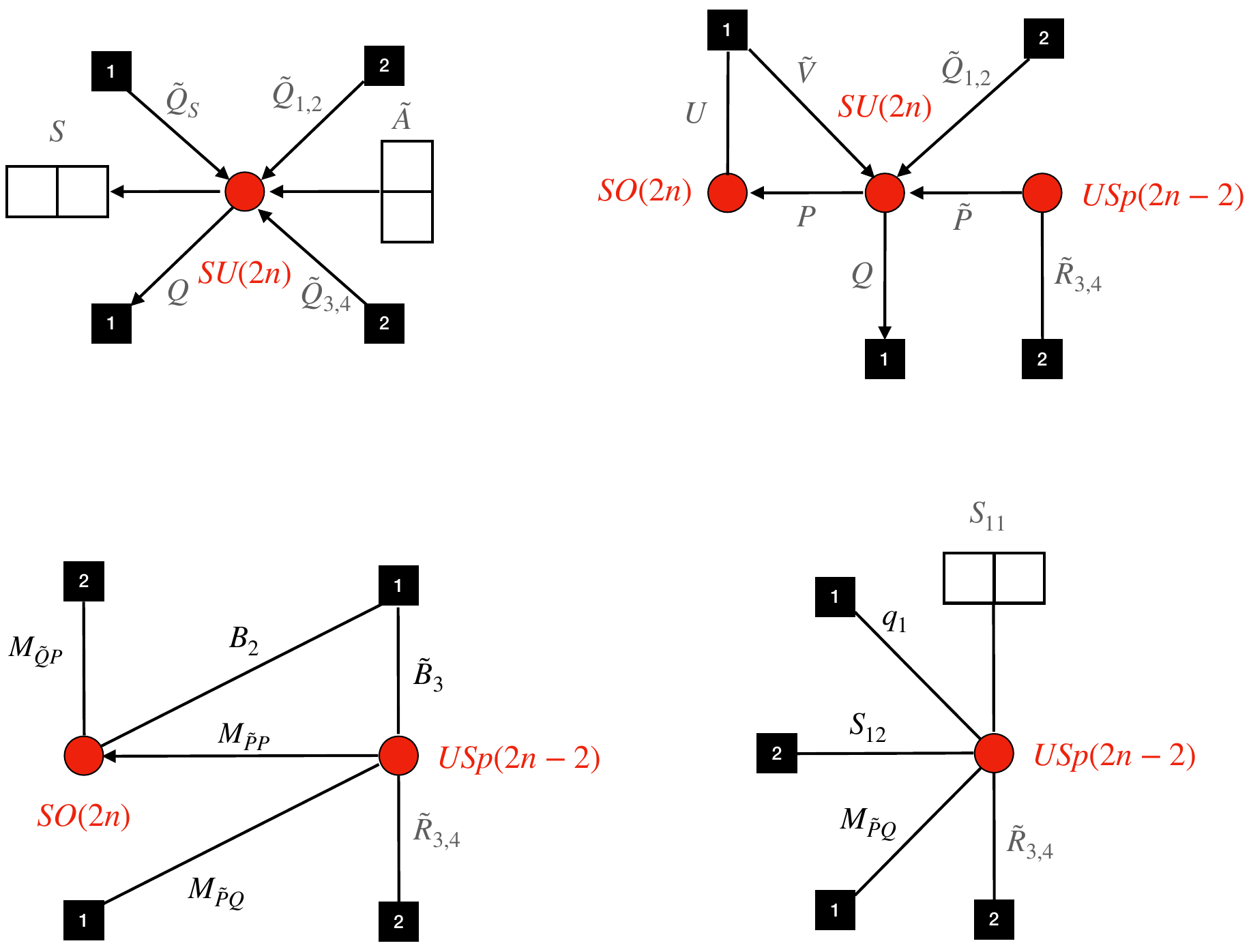}
  \end{center}
  \caption{Scheme of the proof of the duality between $\mathrm{SU}(2n)$ with a symmetric and a conjugate antisymmetric and $\mathrm{USp}(2n-2)$ with an adjoint. In the first quiver we represent the field content of the electric gauge theory. In the second figure we represent the charged fields after deconfining the two tensors using an $\mathrm{SO}(2n)$ and an $\mathrm{USp}(2n-2)$ gauge group. The first quiver is obtained after confining the original $\mathrm{SU}(2n)$ gauge group. The final quiver is obtained by confining the 
  $\mathrm{SO}(2n)$  gauge group and it corresponds to the expected dual model studied from the duplication formula at the level of the three sphere partition function.}
    \label{fig:dualSUUSpSymm1}
\end{figure}

We start our analysis by deconfining the symmetric tensor $S$ and the conjugate antisymmetric tensors $\tilde A$. 
In this way we obtain the second quiver in Figure \ref{fig:dualSUUSpSymm1} with superpotential 

\begin{equation}
W= Y_{\mathrm{SU}(2n)}+Y_{\mathrm{SO}(2n)}^{+}+Y_{\mathrm{USp}(2n)}+
\alpha U^2+ \gamma P^{2n} +P U \tilde V+\sigma \tilde R_3 \tilde R_4.
\end{equation}
Observe that the singlets $\sigma$, $\gamma$ and $\alpha$ are not explicitly shown in the quiver. The combinations $P^2$ and $\tilde P^2$ correspond in the original model to the symmetric $S$ and the conjugate antisymmetric $\tilde A$ respectively. Furthermore the original fields $\tilde Q_{3,4}$ are associated to the combinations $\tilde P \tilde R_{3,4}$ here and the field $\tilde Q_S$ is the baryon $P^{2n-1} U $ of $\mathrm{SO}(2n)$.  
The three linear monopole superpotential terms enforce the constraints on the global charges enforced by (\ref{spot1S}) in the original gauge theory.

Then we confine the $\mathrm{SU}(2n)$ gauge node, in terms of its baryons $\mathcal{B}$, antibaryons $\tilde{ \mathcal{B}} $ and mesons $\mathcal{M}$ defined as
\begin{equation}
\tilde {\mathcal{B}} = 
\left(
\begin{array}{l}
\tilde B_1=\tilde P^{2n-2} \tilde Q_1
\\
\tilde B_2=\tilde P^{2n-2} \tilde V \tilde Q_{1,2}
\\
\tilde B_3=\tilde P^{2n-3} \tilde V \tilde Q_{1}
 \tilde Q_{2}
 \end{array}
 \right)
\!,\, 
 \mathcal{B}^T =
 \left(
\begin{array}{l}
  B_1=P^{2n}\\
 B_2=P^{2n-1} Q
 \end{array}
 \right) \!
,\, 
 \mathcal{M} = \left(
\begin{array}{ccc}
 M_{\tilde P P}
& 
  M_{\tilde Q P}
 &
 M_{\tilde V P}
 \\
 M_{\tilde P Q}
 &
 M_{\tilde Q Q}
 &
 M_{\tilde V Q}
\end{array}
\right).
 \end{equation}

The  model is represented in the third quiver of Figure  \ref{fig:dualSUUSpSymm1}
 and it has superpotential 
 \begin{equation}
 W= Y_{\mathrm{SO}(2n)}^{+}+Y_{\mathrm{USp}(2n)}+\sigma \tilde R_3 \tilde R_4+\alpha U^2+ \gamma B_1+ UM_{P \tilde V}
 +\det \mathcal{M} + \mathcal{B} \mathcal{M}  \tilde{\mathcal{B} }.
\end{equation}
After integrating out the massive fields it becomes
 \begin{eqnarray}
\label{spot3S}
 W&=& Y_{\mathrm{SO}(2n)}^{+}+Y_{\mathrm{USp}(2n)}+\sigma \tilde R_3 \tilde R_4+B_2 M_{\tilde P P } \tilde B_3 + B_2 M_{\tilde Q P} \tilde B_2 + 
\alpha(B_2 \tilde B_1+\nonumber \\
&+&M_{\tilde P Q} M_{\tilde Q P}^2 M_{\tilde P P}^{2n-3}+ 
M_{\tilde Q Q}  M_{\tilde P P}^{2n-2} M_{\tilde Q P})^2+M_{\tilde V Q } M_{\tilde P P }^{2n-2} M_{\tilde Q P}^2.
\end{eqnarray}

The last step consists of confining the $\mathrm{SO}(2n)$ node. In this case we have to consider the symmetric meson $\mathcal{S}$ with components
\begin{equation}
S_{11} = M_{\tilde P P}^2, \, 
S_{12} = M_{\tilde P P} M_{\tilde Q P},\,  
S_{13} = M_{\tilde P P} B_2,\,
S_{22} =  M_{\tilde Q P}^2,\,
S_{23} =M_{\tilde Q P} B_2, \,
S_{33}= B_2^2
 \end{equation}
and the baryons 
\begin{equation}
q_1 = M_{\tilde P P}^{2n-3} M_{\tilde Q P}^2 B_2, \, 
q_2 = M_{\tilde P P}^{2n-2} M_{\tilde Q P} B_2, \, 
q_3 = M_{\tilde P P}^{2n-2} M_{\tilde Q P}^2.
 \end{equation}
 The confinement of the $\mathrm{SO}(2n)$ gauge group generates a superpotential $W \sim S_{IJ} q_I q_J+\det \mathcal{S}$, that, in addition to the deformations in (\ref{spot3S}) gives rise to
 \begin{eqnarray}
 W &=& \sigma \tilde R_3 \tilde R_4+\alpha(S_{33} B_1^2 
+ 
M_{\tilde P Q}^2 S_{22}^2 S_{11}^{2n-3}  
+
 M_{\tilde Q Q}^2   S_{22} S_{11}^{2n-2}+
 B_1 M_{\tilde P Q} q_1+
\nonumber
 \\
 &+& B_1 M_{\tilde Q Q}  q_2+
S_{12} S_{22} S_{11}^{2n-3} M_{\tilde Q Q}M_{\tilde P Q})+
S_{11} q_1^2+
S_{12} q_1 q_2
+
S_{22}q_2^2 +Y_{\mathrm{USp(2n-4)}}^{(bare)}
\nonumber
 \\
 &+&S_{33} S_{22}^2 S_{11}^{2n-2}+S_{11}^{2n-3} S_{12}^2 S_{22} S_{33} +S_{11}^{2n-4} S_{12}^4  S_{33}, 
\end{eqnarray}
where we already integrated out the massive combinations.
At this point we observe that we have obtained the expected dual $\mathrm{USp}(2n-2)$ gauge theory upon the mapping
\begin{equation}
\begin{array}{llllll}
\sigma \leftrightarrow b,&\quad
\alpha \leftrightarrow \Phi,&\quad
S_{33} \leftrightarrow K,&\quad
B_1 \leftrightarrow L,&\quad
S_{22}\leftrightarrow H,&\quad
M_{\tilde Q Q} \leftrightarrow M\\
q_2 \leftrightarrow J,&\quad
 S_{11}\leftrightarrow X,&\quad
  q_1 \leftrightarrow \phi_{X},& \quad
 S_{12}\leftrightarrow B_V,&\quad
 M_{\tilde P Q}\leftrightarrow \phi, &\quad
  \tilde R_{3,4} \leftrightarrow   \tilde R_{3,4}\,.
\end{array}
\end{equation}

We conclude the analysis of this model by studying two real mass flows. The first one eliminates the 
linear monopole superpotentials and provides a ``pure" 3d duality. The second real mass flow gives rise to a 3d confining duality, previously claimed in the literature to lack a 4d origin.

\begin{itemize}
\item {\bf Real mass flow (I): a pure 3d SU/USp duality}. 
We can also remove the linear monopole superpotential from the duality studied above by a real mass flow deformation. 
Here we focus on the flow triggered by two large opposite masses to the antifundamentals $\tilde Q_1$ and $\tilde Q_2$.

The electric theory in this case becomes $\mathrm{SU}(2n)$  with a symmetric tensor $S$, a conjugate antisymmetric $\tilde A$, two antifundamentals $\tilde Q_{3,4}$, a further antifundamental $\tilde Q_S$ and a fundamental $Q$, with superpotential 
\begin{equation}
W =  S \tilde Q_S^2 + A^{n-1} \tilde Q_3 \tilde Q_4.
\end{equation}

The dual model on the other hand corresponds to 
$\mathrm{USp}(2n-2)$ with an adjoint $X$, four fundamentals $\phi_{X}$, $\tilde R_{3,4}$ and $\phi$ and superpotential
\begin{equation}
\label{frommassflow}
W = \sigma \tilde R_3 \tilde R_4 +\Phi (K Y_L^2 +Y_L \phi \phi_{X} )
+ K Y_H^2 X^{2n-2}
+
X \phi_{X}^2,
\end{equation}
where the fields $Y_{L,H}$ are dressed monopoles of the electric phase acting as singlet in the dual picture. Such fields originate from the  singlets $L$ and $H_{1,2}$ respectively, after performing the real mass flow.
They correspond to the combinations denoted as $\Psi_{6,7}$ in {\bf Table 8} of \cite{Amariti:2024gco}.

The flow can be studied at the level of the three-sphere partition function by assigning the parameterization
$\nu_1 = m_A + s$ and $\nu_2 = m_A - s$ and taking the limit $s\rightarrow \infty$. This removes the first balancing condition 
in (\ref{balafterdup}), consistently with the claim that the monopole superpotential is lifted by the real mass flow.
Furthermore, we checked that in the dual $\mathrm{USp}(2n-2)$ theory the divergent terms cancel against the ones obtained at large $s$
on the electric side by simply performing the limit on the vacuum for the unbroken gauge symmetry.
The singlets $M,J,H_{11}$ and  $H_{22}$ are massive, while the fields $Y_L$ and $Y_{H_{12}}$ contribute to the dual partition function as
$\Gamma_h (\omega-\left(2n-\frac{1}{2}\right) \tau_S-\mu)$ and $\Gamma_h( \omega -(n-1)\tau_{\tilde A}-\left(2n-\frac{3}{2}\right) \tau_S-\mu)$ respectively.

In order to corroborate the validity of the duality just proposed, in the following we are going to obtain it from tensor deconfinement.

We start by deconfining the conjugate antisymmetric and the symmetric as in the second quiver of figure \ref{fig:decSO(2n)massive1}. The superpotential for this phase is
\begin{equation}
W= Y_{\mathrm{SO}(2n)}^{+}+Y_{\mathrm{USp}(2n-2)}+
\alpha U^2+ \gamma P^{n} +P U \tilde V+\sigma \tilde R_3  \tilde R_4.
\end{equation}
Then we observe that the $\mathrm{SU}(2n)$ gauge group has $2n-1$ antifundamentals and $2n+1$ fundamentals. It is then confining, as discussed in \cite{Nii:2018bgf} and further investigated in \cite{Amariti:2024gco}. The dual theory is described by the mesons $M_{P \tilde P}$, $M_{\tilde V P}$, $M_{\tilde P Q}$ and  $M_{\tilde V Q}$ and the baryons $B_1=P^{2n}$ and $B_2 = P^{2n-2} Q$. In addition, we have two minimal dressed monopoles that we denote\footnote{These labels are given in order to map such monopoles with the one studied from the real mass flow discussed above.} as
$
Y_{\tilde B_3} \equiv Y_1 \dots Y_{2n-1} \tilde P^{2n-3} \tilde V$
and
$Y_{\tilde B_1} \equiv P^{2n} \tilde P^{2n-2}$.
After confining the $\mathrm{SU}(2n)$ gauge node the superpotential for the third phase, corresponding to the $\mathrm{SO}(2n)\times \mathrm{USp}(2n-2)$ quiver is 
\begin{eqnarray}
\label{Wwithdyn}
W &=&
\alpha U^2+ \gamma B_1 +M_{\tilde V P} U +\sigma \tilde R_3  \tilde R_4+B_2 M_{\tilde P P } Y_{\tilde B_3}   \nonumber \\&+&B_1 M_{\tilde P Q} Y_{\tilde B_3}  + B_2 M_{\tilde V P } Y_{\tilde B_1} + B_1 M_{\tilde V Q} Y_{\tilde B_1} + M_{\tilde V P }  Y_{\mathrm{SO}(2n) \epsilon\cdot \tilde  M_{\tilde P P }^{2n-3}  B_2}^- ,
\end{eqnarray}
where we claim that the interaction is dynamically generated by the duality.
Observe that in the dual phase there is no linear monopole superpotential  associated to the symplectic gauge group anymore.
After integrating out the massive fields it becomes
\begin{equation}
W =
\alpha (B_2  Y_{\tilde B_1} )^2+\sigma \tilde R_3  \tilde R_4+B_2 M_{\tilde P P } Y_{\tilde B_3} + M_{\tilde V P } Y_{\mathrm{SO}(2n) \epsilon\cdot \tilde P^{2n-3}\tilde V }^-.
\end{equation}
The last step of the derivation consists of dualizing the $\mathrm{SO}(2n)$ gauge node with $2n-1$ vectors. 
The gauge invariant combinations correspond in this case to the symmetric tensor $\mathcal{S}$, 
the baryon monopole $q$ and the monopole $\Sigma$.
The components of the symmetric $\mathcal{S}$ are $S_{11}= M_{\tilde P P}^2$, $S_{12}= M_{\tilde P P} B_2$ and $S_{22}= B_2^2$. The baryon monopoles are $q_1 = Y_{\mathrm{SO}(2n) \epsilon\cdot M_{\tilde P P }^{2n-2}}^-$ and $q_2 = Y_{\mathrm{SO}(2n) \epsilon\cdot \tilde  M_{\tilde P P }^{2n-3}  B_2}^-$ .
The superpotential, after integrating out the massive fields, is
\begin{equation}
\label{fromdecsomass1}
W =
\alpha S_{22}  Y_{\tilde B_1}^2+\sigma \tilde R_3  \tilde R_4+
S_{11} q_1^2+\Sigma^2 S_{22} \det S_{11}.
\end{equation}
We conclude by comparing the superpotential (\ref{fromdecsomass1}) with the one found from the real mass flow in (\ref{frommassflow}).
Using the dictionary
$X\leftrightarrow S_{11}$,
$\Phi\leftrightarrow \alpha $,
$K\leftrightarrow S_{22}$,
$Y_L\leftrightarrow Y_{\tilde B_1}$,
$\phi_X \leftrightarrow q_1$ and
$Y_H \leftrightarrow \Sigma$
we have reproduced all the interactions except $W \subset \Phi Y_L X \phi_X$ that in the language at hand corresponds to $W \subset \alpha Y_{\tilde B_1} S_{11} q_1$.
We claim that this mismatch is due to the fact that in the superpotential (\ref{Wwithdyn}) also the term 
$W \subset  \alpha Y_{\tilde B_1} M_{P \tilde P}^2 Y_{\mathrm{SO}(2n) \epsilon\cdot M_{\tilde P P }^{2n-2}}^-$ is dynamically generated. This claim is consistent with the global symmetry structure and with the fact that the baryon monopoles emerge in this phase also by applying the deconfinement techniques to the original duality and performing the real mass flow on the  $\mathrm{SO}(2n) \times \mathrm{USp}(2n-2)$ quiver.
The analysis can be performed also at the level of the partition function,
and we leave the details of the analysis to the interested reader. 

\begin{figure}
\begin{center}
  \includegraphics[width=10.5cm]{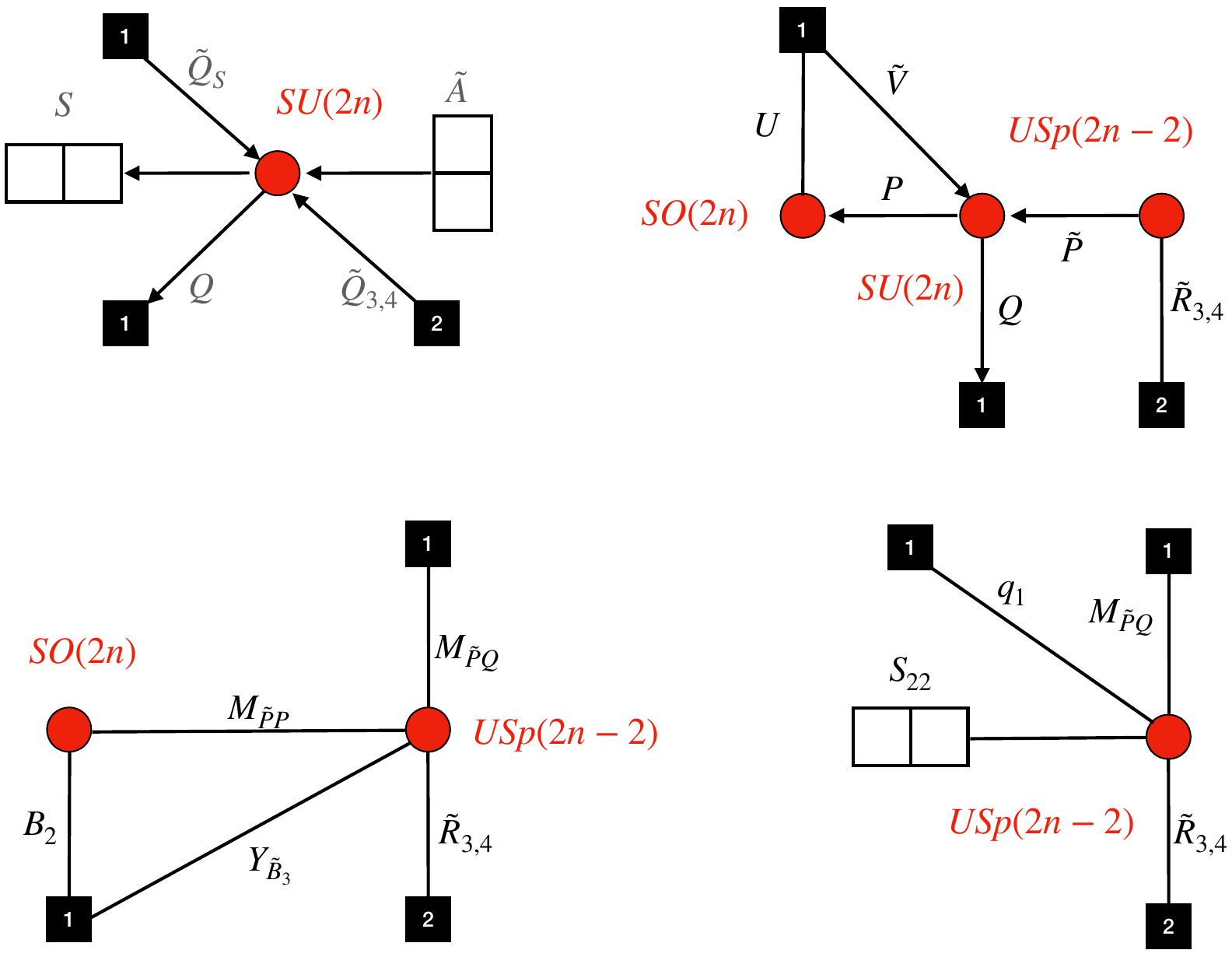}
  \end{center}
  \caption{In this figure we show the steps to prove the duality between $\mathrm{SU}(2n)$ and $\mathrm{USp}(2n-2)$ through tensor deconfinement and ordinary dualities.}
    \label{fig:decSO(2n)massive1}
\end{figure}

\item {\bf Real mass flow (II): recovering a 3d confining duality}. 
The second  real mass flow  removes the superpotential deformation $\tilde A^{n-1} \tilde Q_3 \tilde Q_4$ and gives origin to a confining $\mathrm{USp}(2n-2)$ gauge theory that has been studied already in \cite{Benvenuti:2021nwt}.

The discussion is very similar to the one above and for this reason here we will be more sketchy.
Again we need to scale the masses of the fields $\tilde Q_{3,4}$ consistently with the global symmetries. At the level of the three-sphere partition function these masses scale as in formula (\ref{flow34}).
This scaling is enough to cancel the divergences between the electric and the magnetic theory, without further Higgs flows.

At the level of the deconfinement discussed above we observe that in the second quiver of Figure \ref{fig:decSO(2n)massive1}
the fields denoted as  $\tilde R$ are massive, and they are integrated out, while  the flipper $\sigma$ corresponds to 
 $\mathrm{Pf} \tilde A$.
 At this point we can make contact with the discussion of \cite{Amariti:2024gco}, where this field flips the $Y_{\mathrm{USp}(2n-2)}$ monopole in the deconfined phase.
Once we take this dynamical interaction into account the rest of the analysis is straightforward, because it coincides with the one of \cite{Amariti:2024gco}.

\end{itemize}

%
%
%
%
\subsubsection{$\mathrm{SU}(2n)$ with the deformation $W = \tilde A^{n-2} \tilde Q^4$}
%
%
%
%

Here we consider the $\mathrm{SU}(2n)/\mathrm{USp}(2n-2)$ duality obtained after deforming the electric theory by 
the superpotential (\ref{Wdef1SU2n}) and then reducing on $S^1$.
The starting point is then the identity (\ref{onS1defo1}) provided the validity of the balancing conditions (\ref{BCsuusp1}).
We proceed then by freezing the vector associated to the  masses $\mu_a$ as in (\ref{freddo}), again  defining
$\mu_4$ as $\mu$ and $\tau_A$ as $\tau_S$.
Freezing the masses in this way in the identity (\ref{onS1defo1}) and applying the duplication formula we arrive at
\begin{eqnarray}
\label{identity91}
&&
Z_{\mathrm{SU}(2n)}^{(1;5;\cdot;\cdot;1;1;\cdot)}\left(\mu;\vec \nu,\omega-\frac{\tau_S}{2};\cdot;\cdot;\tau_{\tilde A};\tau_S;\cdot\right)
=
\Gamma_h((2n-1)\tau_S+2\mu)\Gamma_h(2n\tau_S)\Gamma_h(n \tau_{\tilde A})
\nonumber \\
&&
\prod_{a<b} \Gamma_h(2\omega-\hat \nu_a- \hat \nu_b)
Z_{\mathrm{USp}(2n)}^{(6;\cdot;1)}
\left(\mu+\frac{\tau_{\tilde A}}{2},\vec \nu-\frac{\tau_{\tilde A}}{2} ,\omega-\frac{\tau_S+\tau_{\tilde A}}{2}    ;\cdot;\tau_S+\tau_{\tilde A}\right).
\end{eqnarray}
This identity is valid provided the two constraints
\begin{equation}
\label{BC91}
(2n-2)\tau_{\tilde A}+\left(2n-\frac{1}{2}\right) \tau_S+\mu+\sum_{a=1}^4\nu_a=3\omega,\quad
(n-2)\tau_{\tilde A}+\sum_{a=1}^4\nu_a=2\omega
\end{equation}
are satisfied.
The field theory interpretation of the identity (\ref{identity91}) together with the constraints (\ref{BC91})  is that there is a duality between:
\begin{itemize}
\item A $\mathrm{SU}(2n)$ gauge theory with a symmetric $S$, an antisymmetric $\tilde A$, four antifundamentals $\tilde Q$, an antifundamental $\tilde Q_S$ and a fundamental $Q$, with the superpotential 
\begin{equation}
\label{spot626}
W = S \tilde Q_S^2 + \tilde A^{n-2} \tilde Q^4+ Y_{\mathrm{SU}(2n-2)}^{(bare)}.
\end{equation}

\item A $\mathrm{USp}(2n)$ gauge theory with a symmetric (adjoint) $X$, four fundamentals $\tilde R$, one fundamental $U$ and 
one fundamental $Q_X$ in addition to the singlets $\sigma = S^{n-1} Q^2$, $\tilde B = \mathrm{Pf} \, \tilde A$, $ \Phi=\det S$ and $C=\tilde A^{n-1} \tilde Q^2$, interacting with superpotential 
\begin{equation}
\label{expectedusp22}
W = C R^2 + \Phi X^{2n-1} U^2 + \Phi \sigma \tilde B^2 + \Phi Q_X \tilde B U + X Q_X^2+ Y_{\mathrm{USp}(2n-2)}^{(bare)}.
\end{equation}
\end{itemize} 

\begin{figure}
\begin{center}
  \includegraphics[width=11cm]{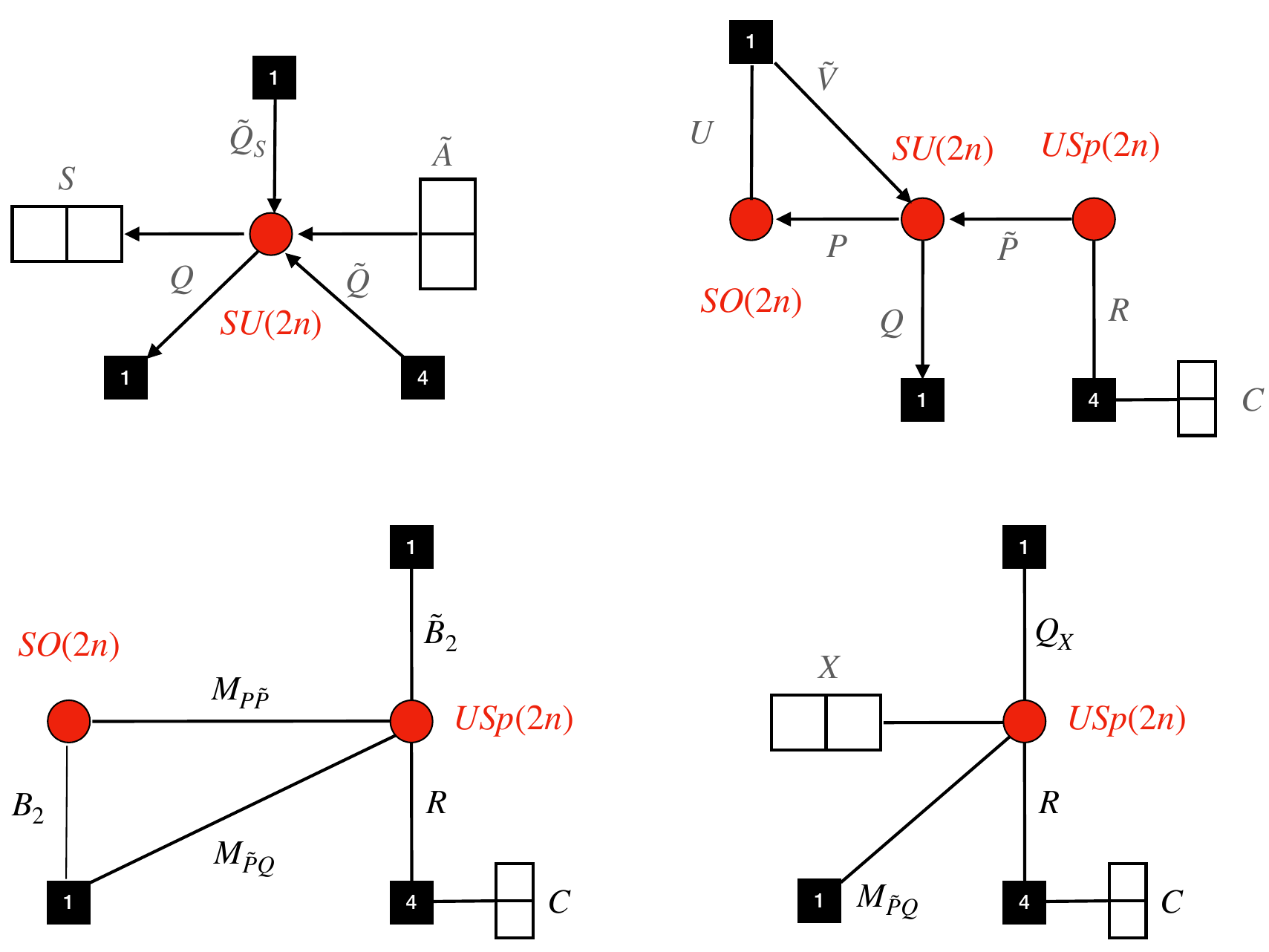}
  \end{center}
  \caption{ 
    In this figure we show the steps to prove the duality between $\mathrm{SU}(2n)$with superpotential (\ref{spot626}) and $\mathrm{USp}(2n)$ with superpotential (\ref{expectedusp22}) through tensor deconfinement and ordinary dualities.}
    \label{fig:decusp2nm2}
\end{figure}

In the following we provide the proof of this duality by using tensor deconfinement. The various steps are summarized in Figure \ref{fig:decusp2nm2}.
The first step consists of deconfining the symmetric $S$ and the conjugate antisymmetric $\tilde A$, obtaining the second quiver in Figure \ref{fig:decusp2nm2}.
The superpotential for this model is
\begin{equation}
W= Y_{\mathrm{SU}(2n)}+Y_{\mathrm{SO}(2n)}^{+}+Y_{\mathrm{USp}(2n)}+
\alpha U^2+ \gamma P^{n} +P U \tilde V+C R^2,
\end{equation}
and the original fields $S$ and $\tilde A$ correspond to the combinations $P^2$ and $\tilde P^2$ in this deconfined model.
Furthermore, the field $\tilde Q_S$ correspond in this phase to the $\mathrm{SO}(2n)$ baryon $\epsilon_{2n} P^{2n-1} U$. 
The other  $\mathrm{SO}(2n)$ baryon $\epsilon_{2n} P^{2n}$ is instead flipped by the singlet $\gamma$.
The singlet $C$ corresponds in the original model to the operator $\tilde A^{n-1} \tilde Q^2$, while the singlet $\alpha$ corresponds to the operator $\det S$.

The next step consists of confining the $\mathrm{SU}(2n)$ gauge group with $2n+1$ flavors and a linear monopole superpotential. The confined degrees of freedom are the mesonic combinations
$M_{\tilde P P} $, $M_{ \tilde P Q} $, $M_{ \tilde V Q}$ and $M_{ \tilde V P}$ and the baryonic ones
$B_1=P^{2n}$,  $\tilde B_1=\tilde P^{2n}$,  $B_2=P^{2n-1} Q$ and $\tilde B_2=\tilde P^{2n-1} \tilde V$.
The charged matter content is summarized in the third quiver in Figure \ref{fig:decusp2nm2} and the superpotential for this phase is 
\begin{equation}
W = M_{\tilde P P} B_2 \tilde B_2+M_{ \tilde V Q} B_1 \tilde B_1+M_{ \tilde V P} B_2 \tilde B_1+M_{ \tilde P Q} B_1 \tilde B_2+\alpha U^2+ \gamma B_1+ U M_{ \tilde V P}+C R^2.
\end{equation}
This superpotential, by integrating out the massive fields, simplifies to  
\begin{align}
W = & M_{ \tilde P P} B_2 \tilde B_2+M_{ \tilde V P} B_2 \tilde B_1+C R^2+M_{ \tilde V Q} M_{ \tilde P P}^{2n} 
+
\alpha(B_2 \tilde B_1 + M_{ \tilde P P}^{2n-1}M_{ \tilde P Q})^2
+Y_{\mathrm{SO}(2n)}^+ + \nonumber \\
&+
Y_{\mathrm{USp}(2n)}.
\end{align}

We conclude by confining the $\mathrm{SO}(2n)$ gauge node with $2n+1$ fundamentals and a linear monopole superpotential $Y_{\mathrm{SO}(2n)}^+$.
The symmetric meson of this confining duality is split into the three components 
$X= M_{P \tilde P}^2$, $Q_{m}= M_{P \tilde P} B_2$
and $\sigma= B_2^2$. Furthermore, the baryons of this duality are denoted as 
$Q_X=M_{P \tilde P}^{2n-1} B_2$ and $s=M_{P \tilde P}^{2n}$. The superpotential for the leftover $\mathrm{USp}(2n)$ gauge group is
\begin{eqnarray}
W &=&  Y_{\mathrm{USp}(2n-2)}^{(bare)}+Q_m \tilde B_2+C R^2
+
M_{ \tilde V Q} s
+
\alpha X^{2n-1} M_{ \tilde P Q}^2
\nonumber \\
&+&
\alpha \sigma \tilde B_1^2 +\alpha Q_X \tilde B_1 M_{ \tilde P Q} + X Q_X^2 + Q_{m} Q_X s + \sigma s^2.
\end{eqnarray}
By integrating out the massive fields and identifying the fields 
$\{\alpha, M_{\tilde P Q},\tilde B_1 \}$ with the fields $\{\Phi, U,\tilde B \}$ we obtain the expected superpotential (\ref{expectedusp22}).

%
%
%
%
\subsubsection{$\mathrm{SU}(2n+1)$ with the deformation $W = \tilde A^{n-1} \tilde Q^3$}
%
%
%
%

Here we consider the $\mathrm{SU}(2n+1)/\mathrm{USp}(2n)$ duality obtained after deforming the electric theory by 
the superpotential (\ref{Wdef1SU2np1}) and then reducing on $S^1$.
The starting point is then the identity (\ref{id2np1fsecondonS1}) provided the validity of the balancing conditions (\ref{BC2np1secondonS1}).
We proceed then by freezing the vector associated to the  masses $\mu_a$ as in (\ref{freddo}), again  defining
$\mu_4$ as $\mu$ and $\tau_A$ as $\tau_S$.
Freezing the masses in this way in the identity (\ref{id2np1fsecondonS1}) and applying the duplication formula we arrive at
\begin{eqnarray}
\label{identity2np13}
Z_{\mathrm{SU}(2n+1)}^{(1;5;\cdot;\cdot;1;1;\cdot)}\left(\mu;\vec \nu,\omega-\frac{\tau_S}{2};\cdot;\cdot;\tau_{\tilde A};\tau_S;\cdot\right)=
\prod_{a=1}^4 \Gamma_h(n \tau_{\tilde A}+\nu_a)
&&
\nonumber \\
 \Gamma_h(
2 \nu_1 + \tau_S,2n \tau_S+2 \mu,\omega-\nu_1-\frac{\tau_S}{2},(2n+1)\tau_S,\nu_1+\mu)
&&
\nonumber \\
Z_{\mathrm{USp}(2n)}^{(6;\cdot;1)}
\left(
\mu+\frac{\tau_{\tilde A}}{2},
\nu_1+\tau_{S}+\frac{\tau_{\tilde A}}{2},
\nu_{2,3,4}-\frac{\tau_{\tilde A}}{2},
\omega-\frac{\tau_{\tilde A}+\tau_{S}}{2}  
;\cdot;\tau_S+\tau_{\tilde A}\right).
&&
\end{eqnarray}

This identity is valid provided the two constraints
\begin{equation}
\label{BC2np13}
(2n-1)\tau_{\tilde A}+\left(2n+\frac{1}{2}\right) \tau_S+\mu+\sum_{a=1}^4\nu_a=3\omega,\quad
(n-1)\tau_{\tilde A}+\nu_2+\nu_3+\nu_4=2\omega
\end{equation}
are satisfied.
The field theory interpretation of the identity (\ref{identity2np13}) together with the constraints (\ref{BC2np13})  is that there is a duality between
\begin{itemize}
\item An $\mathrm{SU}(2n+1)$ gauge theory with a symmetric $S$, a conjugate antisymmetric $\tilde A$, four antifundamentals $\tilde Q$, an antifundamental $\tilde Q_S$ and a fundamental $Q$, with the superpotential 
\begin{equation}
\label{eleusp2nsu2np1}
W = \tilde A^{n-1} \tilde Q^3+S \tilde Q_S^2+Y_{\mathrm{SU}(2n-1)}^{(bare)}
\end{equation}

\item An $\mathrm{USp}(2n)$ gauge theory with a symmetric (adjoint) $X$, three fundamentals $ R$, one fundamental $U$, one fundamental V  and 
one fundamental $Q_X$ in addition to the singlets 
$
K=S \tilde Q_1^2$, 
$J=S^{2n} Q^2$, 
$H=S^{2n} A^n Q$, 
$\sigma=\det S$, 
$M=Q \tilde Q_1$, 
$\tilde B_n=\tilde A^n \tilde Q_{1}$, and $C=\tilde A^n \tilde Q_{2,3,4}$, interacting with a superpotential 
\begin{eqnarray}
\label{expectedusp2nsu2np1}
W &=& C R^2 + X Q_X^2 + K H^2 +  H V Q_X+\sigma \tilde B_n J+\sigma M X^{2n} \nonumber \\
&+& \sigma K X^{2n-1} U^2
+\sigma \tilde B_n M J + \sigma \tilde B_n U Q_X 
+Y_{\mathrm{USp}(2n-2)}^{(bare)}
\end{eqnarray}
\end{itemize}

\begin{figure}
\begin{center}
  \includegraphics[width=10cm]{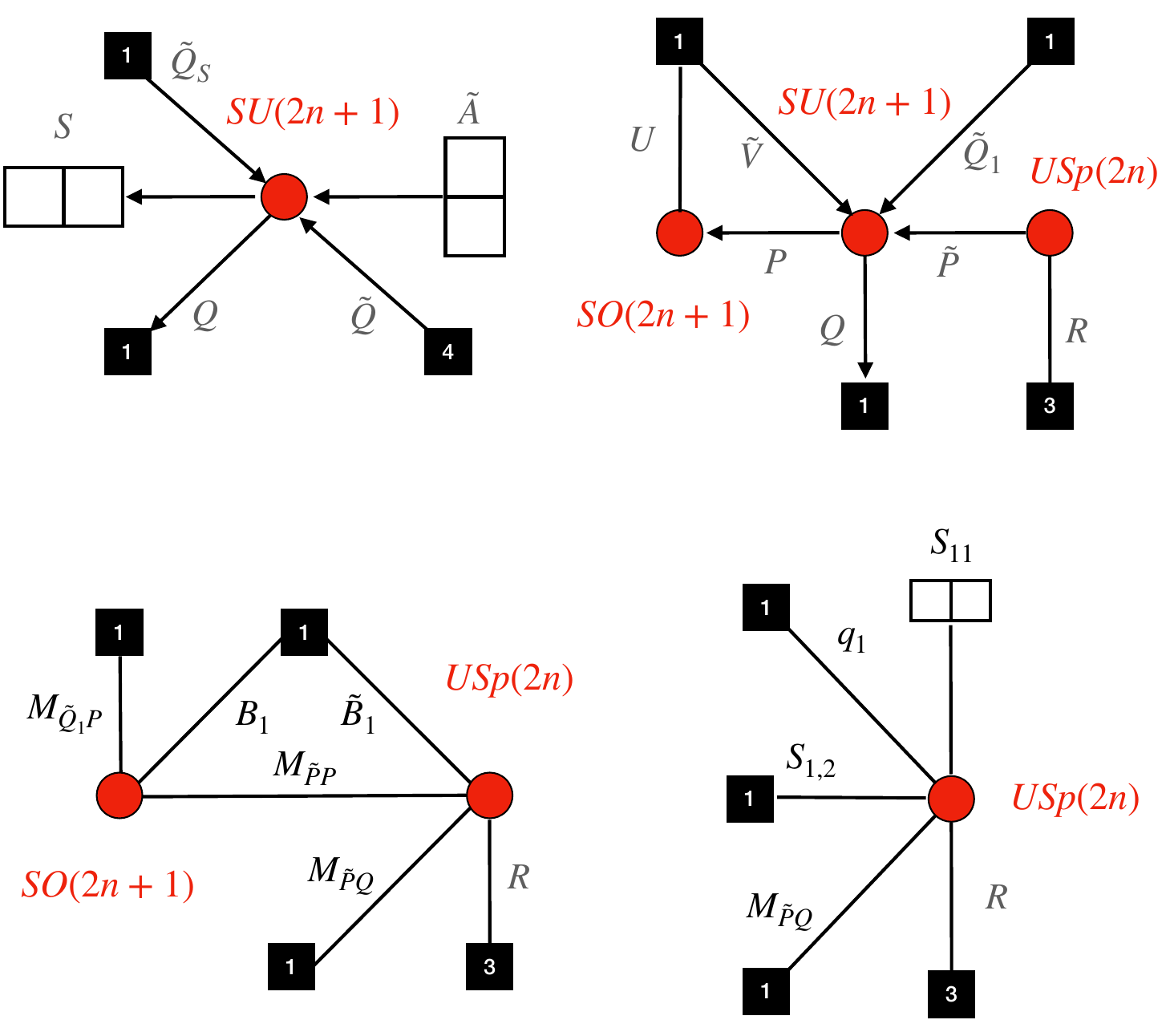}
  \end{center}
  \caption{In this figure we show the steps to prove the duality between $\mathrm{SU}(2n+1)$with superpotential (\ref{eleusp2nsu2np1}) and $\mathrm{USp}(2n)$ with superpotential (\ref{expectedusp2nsu2np1}) through tensor deconfinement and ordinary dualities.} 
    \label{fig:DEC2NP13}
\end{figure}

In the following we provide the proof of this duality by using tensor deconfinement. The various steps are summarized in Figure \ref{fig:DEC2NP13}.
The first step consists of deconfining the symmetric $S$ and the conjugate antisymmetric $\tilde A$, obtaining the second quiver in Figure \ref{fig:DEC2NP13}.
  
The superpotential for this model is
\begin{equation}
W= Y_{\mathrm{SU}(2n+1)}+Y_{\mathrm{SO}(2n+1)}^{+}+Y_{\mathrm{USp}(2n)}+
\alpha U^2+ \gamma P^{2n+1} +P U \tilde V+C R^2
\end{equation}
and the original fields $S$ and $\tilde A$ correspond to the combinations $P^2$ and $\tilde P^2$ in this deconfined model.
Furthermore, the field $\tilde Q_S$ correspond in this phase to the $\mathrm{SO}(2n+1)$ baryon $\epsilon_{2n+1} P^{2n} U$. 
The other  $\mathrm{SO}(2n+1)$ baryon $\epsilon_{2n+1} P^{2n+1}$ is instead flipped by the singlet $\gamma$.
The singlet $C$ corresponds in the original model to the operator $\tilde A^{n} \tilde Q_{2,3,4}$, while the singlet $\alpha$ corresponds to the operator $\det S$.

The next step consists of confining the $\mathrm{SU}(2n+1)$ gauge group with $2n+2$ flavors and a linear monopole superpotential. The confined degrees of freedom are the mesonic combinations
$M_{\tilde P P} $, $M_{ \tilde P Q} $,
$M_{\tilde Q_1 P} $, $M_{ \tilde Q_1 Q} $,
$M_{ \tilde V Q}$ and $M_{ \tilde V P}$ 
and the baryonic ones
$B_1=P^{2n} Q$,  $\tilde B_1=\tilde P^{2n-1} \tilde V \tilde Q_1$, 
 $B_2=P^{2n+1} $, $\tilde B_2=\tilde P^{2n} \tilde V$ and  $\tilde B_3=\tilde P^{2n} \tilde Q_1$ .
The  $\mathrm{SO}(2n+1) \times \mathrm{USp}(2n)$ charged matter content is summarized in the third quiver in Figure \ref{fig:DEC2NP13} and the superpotential for this phase, after integrating pout the massive fields, is
\begin{eqnarray}
W&=& B_1 M_{ \tilde P P} \tilde B_1+B_1 M_{ \tilde Q_1 P} \tilde B_2+B_1 M_{\tilde V P} \tilde B_3
+M_{\tilde V Q} M_{\tilde P P}^{2n} M_{\tilde Q_1 P} \\
&+&
\alpha(B_1 \tilde B_3
+
M_{\tilde Q_1 Q} M_{\tilde P P}^{2n}+
M_{\tilde P Q} M_{\tilde P P}^{2n-1} M_{\tilde Q_1 P} )^2
+ Y_{\mathrm{SO}(2n+1)}^{+}+Y_{\mathrm{USp}(2n)}+C R^2. \nonumber 
\end{eqnarray}

We conclude by confining the $\mathrm{SO}(2n+1)$ gauge node with $2n+2$ fundamentals and a linear monopole superpotential $Y_{\mathrm{SO}(2n+1)}^+$.
The symmetric meson of this confining duality is split into the  components 
$S_{11} = M_{\tilde P P}^2$, $S_{12} = M_{\tilde P P} M_{\tilde Q_1 P}$, 
$S_{13} = M_{\tilde P P} B_1$, $S_{22} = M_{\tilde Q_1 P}^2$, $S_{23} =  M_{\tilde Q_1 P} B_1$ and
$S_{33} = B_1^2$.
 Furthermore, the baryons of this duality are denoted as 
$q_1 = M_{\tilde P P}^{2n-1} M_{\tilde Q_1 P} B_1$, $q_2 = M_{\tilde P P}^{2n}  B_1$ and 
$q_3 = M_{\tilde P P}^{2n} M_{\tilde Q_1 P}$. 
The superpotential for the leftover $\mathrm{USp}(2n)$ gauge group, after integrating out the massive fields, coincides  with  (\ref{expectedusp2nsu2np1})
provided the identifications among the $\mathrm{USp}(2n)$ charged fields
\begin{eqnarray}
X \leftrightarrow S_{11},\quad
q_X  \leftrightarrow q_1,\quad
V \leftrightarrow S_{12},\quad
U\leftrightarrow M_{\tilde P Q}
\end{eqnarray}
and the $\mathrm{USp}(2n)$ singlets
\begin{eqnarray}
K \leftrightarrow   S_{22},\quad
J \leftrightarrow  S_{33}  ,\quad
H \leftrightarrow   q_2,\quad
\sigma \leftrightarrow \alpha  ,\quad
M \leftrightarrow M_{\tilde Q_1 Q}   ,\quad
\tilde B_n \leftrightarrow   \tilde B_3
\end{eqnarray}
while $R$ and $C$ are unchanged.

%
%
%
%
\subsubsection{$\mathrm{SU}(2n+1)$ with the deformation $W = \tilde A^{n} \tilde Q_4$}
%
%
%
%

\begin{figure}
\begin{center}
  \includegraphics[width=12cm]{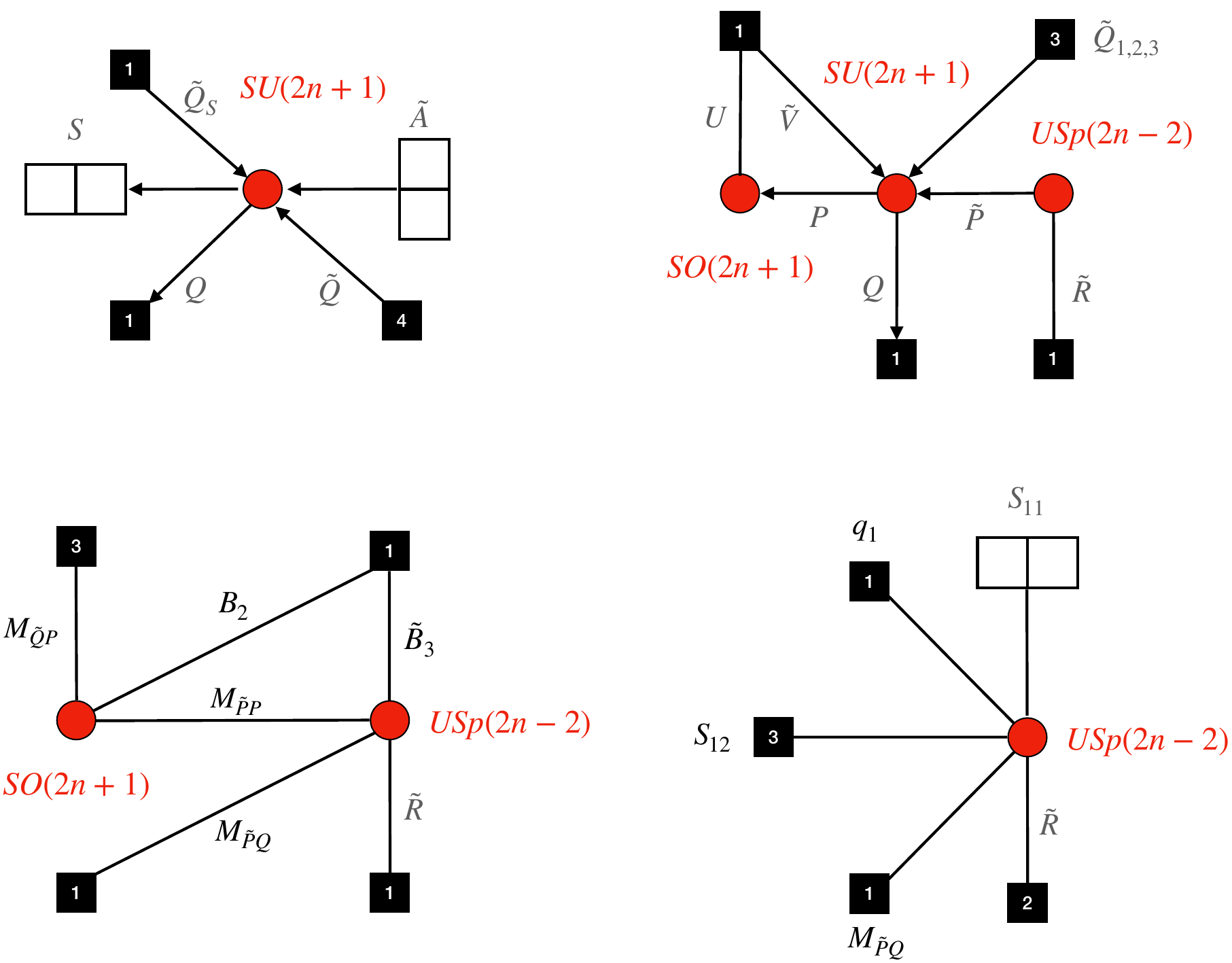}
  \end{center}
  \caption{In this figure we show the steps to prove the duality between $\mathrm{SU}(2n+1)$with superpotential (\ref{eleexpectedusp2nsu2np1bis}) and $\mathrm{USp}(2n-2)$ with superpotential (\ref{expectedusp2nsu2np1bis}) through tensor deconfinement and ordinary dualities.} 
    \label{fig:DEC2NP14}
\end{figure}

Here we consider the $\mathrm{SU}(2n+1)/\mathrm{USp}(2n-2)$ duality obtained after deforming the electric theory by 
the superpotential (\ref{Wdef2SU2np1}) and then reducing on $S^1$.
The starting point is then the identity (\ref{id2np1fsecondonS1}) provided the validity of the balancing conditions (\ref{BC2np1secondonS1}).
We proceed then by freezing the vector associated to the  masses $\mu_a$ as in (\ref{freddo}), again  defining
$\mu_4$ as $\mu$ and $\tau_A$ as $\tau_S$.
Freezing the masses in this way in the identity (\ref{id2np1fsecondonS1}) and applying the duplication formula we arrive at
\begin{eqnarray}
\label{identity2npp14}
&&
Z_{\mathrm{SU}(2n+1)}^{(1;5;\cdot;\cdot;1;1;\cdot)}\left(\mu;\vec \nu,\omega-\frac{\tau_S}{2};\cdot;\cdot;\tau_{\tilde A};\tau_S;\cdot\right)=
\prod_{a=1}^3 \Gamma_h\left(\nu_a+\mu,\omega-\nu_a-\frac{\tau_S}{2}\right)
\nonumber \\
\times
&&
\prod_{1\leq a\leq b \leq 3}\Gamma_h(\nu_a +\nu_b +\tau_S ) \cdot
\Gamma_h\left(2n \tau_S+2\mu,(2n+1) \tau_S,
(n-1)\tau_A+ \sum_{a=1}^3 \nu_a\right)
\nonumber \\
\times &&
Z_{\mathrm{USp}(2n-2)}^{(6;\cdot;1)}
\left(
\mu+\frac{\tau_{\tilde A}}{2},
\nu_{1,2,3}+\tau_S+\frac{\tau_{\tilde A}}{2},
\nu_{4}-\frac{\tau_{\tilde A}}{2},
\omega -\frac{\tau_S+ \tau_{\tilde A}}{2};
\cdot;\tau_S+\tau_{\tilde A}\right).
\end{eqnarray}

This identity is valid provided the two constraints
\begin{eqnarray}
\label{BC2np14}
(2n-1)\tau_{\tilde A}+\left(2n+\frac{1}{2}\right) \tau_S+\mu+\sum_{a=1}^4\nu_a=3\omega
\quad \& \quad
n\tau_{\tilde A}+\nu_4=2\omega
\end{eqnarray}
are satisfied.
The field theory interpretation of the identity (\ref{identity2npp14}) together with the constraints (\ref{BC2np14})  is that there is a duality between
\begin{itemize}
\item An $\mathrm{SU}(2n+1)$ gauge theory with a symmetric $S$, a conjugate antisymmetric $\tilde A$, four antifundamentals $\tilde Q$, an antifundamental $\tilde Q_S$ and a fundamental $Q$, with the superpotential 
\begin{equation}
\label{eleexpectedusp2nsu2np1bis}
W = \tilde A^{n} \tilde Q+S \tilde Q_S^2+Y_{\mathrm{SU}(2n-1)}^{(bare)}.
\end{equation}

\item An $\mathrm{USp}(2n-2)$ gauge theory with a symmetric (adjoint) $X$, one fundamental $\tilde R$, one fundamental $q_X$, one fundamental $U$ and three fundamentals $V$
 in addition to the singlets 
$K=S \tilde Q^2$, $H=S^{2n} Q^2$,
$\sigma=\det S$,  
$\tilde B_{n-1}=\tilde A^{n-1} \tilde Q_1 \tilde Q_2\tilde Q_3$,
$M=\tilde Q Q$
and $J=S^{2n} \tilde A^{n-1} Q \tilde Q^2$
interacting with a superpotential 
\begin{eqnarray}
\label{expectedusp2nsu2np1bis}
W &=& 
\sigma (J \tilde B_{n-1}+U^2 X^{2n-3} K^3+K^2 X^{2n-2} M^2+\tilde B_{n-1} U q_X
 \\
&+&
\tilde B_{n-1} M J+V  U M K^2 X^{2n-3})+X q_X^2+J q_X V+K J^2
+ Y_{\mathrm{USp}(2n-4)}^{(bare)} \nonumber.
\end{eqnarray}
\end{itemize} 

In the following we provide the proof of this duality by using tensor deconfinement. The various steps are summarized in Figure \ref{fig:DEC2NP14}.
The first step consists of deconfining the symmetric $S$ and the conjugate antisymmetric $\tilde A$, obtaining the second quiver in Figure \ref{fig:DEC2NP14}.

The superpotential for this model is
\begin{equation}
W= Y_{\mathrm{SU}(2n+1)}+Y_{\mathrm{SO}(2n+1)}^{+}+Y_{\mathrm{USp}(2n-2)}+
\alpha U^2+ \gamma P^{2n+1} +P U \tilde V
\end{equation}
and the original fields $S$ and $\tilde A$ correspond to the combinations $P^2$ and $\tilde P^2$ in this deconfined model.
Furthermore, the field $\tilde Q_S$ correspond in this phase to the $\mathrm{SO}(2n+1)$ baryon $\epsilon_{2n+1} P^{2n} U$. 
The other  $\mathrm{SO}(2n+1)$ baryon $\epsilon_{2n+1} P^{2n+1}$ is instead flipped by the singlet $\gamma$.
The singlet $\alpha$ corresponds to the operator $\det S$.

The next step consists of confining the $\mathrm{SU}(2n+1)$ gauge group with $2n+2$ flavors and a linear monopole superpotential. The confined degrees of freedom are the mesonic combinations
$M_{\tilde P P}$, $M_{\tilde P Q}$, $M_{\tilde V P}$, $M_{\tilde V Q}$, $M_{\tilde Q P}$ and $M_{\tilde Q Q}$,
the baryonic ones
$B_1=P^{2n+1}$ and $B_2=P^{2n} Q$
and the anti-baryonic ones
$\tilde B_1=\tilde P^{2n-2} \tilde Q_1 \tilde Q_2\tilde Q_3$,  
$\tilde B_2=\tilde P^{2n-2} \tilde V \tilde Q_{a} \tilde Q_{b}$ and 
$\tilde B_3=\tilde P^{2n-3} \tilde V \tilde Q_{1} \tilde Q_{2}\tilde Q_3$.
The  $\mathrm{SO}(2n+1) \times \mathrm{USp}(2n-2)$ charged matter content is summarized in the third quiver in Figure \ref{fig:DEC2NP14} and the superpotential for this phase, after integrating out the massive fields, is
\begin{eqnarray}
W &=& B_2 M_{\tilde P P } \tilde B_3 + B_2 M_{\tilde Q P} \tilde B_2 + 
\alpha(B_2 \tilde B_1+M_{\tilde P Q} M_{\tilde Q P}^3 M_{\tilde P P}^{2n-3}\nonumber \\
&+& M_{\tilde Q Q}  M_{\tilde P P}^{2n-2} M_{\tilde Q P}^2)^2+M_{\tilde V Q } M_{\tilde P P }^{2n-2} M_{\tilde Q P}^3+ Y_{\mathrm{SO}(2n+1)}^{+}+Y_{\mathrm{USp}(2n-2)}.
\end{eqnarray}

We conclude by confining the $\mathrm{SO}(2n+1)$ gauge node with $2n+2$ fundamentals and a linear monopole superpotential $Y_{\mathrm{SO}(2n+1)}^+$.
The symmetric meson of this confining duality is split into the  components 
$S_{11} = M_{\tilde P P}^2$, $S_{12} = M_{\tilde P P} M_{\tilde Q P}$, 
$S_{13} =M_{\tilde P P} B_2$, $S_{22} =  M_{\tilde Q P}^2$, $S_{23} =M_{\tilde Q P} B_2$ and
$ S_{33}= B_2^2$.
 Furthermore the baryons of this duality are denoted as 
$q_1 = M_{\tilde P P}^{2n-3} M_{\tilde Q P}^3 B_2$,
$q_2 = M_{\tilde P P}^{2n-2} M_{\tilde Q P}^2 B_2$ and 
$q_3 = M_{\tilde P P}^{2n-2} M_{\tilde Q P}^3$. 

The superpotential for the leftover $\mathrm{USp}(2n-2)$ gauge group, after integrating out the massive fields, coincides  with  (\ref{expectedusp2nsu2np1bis})
provided the identifications among the $\mathrm{USp}(2n-2)$ charged fields
\begin{eqnarray}
X \leftrightarrow S_{11},\quad
q_X  \leftrightarrow q_1,\quad
V \leftrightarrow S_{12},\quad
U\leftrightarrow M_{\tilde P Q}
\end{eqnarray}
and the $\mathrm{USp}(2n)$ singlets
\begin{eqnarray}
K \leftrightarrow   S_{22},\quad
H \leftrightarrow  S_{33}  ,\quad
J \leftrightarrow   q_2,\quad
\sigma \leftrightarrow \alpha  ,\quad
M \leftrightarrow M_{\tilde Q Q}   ,\quad
\tilde B_{n-1} \leftrightarrow   \tilde B_1,
\end{eqnarray}
while $\tilde R$ is unchanged.

\subsection{SU/SO dualities}
\label{subsec:SO}

Here we discuss an alternative freezing involving the masses of the antifundamentals, which gives rise to  effective dualities between  $\mathrm{SU}(N)$ and $\mathrm{SO}(M)$ gauge theories.
Again we consider the identities of Section \ref{sec:3d} and fix the parameters associated to the antifundamentals 
as
\begin{equation}
\label{freezinganti}
\vec \nu = \left\{\nu,\frac{\tau_{ \tilde S}}{2}, \frac{\omega_1}{2}+\frac{\tau_{\tilde S}}{2},\frac{\omega_2}{2}+\frac{\tau_{\tilde S}}{2} \right\},
\end{equation}
and we further redefine $\tau_{\tilde A}$ as $\tau_{\tilde S}$.
By applying the duplication formula on the LHS of the identities of Section \ref{sec:3d} we convert the contribution to  the three-sphere partition function of a $\mathrm{SU}(N)$  conjugate antisymmetric  and four $\mathrm{SU}(N)$ antifundamentals into the contribution of a $\mathrm{SU}(N)$ conjugate antisymmetric $\tilde S$, one  $\mathrm{SU}(N)$ antifundamentals $\tilde Q$ and one $\mathrm{SU}(N)$ fundamentals $Q_S$, again compatibly with a superpotential $W \subset \tilde S Q_S^2$ interaction.

For each model under investigation other superpotential terms, either involving the charged fields
or the monopoles, are allowed, as will see in the various examples below. 
In the following, we will study the fate of the effective  dualities of section \ref{sec:3d} under the application of the freezing (\ref{freezinganti}) and of the duplication formula.

Here we focus only on two models, corresponding to $\mathrm{SU}(2n)$ with the superpotential (\ref{Wdef2SU2n}) or  (\ref{Wdef1SU2n})
and  $\mathrm{SU}(2n+1)$ with the superpotential (\ref{Wdef1SU2np1}).
The reason is that the other  possible cases involving the other deformations are either not independent of the ones found here or they give rise to identities that do not have a clear physical interpretation.

Anyway, there are still four cases to distinguish, one from  $\mathrm{SU}(2n)$ with the superpotential (\ref{Wdef2SU2n}),
one from $\mathrm{SU}(2n)$ with the superpotential (\ref{Wdef1SU2n}) and two from $\mathrm{SU}(2n+1)$ with the superpotential (\ref{Wdef1SU2np1}).
The reason in this case is that when we are freezing three mass parameters as in (\ref{freezinganti}), we are still not specifying if the associated fields are involved in the dangerously irrelevant superpotential deformations.
We have isolated in each case two different possibilities that gives rise to a quite different IR duality.

The symmetric tensors in the cases discussed below are deconfined by using the confining dualities for 3d orthogonal SQCD with vectors worked out in \cite{Benini:2011mf,Aharony:2013kma,Benvenuti:2021nwt,Aharony:2011ci,Kapustin:2011gh,Hwang:2011ht}.

%
%
%
%
%
\subsubsection{$\mathrm{SU}(2n)$ with superpotential (\ref{Wdef2SU2n})}
%
%
%
%
%

In this case we keep the order of the masses as in the freezing (\ref{freezinganti}) and consider the identity  
(\ref{onS1sefnm1}).  We obtain  the three-sphere partition function 
of a $\mathrm{SU}(2n)$ gauge theory with an antisymmetric $A$, a conjugate symmetric $\tilde S$, four fundamentals $Q$, one extra fundamental $Q_S$ and an antifundamental $\tilde Q$. The constraints on the mass parameters are 
\begin{equation}
\label{BCSO}
(2n-2)\tau_A +\left(2n-\frac{1}{2}\right) \tau_{\tilde S} +\sum_{a=1}^{4} \mu_a +\nu=3\omega,
\quad
2 n \tau_{\tilde S}=2 \omega,
\end{equation}
and they are compatible with the presence of a superpotential 
\begin{equation}
\label{eleSOSO}
W = Y_{\mathrm{SU}(2n-2)}^{(bare)} + \det \tilde S +\tilde S Q_S^2.
\end{equation}

On the other hand, the application of the duplication formula on the RHS of  (\ref{onS1sefnm1}) gives a $\mathrm{SO}(2n-1)$ gauge theory with five vectors and an antisymmetric (adjoint) and a series of singlets. 
In order to have a proper understanding of such dual phase we provide the explicit identity obtained from the application of the duplication formula on  (\ref{onS1sefnm1}) by freezing the masses as in (\ref{freezinganti})

\begin{eqnarray}
\label{dupfreesinganti1}
&&
Z_{\mathrm{SU}(2n)}^{(5;1;\cdot;1;\cdot;\cdot;1)} (\vec \mu,\omega-\frac{\tau_{\tilde S}}{2}; \nu;\cdot;\tau_A;\cdot;\cdot;\tau_{\tilde S})
=
\Gamma_h \Big(n \tau_A,(n-2) \tau_A+\sum_{a=1}^4 \mu_a, (n-1/2) \tau_{\tilde S}+\nu\Big) 
 \nonumber \\
 &&
 \prod_{a=1}^4  \Gamma_e(\mu_a +\nu)
\prod_{a<b} \Gamma_h ((n-1) \tau_A+\mu_a +\mu_b )
Z_{\mathrm{SO}(2n-1)}^{(5;1;\cdot)}\left(\vec \mu + \frac{\tau_{\tilde S}}{2},\nu+\tau_A + \frac{\tau_{\tilde S}}{2};\tau_A+\tau_{\tilde S} ;\cdot\right). \nonumber\\
\end{eqnarray}

The singlets associated to the hyperbolic Gamma functions appearing in the RHS of the identity can be interpreted as 
the gauge invariant combinations of the $\mathrm{SU}(2n)$ gauge theory as follows
\begin{itemize}
\item $\Gamma_h((n-2) \tau_A+\sum_{a=1}^4 \mu_a)$: the gauge invariant operator of the electric theory that gives rise to this hyperbolic gamma function corresponds to the combination $B_{n-2}\equiv A^{n-2} Q^4$;
\item $\Gamma_h((n-1) \tau_A+\mu_a +\mu_b)$: the gauge invariant operator of the electric theory that gives rise to this hyperbolic gamma function corresponds to the combination $ B_{n-1}=A^{n-1} Q^2$;
\item $\Gamma_h(n \tau_A)$: the gauge invariant operator of the electric theory that gives rise to this hyperbolic gamma function corresponds to the combination $B_n=\mathrm{Pf} \,A$;
\item $\Gamma_h((n-1/2) \tau_{\tilde S}+\nu)$: the gauge invariant operator of the electric theory that gives rise to this hyperbolic gamma function corresponds to the combination $M_S=Q_S\tilde Q$;
\item $\Gamma_h(\mu_a +\nu)$ : the gauge invariant operator of the electric theory that gives rise to this hyperbolic gamma function corresponds to the combination $M=\tilde Q Q$.
 \end{itemize}
 
 We further denote as $X$ the adjoint of $\mathrm{SO}(2n-1)$,  with $U$ the four vectors with mass parameter $\vec \mu+ \frac{\tau_{\tilde S}}{2}$ and with $U$ the remaining $\mathrm{SO}(2n-1)$ vector, with mass parameter  $\nu+\tau_A + \frac{\tau_{\tilde S}}{2}$.
 The superpotential interaction compatible with this symmetry structure is 
 \begin{eqnarray}
 \label{expsofinal}
 W &=& Y_{\mathrm{SO(2n-3)}}^{(bare)}  + B_{n-1} (U^2 V X^{n-2}  +  M U  X^{n-1})
 +B_n (M U^3  X^{n-2}+U^4 V  X^{n-3})
 \nonumber \\
& +&B_{n-2} X^{n-1} V
 +M_S B_n B_{n-2}
 +M_S B_{n-1}^2,
 \end{eqnarray} 
 where   $Y_{\mathrm{SO(2n-3)}}^{(bare)}$  refers to the symmetry breaking pattern 
$SO(2n-1) \rightarrow SO(2n-3) \times U(1)$. 

In the following we provide a derivation of such duality, read from the application of the duplication formula, by tensor deconfinement.

\begin{figure}
\begin{center}
  \includegraphics[width=13cm]{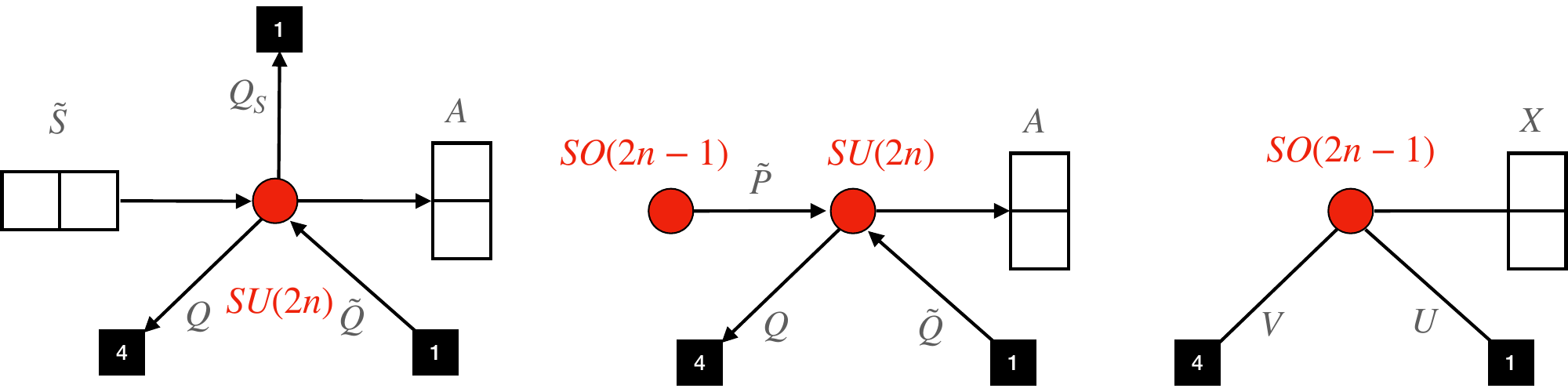}
  \end{center}
 \caption{In this figure we show the steps to prove the duality between $\mathrm{SU}(2n)$with superpotential (\ref{eleSOSO}) and $\mathrm{SO}(2n-1)$ with superpotential (\ref{expsofinal}) through tensor deconfinement and ordinary dualities.} 
    \label{fig:decsec9}
\end{figure}
 In this case we deconfine the tensor $\tilde S$ obtaining the second quiver in Figure \ref{fig:decsec9}. The superpotential 
for this model is 
\begin{equation}
W = Y_{\mathrm{SO}(2n-1)}^+ + Y_{\mathrm{SU}(2n)}.
\end{equation}
 The original symmetric $\tilde S$ in this phase corresponds to the operator $\tilde P^2$, while the field $Q_S$ corresponds to the $\mathrm{SO}(2n-1)$ baryon $\epsilon_{2n-1} P^{2n-1}$. The bare monopoles for $\mathrm{SO}(2n-1)$ and $\mathrm{SU}(2n)$ impose in this phase the two constraints (\ref{BCSO}) in the three-sphere partition function.
 
 Then we observe that, the $\mathrm{SU}(2n)$ gauge group has the field content and the superpotential of a 4d confining duality reduced on $S^1$. We can then confine it, and we arrive to the third quiver in figure \ref{fig:decsec9}, corresponding to the expected $\mathrm{SO}(2n-1)$ dual phase.
  Explicitly the $\mathrm{SU}(2n)$ gauge invariant combinations are
 $ V=A \tilde Q \tilde P$, $ X=A \tilde P^2$, $ U=\tilde P Q$
and $ M_S=\tilde P^{2n-1} \tilde Q$,
in addition to  $B_{n-2}, B_{n-1}$ and  $M$  defined as above. 
The final superpotential  obtained by confining $\mathrm{SU}(2n)$ coincides with (\ref{expsofinal}).

%
%
%
%
%
\subsubsection{$\mathrm{SU}(2n)$ with superpotential (\ref{Wdef1SU2n})}
%
%
%
%
%

In this case we consider the  freezing (\ref{freezinganti}) and consider the identity  
(\ref{onS1defo1}).
We obtain the identity
\begin{eqnarray}
\label{dupfreesinganti2}
&&
Z_{\mathrm{SU}(2n)}^{(5;1;\cdot;1;\cdot;\cdot;1)} (\vec \mu,\omega-\frac{\tau_{\tilde S}}{2}; \nu;\cdot;\tau_A;\cdot;\cdot;\tau_{\tilde S})
=
\Gamma_h\left((n-2)\tau_A +\sum_{a=1}^4 \mu_a\right)
\nonumber
\\
&&
\prod_{a<b} \Gamma_h((n-1)\tau_A+\mu_a+\mu_b)
\Gamma_h(n\tau_A, 2 n \tau_{\tilde S})
Z_{\mathrm{SO}(2n)}^{(5;1;\cdot)}\left(\vec \mu + \frac{\tau_{\tilde S}}{2},\nu-\frac{\tau_{\tilde S}}{2} ;\tau_A+\tau_{\tilde S} ;\cdot\right), \nonumber\\
\end{eqnarray}
holding provided the constraints 
\begin{equation}
(2n-2)\tau_{A} + \left(2n-\frac{1}{2} \right)\tau_{\tilde S} +\sum_{a=1}^4 \mu_a + \nu = 3 \omega
, \quad
(2n-1)\tau_{\tilde S} + 2\nu = 2 \omega
\end{equation}
are satisfied.

The electric gauge theory corresponds to $\mathrm{SU}(2n)$ with a conjugate symmetric $\tilde S$, an antisymmetric $\tilde A$, four fundamentals $Q$ and one fundamental $Q_S$ and one antifundamental $\tilde Q$, with the superpotential 
\begin{equation}
\label{threeconstrW}
W = Y_{\mathrm{SU}(2n-2)} + \tilde S^{2n-1} \tilde Q^2 +\tilde S Q_S^2.
\end{equation}
The dual theory corresponds to an $\mathrm{SO}(2n)$ gauge theory with an adjoint $X$ and five fundamentals, four denoted as $V$ and one denoted as $\tilde R$.
In this case there are also various singlets, that can be related to the gauge invariant combinations in the chiral ring of the electric phase and are read from the identity among the three-sphere partition functions. 

The singlets of the electric phase that appear in the dual description  read from the RHS of (\ref{dupfreesinganti2}) are
$J=A^{n-2} Q^4$,
$H= A^{n-1} Q^2$,
$K = \mathrm{Pf} A$ and $\sigma=\det S$,
where we followed the same ordering as in (\ref{dupfreesinganti2}).
Observe also that in this case in order to reconstruct the dimension of the Weyl group for $\mathrm{SO}(2n)$  we have used the relations $\Gamma_h\left(\omega+\frac{\omega_{1,2}}{2}\right)=\sqrt 2$.
From the duality map we claim that the dual superpotential is
\begin{equation}
\label{expectedSO2}
W =Y_{\mathrm{SO(2n-2)}}^{(bare)} +J \mathrm{Pf} X+ H X^{n-1}  V^2    + K X^{n-2} V^4 +   \sigma \tilde R^2,
\end{equation}
where   $Y_{\mathrm{SO(2n-2)}}^{(bare)}$  refers to the symmetry breaking pattern 
$SO(2n) \rightarrow SO(2n-2) \times U(1)$.

\begin{figure}
\begin{center}
  \includegraphics[width=11cm]{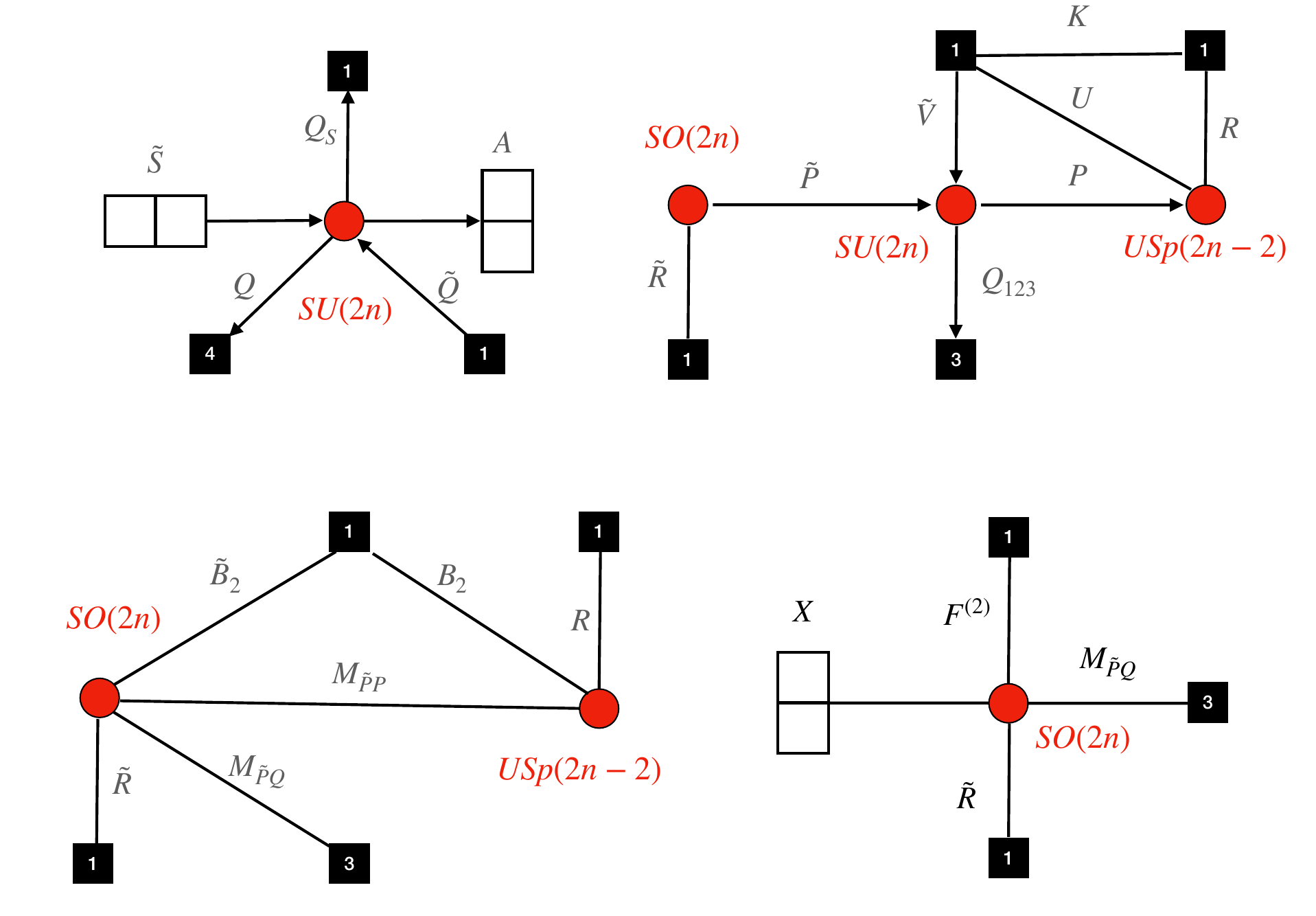}
  \end{center}
    \caption{In this figure we show the steps to prove the duality between $\mathrm{SU}(2n)$with superpotential (\ref{threeconstrW}) and $\mathrm{SO}(2n)$ with superpotential (\ref{expectedSO2}) through tensor deconfinement and ordinary dualities.} 
   \label{fig:decsuuspnm2}
\end{figure}

In the following we provide a derivation of such duality, read from the application of the duplication formula, by tensor deconfinement.
In this case we deconfine the tensors $\tilde S$ and $A$, obtaining the second quiver in Figure \ref{fig:decsuuspnm2}. 
The superpotential 
for this model is 
\begin{equation}
\label{SecondWSUSO2}
W = Y_{\mathrm{SO}(2n)}^++Y_{\mathrm{USp}(2n-2)} + Y_{\mathrm{SU}(2n)} + P U \tilde V + U R K + \sigma \tilde R^2+\gamma \tilde P^{2n}.
\end{equation}
 The original conjugate symmetric $\tilde S$ in this phase corresponds to the operator $\tilde P^2$,  the original antisymmetric $A$ in this phase corresponds to the operator $P^2$.
 The field $Q_S$ corresponds to the $\mathrm{SO}(2n)$ baryon $\epsilon_{2n} \tilde P^{2n-1} \tilde R$. On the other hand the baryon $\mathrm{SO}(2n)$ baryon $\epsilon_{2n} \tilde P^{2n}$ is flipped by $\gamma$.
  On the other hand, a crucial aspect of this deconfinement is that we have also (apparently) broken the non-abelian $\mathrm{SU}(4)$ flavor symmetry in this phase.  The fourth fundamental $Q_4$ in the deconfined phase corresponds to the combination $Q_4 = P R$. Observe that the three constraints on the global symmetries imposed by the whole electric superpotential (\ref{threeconstrW}) are imposed here from the three linear monopole  superpotentials in (\ref{SecondWSUSO2}).

The next step consists of dualizing the $\mathrm{SU}(2n)$ gauge node by treating the other gauge symmetry as flavor. In this way the $\mathrm{SU}(2n)$ gauge theory has $2n+1$ pairs of fundamentals and antifundamentals and linear monopole superpotential. The theory is then confining and the $\mathrm{SU}(2n)$ gauge invariant degrees of freedom of this phase are the meson $\mathcal{M}$
\begin{equation}
\mathcal{M}= \left(
\begin{array}{cc}
M_{\tilde V Q } & M_{\tilde P Q } \\
M_{\tilde V P }  & M_{\tilde P P }
\end{array}
\right),
\end{equation}
the baryons $B_1 = P^{2n-2} Q^2$ and $B_2 = P^{2n-3} Q^3$ and the antibaryons
$\tilde B_1 = \tilde P^{2n}$ and $\tilde B_2 = \tilde P^{2n-1} \tilde V$.
The superpotential for the model obtained after confining the $\mathrm{SU}(2n)$ gauge node becomes
\begin{eqnarray}
&W& = B_1 \tilde B_1 M_{\tilde V Q}+B_1 \tilde B_2 M_{\tilde P Q}+B_2 \tilde B_2 M_{\tilde P P} + B_2 \tilde B_1 M_{\tilde V P} + M_{\tilde P P }^{2n-2}M_{\tilde P Q }^2M_{\tilde V Q }\nonumber \\
&+& 
M_{\tilde V P } M_{\tilde P P }^{2n-3}M_{\tilde P Q }^3+Y_{\mathrm{SO}(2n)}^++Y_{\mathrm{USp}(2n-2)} +  U M_{ \tilde V P} + U R K + \sigma \tilde R^2+\gamma \tilde B_1 , 
\end{eqnarray}
where the charged fields for this phase are depicted explicitly in the third quiver of Figure \ref{fig:decsuuspnm2}.
The superpotential  is simplified by integrating out the massive fields, and it becomes
\begin{eqnarray}
W &=& B_1 \tilde B_2 M_{\tilde P Q}+B_2 \tilde B_2 M_{\tilde P P}  +  M_{\tilde P P }^{2n-2}M_{\tilde P Q }^2M_{\tilde V Q }
\nonumber \\
&+& 
R K M_{\tilde P P }^{2n-3}M_{\tilde P Q }^3+Y_{\mathrm{SO}(2n)}^++Y_{\mathrm{USp}(2n-2)} +   \sigma \tilde R^2.
\end{eqnarray}
The last step consists of confining the $\mathrm{USp}(2n-2)$ gauge node. This gauge theory is indeed confining because 
there are $2n+2$ fundamentals and linear monopole superpotential. The $\mathrm{USp}(2n-2)$ gauge invariant degrees of freedom
are $X= M_{P \tilde P}^2$,  $F^{(1)} = M_{P \tilde P} B_2$, $F^{(2)} = M_{P \tilde P} R$ and 
$J = B_2 R$ and the superpotential for the leftover $\mathrm{SO}(2n)$ gauge group is
\begin{eqnarray}
W &=&
J \mathrm{Pf} X + X^{n-1} F^{(1)} F^{(2)} +B_1 \tilde B_2 M_{\tilde P Q}+\tilde B_2 F^{(1)}   +  X^{n-1} M_{\tilde P Q }^2M_{\tilde V Q }
\nonumber \\
&+& 
K F^{(2)}  X^{n-2}M_{\tilde P Q }^3+   \sigma \tilde R^2+Y_{\mathrm{SO(2n-2)}}^{(bare)} ,
\end{eqnarray}
where the $\mathrm{SO}(2n)$ adjoint $X$ and the five vectors $\tilde R$, $F^{(2)}$ and $M_{\tilde P Q}$ are represented in the last quiver in Figure \ref{fig:decsuuspnm2}.
Integrating out the massive fields the superpotential becomes
\begin{eqnarray}
\label{klastfromdec}
W &=&Y_{\mathrm{SO(2n-2)}}^{(bare)} +J \mathrm{Pf} X+ X^{n-1} B_1 M_{\tilde P Q} F^{(2)}  +  X^{n-1} M_{\tilde P Q }^2M_{\tilde V Q } \nonumber \\
&+& K F^{(2)}  X^{n-2}M_{\tilde P Q }^3 +   \sigma \tilde R^2.
\end{eqnarray}

Observe that the $\mathrm{SU}(4)$ flavor symmetry in this last phase is manifest. Indeed, by redefining $V \equiv \{ M_{\tilde P Q},F^{(2)} \} $ and $H\equiv \{  B_1, M_{\tilde V Q } \}$ the  superpotential (\ref{klastfromdec}) coincides with (\ref{expectedSO2}).

%
%
%
%
%
\subsubsection{$\mathrm{SU}(2n+1)$ with superpotential (\ref{Wdef1SU2np1}) and $\mathrm{SO}(2n)$ dual}
%
%
%
%
%
\begin{figure}[H]
\begin{center}
  \includegraphics[width=12.5cm]{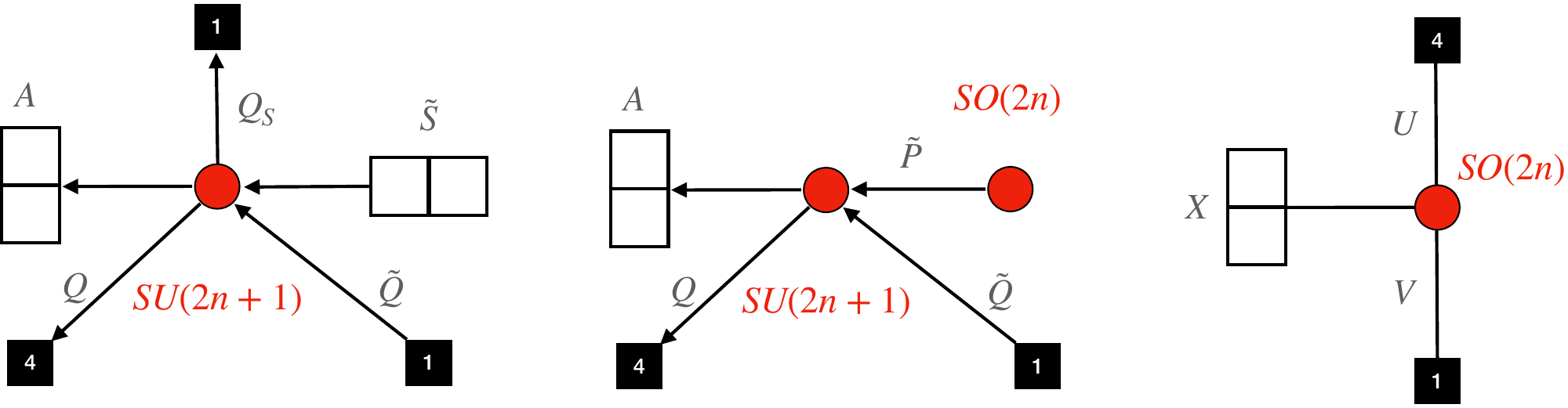}
  \end{center}
 \caption{In this figure we show the steps to prove the duality between $\mathrm{SU}(2n+1)$ with superpotential (\ref{so2np1elesoso}) and $\mathrm{SO}(2n)$ with superpotential (\ref{WdatrovareSOev}) through tensor deconfinement and ordinary dualities.} 
    \label{figDecoSUoddSOeven}
\end{figure}


In this case we consider the  freezing (\ref{freezinganti}) and consider the identity  
(\ref{id2np1fsecondonS1}).
After applying the duplication formula we arrive at the identity 
\begin{eqnarray}
\label{soevfirst}
&&
Z_{\mathrm{SU}(2n+1)}^{(5;1;\cdot;1;\cdot;\cdot;1)}(\vec \mu,\omega-\frac{\tau_{\tilde S}}{2};\nu;\cdot;\tau_A;\cdot;\cdot;\tau_{\tilde S})
=
\!\!\!\!\!
\prod_{1\leq a<b<c\leq 4 }
\!\!\!\!\! \Gamma_h(
(n-1) \tau_{ A} + \mu_a+ \mu_b+ \mu_c)
\nonumber \\
&&
\prod_{a=1}^{4} \Gamma_h(\mu_a+\nu)
\Gamma_h\left(n \tau_{\tilde A} + \nu,\omega-\frac{\tau_{\tilde S}}{2}+\nu \right)
Z_{\mathrm{SO}(2n)}^{(5;1;\cdot)}
\left(\vec \mu+
\frac{\tau_{\tilde S}}{2} ,\nu +\tau_{A}+\frac{\tau_{\tilde S}}{2};\tau_{\tilde S}+\tau_{ A};\cdot
\right),  \nonumber \\
\end{eqnarray}
which is valid provided the relations
\begin{equation}
(2n-1)\tau_A + \left(2n+\frac{1}{2}\right)\tau_{\tilde S} +\nu+\sum_{a=1}^4 \mu_a=3 \omega
 \quad \& \quad
(2n+1) \tau_{\tilde S} =2 \omega
\end{equation}
are satisfied. At physical level we interpret the identity (\ref{soevfirst}) as a duality between $\mathrm{SU}(2n+1)$ and $\mathrm{SO}(2n)$. 
More precisely the two dual models correspond to 
\begin{itemize}
\item On the electric side we have an $\mathrm{SU}(2n+1)$ gauge theory with an antisymmetric $A$, a conjugate symmetric $\tilde S$, four fundamentals $Q$, one fundamental $Q_S$ and one antifundamental $\tilde Q$ with superpotential
\begin{equation}
\label{so2np1elesoso}
W = Y_{\mathrm{SU}(2n-1)}^{(bare)} + \tilde S Q_S^2+\det \tilde S.
\end{equation}
\item On the magnetic side we have an $\mathrm{SO}(2n)$ gauge theory with an antisymmetric (adjoint) $X$, four vectors  $U$ and one vector $V$. In this case there are also four singlets $M = \tilde Q Q$, $B_n =  A^{n} Q$, $B_{n-1} = A^{n-1} Q^3$ and $ \tilde B =\tilde Q Q_S$, where we specified their relation with the electric gauge invariant combinations.
In this case the constraints from the global charges are compatible with a dual superpotential 
\begin{equation}
\label{WdatrovareSOev}
W = 
B_n( M U^2 X^{n-1} + U^3 X^{n-2} V)+
B_{n-1} (M X^n + U X^{n-1} V)+
B_n B_{n-1} \tilde B +
Y_{\mathrm{SO(2n-2)}}^{(bare)},
\end{equation}
where   $Y_{\mathrm{SO(2n-2)}}^{(bare)}$  refers to the symmetry breaking pattern 
$SO(2n) \rightarrow SO(2n-2) \times U(1)$. 
\end{itemize}

In the following we want to find a proof of the duality just proposed using tensor deconfinement. We start our analysis by deconfining the conjugate symmetric tensor  $\tilde S$ as in the second quiver of figure \ref{figDecoSUoddSOeven}. This deconfinement implies that the conjugate symmetric tensor  $\tilde S$  corresponds to the $\mathrm{SO}(2n)$  invariant contraction $\tilde P^2$ in this deconfined picture. 
The superpotential of this quiver is given by 
\begin{equation}
W = Y_{\mathrm{SU}(2n+1)} + Y_{\mathrm{SO}(2n)}^+
\end{equation}

The next step consists of confining the $\mathrm{SU}(2n+1)$ gauge theory. There are two types of fields that survive this confinement, i.e. 
$\mathrm{SO}(2n)$ singlets and $\mathrm{SO}(2n)$  charged fields, either vectors or adjoint(s). 
The singlets are
\begin{equation}
B_{n-1}=A^{n-1} Q^3, \quad
B_n=A^n Q, \quad
\tilde B=\tilde P^{2n} \tilde Q, \quad
M=\tilde Q Q,
\end{equation}
while the charged fields (represented in the third quiver in Figure \ref{figDecoSUoddSOeven}) are
\begin{equation}
V=A \tilde Q \tilde P, \quad
X=A \tilde P^2,\quad
U=\tilde P Q.
\end{equation}
By inspection we see that after confining the $\mathrm{SU}(2n+1)$ gauge node the final superpotentials becomes (\ref{WdatrovareSOev}).

%
%
%
%
%
\subsubsection{$\mathrm{SU}(2n+1)$ with superpotential (\ref{Wdef1SU2np1}) and $\mathrm{SO}(2n+1)$ dual}
%
%
%
%
%
We conclude this survey by considering the  freezing (\ref{freezinganti}) with $\nu_1 \leftrightarrow \nu_2$.
We then consistently freeze the masses in the identity  
(\ref{id2np1firstonS1}) and apply the duplication formula, obtaining
\begin{figure}
\begin{center}
  \includegraphics[width=13.5cm]{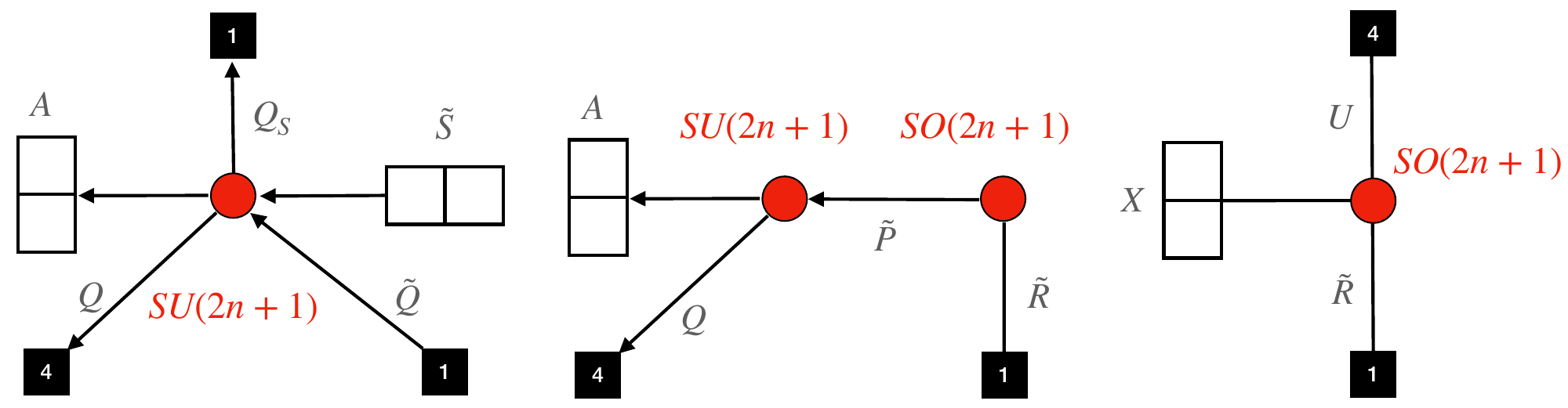}
  \end{center}
 \caption{In this figure we show the steps to prove the duality between $\mathrm{SU}(2n+1)$ with superpotential (\ref{elesoso2np1}) and $\mathrm{SO}(2n)$ with superpotential (\ref{WdatrovareSOodd}) through tensor deconfinement and ordinary dualities.} 
    \label{figDecoSUoddSOodd}
\end{figure}
\begin{eqnarray}
\label{soddsecond}
&&
Z_{\mathrm{SU}(2n+1)}^{(5;1;\cdot;1;\cdot;\cdot;1)}(\vec \mu,\omega-\frac{\tau_{\tilde S}}{2};\nu;\cdot;\tau_A;\cdot;\cdot;\tau_{\tilde S})
=
\Gamma_h((2n+1) \tau_{\tilde S})
\prod_{a=1}^{4}  \Gamma_h(n \tau_{ A} + \mu_a)
\nonumber \\
&&
\!\!\!\!\! 
\prod_{1\leq a<b<c\leq 4 }
\!\!\!\!\! \Gamma_h(
(n-1) \tau_{ A} + \mu_a+ \mu_b+ \mu_c)
Z_{\mathrm{SO}(2n+1)}^{(5;1;\cdot)}
\left(\vec \mu+
\frac{\tau_{\tilde S}}{2} ,\nu -\frac{\tau_{\tilde S}}{2};\tau_{\tilde S}+\tau_{ A};\cdot
\right), \nonumber \\
\end{eqnarray}
which is valid provided the relations
\begin{equation}
(2n-1)\tau_A + \left(2n+\frac{1}{2}\right)\tau_{\tilde S} +\nu+\sum_{a=1}^4 \mu_a=3 \omega
 \quad \& \quad
2n \tau_{\tilde S}+2\nu =2 \omega
\end{equation}
are satisfied. At physical level we interpret the identity (\ref{soddsecond}) as a duality between $\mathrm{SU}(2n+1)$ and $\mathrm{SO}(2n+1)$. 
More precisely the two dual models correspond to 
\begin{itemize}
\item On the electric side we have a $\mathrm{SU}(2n+1)$ gauge theory with an antisymmetric $A$, a conjugate symmetric $\tilde S$, four fundamentals $Q$, one fundamental $Q_S$ and one antifundamental $\tilde Q$ with superpotential
\begin{equation}
\label{elesoso2np1}
W = Y_{\mathrm{SU}(2n-1)}^{(bare)} + \tilde S Q_S^2+\tilde S^{2n} \tilde Q^2.
\end{equation}
\item On the magnetic side we have an $\mathrm{SO}(2n+1)$ gauge theory with an antisymmetric (adjoint) $X$, four vectors  $U$ and one vector $R$. In this case there are also three singlets  $B_n =  A^{n} Q$, $B_{n-1} = A^{n-1} Q^3$ and $ \sigma = \det \tilde S$, where we specified their relation with the electric gauge invariant combinations.
In this case the constraints from the global charges are compatible with a dual superpotential 
\begin{equation}
\label{WdatrovareSOodd}
W = 
Y_{\mathrm{SO(2n-1)}}^{(bare)}+ \sigma \tilde R^2+
B_n  U^3 X^{n-1} +B_{n-1}  U X^{n},
\end{equation}
where $Y_{\mathrm{SO(2n-1)}}^{(bare)}$  refers to the symmetry breaking pattern 
$SO(2n+1) \rightarrow SO(2n-1) \times U(1)$. 
\end{itemize}
In the following we want to find a proof of the duality just proposed using tensor deconfinement. We start our analysis by deconfining the conjugate symmetric tensor  $\tilde S$ as in the second quiver of figure \ref{figDecoSUoddSOodd}. This deconfinement implies that the conjugate symmetric tensor  $\tilde S$  corresponds to the $\mathrm{SO}(2n+1)$  invariant contraction $\tilde P^2$ in this deconfined picture. 
The superpotential of this quiver is given by 
\begin{equation}
W = Y_{\mathrm{SO}(2n)}^+ + Y_{\mathrm{SU}(2n+1)} + \sigma \tilde R^2+\gamma \tilde P^{2n+1}.
\end{equation}
Observe that reconfining the conjugate symmetric the F-terms impose the dictionary $\sigma = \det \tilde S$.

The next step consists of confining the $\mathrm{SU}(2n+1)$ gauge theory. There are two types of fields that survive this confinement, i.e. 
$\mathrm{SO}(2n+1)$ singlets and $\mathrm{SO}(2n+1)$  charged fields, either vectors or adjoint(s). 
The singlets are
\begin{equation}
B_{n-1}=A^{n-1} Q^3 \quad \text{and} \quad
B_n=A^n Q,
\end{equation}
while the charged fields, represented in the third quiver in Figure \ref{figDecoSUoddSOodd}, are
\begin{equation}
X=A \tilde P^2 \quad \text{and} \quad
U=\tilde P Q.
\end{equation}
By inspection we see that after confining the $\mathrm{SU}(2n+1)$ gauge node the final superpotentials becomes 
\begin{eqnarray}
W = Y_{\mathrm{SO(2n-1)}}^{(bare)}+ \sigma \tilde R^2+\gamma \tilde B+
B_n  U^3 X^{n-1} +B_{n-1}  U X^{n} +B_n B_{n-1} \tilde B,
\end{eqnarray}
that coincides with (\ref{WdatrovareSOodd}) after integrating out the massive fields.

\subsection{$\mathrm{SU}(N)$ with a symmetric flavor and the $\mathrm{SO}(N)$ dual with a symmetric}

We conclude our survey by considering $\mathrm{SU}(N)$ with a symmetric and two fundamental flavors.

The model is obtained by applying the freezing and duplication formula to $\mathrm{SU}(2n)$ with $W=\tilde A^{n-2} \tilde Q^4$ and to $\mathrm{SU}(2n+1)$ with $W=\tilde A^{n-1} \tilde Q_2  \tilde Q_3  \tilde Q_4$.
In the second case we  freeze the masses of the fields $\tilde Q_{1,2,3}$ as
$\{\nu_1,\nu_2,\nu_3\} = \frac{ \tau_{\tilde S}}{2}+\frac{1}{2}
\{\omega_1,\omega_2,0\}$, leaving $\nu_4\equiv \nu$ free. 
We further freeze the masses of the fundamentals as in (\ref{freddo}).

We then apply the duplication formula to the identities \eqref{onS1defo1} and
\eqref{id2np1fsecondonS1}, and we obtain a unified formula, corresponding to 
\begin{eqnarray}
&&
Z_{\mathrm{SU}(N)}^{(2;2;\cdot;\cdot;1;1)}
\left(
\mu,\omega-\frac{\tau_{\tilde S}}{2};
\nu,\omega-\frac{\tau_{ S}}{2};
\cdot;\cdot;
\tau_{ S};\tau_{ \tilde S}
\right)=\Gamma_h(N \tau_S,N \tau_{\tilde S})
\nonumber
\\
&&
\Gamma_h((N-1) \tau_S+2\mu)
Z_{\mathrm{SO}(N)}^{(3;\cdot;1)}
\left(
\mu+\frac{\tau_{\tilde S}}{2},
\nu-\frac{\tau_{ \tilde{S}}}{2},
\omega-\frac{\tau_{ S}+\tau_{ \tilde S}}{2};
\cdot;
\tau_{ S}+\tau_{ \tilde S}
\right),
\end{eqnarray}
with the balancing conditions
\begin{equation}
\label{BCSOSY}
\left(N-\frac{1}{2}\right) (\tau_{ S}+\tau_{ \tilde S})+\mu+\nu=2 \omega
,\quad
N \tau_{\tilde S} + 2 \nu = 2 \omega.
\end{equation}

We can interpret this identity as a duality between
\begin{itemize}
\item An $\mathrm{SU}(N)$ theory with a symmetric tensor $S$ and a conjugate symmetric tensors $\tilde S$, two fundamentals denoted as $Q$ and $Q_{\tilde S}$ and two 
antifundamentals denoted as $\tilde Q$ and  $\tilde Q_S$.
This model has superpotential 
\begin{equation}
W = Y_{\mathrm{SU}(N-2)}^{(bare)} + S \tilde Q_S^2 + \tilde S Q_{\tilde S}^2 + \tilde S^{N-1}\tilde Q^2.
\end{equation}
\item An $\mathrm{SO}(N)$ dual theory with a reducible symmetric $X$, three fundamentals that we denote as $\phi_X$, $U$ and $V$. In this dual phase there are also three singlets
$H,\tilde H$ and $J$, that correspond to the electric gauge invariant combinations $\det S$, $\det \tilde S$ and $S^{N-1} Q^2$ respectively.
The superpotential of this dual theory, compatible with the relations above, is
\begin{equation}
\label{Åttesososym}
W =Y_{\mathrm{SO(N-2)}}^{(bare)} + 
H U^2 X^{N-1}+
\tilde H V^2
+X \phi_X^2+
J X^{N-1},
\end{equation}
\end{itemize}
where $Y_{\mathrm{SO(N-2)}}^{(bare)}$  refers to the symmetry breaking pattern 
$SO(N) \rightarrow SO(N-2) \times U(1)$. In this case, i.e. in presence of a symmetric $SO(N)$ tensor in the low energy spectrum, this monopole is gauge invariant and the presence of the linear monopole superpotential term in \eqref{Åttesososym} forces the first constraint \eqref{BCSOSY} in the dual theory.

\begin{figure}
\begin{center}
  \includegraphics[width=12.5cm]{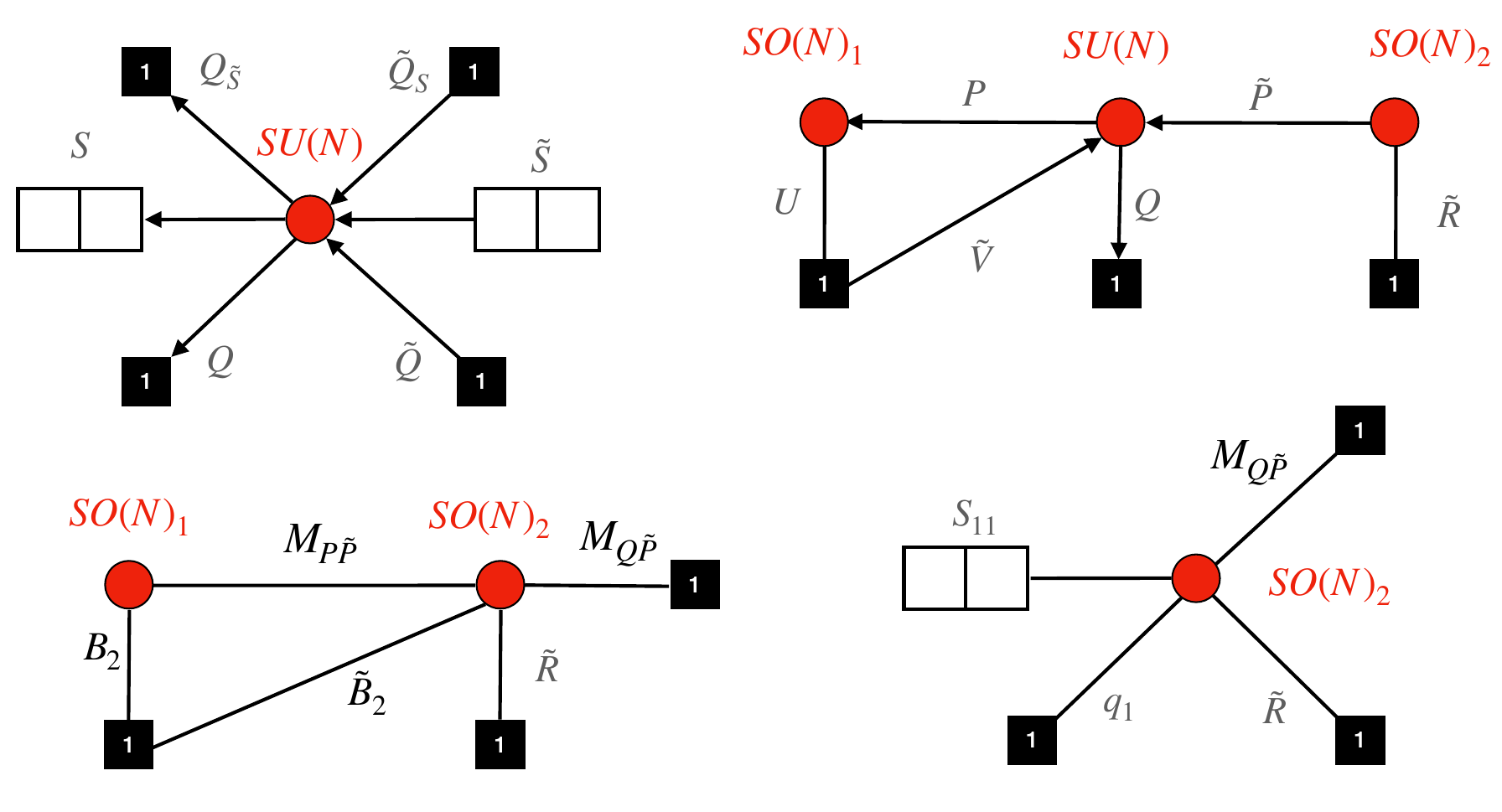}
  \end{center}
 \caption{In this figure we show the steps to prove the duality between $\mathrm{SU}(N)$ with superpotential (\ref{elesSUN}) and $\mathrm{SO}(N)$ with superpotential (\ref{spotsymmso}) through tensor deconfinement and ordinary dualities.} 
    \label{figfigSStilde}
\end{figure}

In the following, we want to find a proof of the duality just proposed using tensor deconfinement. We start our analysis by deconfining the symmetric and the conjugate symmetric tensor, $ S$ $\tilde S$ respectively, as in the second quiver of Figure \ref{figfigSStilde}. This deconfinement implies that the  symmetric tensor  $ S$  corresponds to the $\mathrm{SO}(2n+1)$  invariant contraction $ P^2$ in this deconfined picture and that the
conjugate symmetric tensor  $\tilde S$  corresponds to $\tilde P^2$.

The superpotential of this quiver is given by 
\begin{equation}
\label{elesSUN}
W = Y_{\mathrm{SO}(N)_1}^+ +   Y_{\mathrm{SO}(N)_2}^+ + Y_{\mathrm{SU}(N)} + \sigma \tilde R^2+ \gamma  P^{N}+\tilde \gamma \tilde P^{N}+\alpha U^2 + U \tilde V P.
\end{equation}

Then we observe that $\mathrm{SU}(2n+1)$ is confining.  The confined degrees of freedom correspond to the mesonic components $M_{P \tilde P}, M_{P \tilde V}, M_{Q \tilde P}$ and $M_{Q \tilde V}$ while the baryonic components are $B_1 = P^N$,
$B_2 = P^{N-1} Q$, $\tilde B_1 = \tilde P^N$ and $\tilde B_2 = \tilde P^{N-1} \tilde V$.
The model after this confining duality corresponds to the third quiver in Figure 
\ref{figfigSStilde} and the superpotential, after integrating out the massive fields, is
\begin{equation}
\label{spotsymmso}
W = Y_{\mathrm{SO}(N)_1}^+ +   Y_{\mathrm{SO}(N)_2}^+ + \sigma \tilde R^2+M_{P \tilde P} B_2 \tilde B_2 + M_{P \tilde P}^{N} M_{Q \tilde V} + \alpha (M_{P \tilde P}^{N-1} M_{Q \tilde P})^2.
\end{equation}
The last step consists of confining the $\mathrm{SO}(N)_1$ node, in terms of the gauge invariant combinations $S_{11} = M_{P \tilde P}^2$, 
$S_{12} = M_{P \tilde P} B_2$, $S_{22} = B_2^2$, $q_1 =  M_{P \tilde P}^{N-1} B_2$ and  $q_2 =  M_{P \tilde P}^{N} $.

The model is represented in the last quiver of Figure \ref{figfigSStilde} and the superpotential coincides with \eqref{Åttesososym} after the identifications
\begin{equation}
\alpha  \leftrightarrow \det S,\quad
\sigma \leftrightarrow \det \tilde S,\quad
M_{Q \tilde P}  \leftrightarrow U,\quad
S_{11} \leftrightarrow X,\quad
S_{22} \leftrightarrow J,
\quad
\tilde R \leftrightarrow V,\quad
q_1 \leftrightarrow \phi_X.
\end{equation}

%
%
%
%
%
\section{Conclusions}
\label{sec:conc}
%
%
%

In this paper we have studied 4d and 3d IR dualities involving a $\mathrm{SU}(N)$ gauge theory with tensorial matter and a non-trivial superpotential.
We started our analysis from $\mathrm{SU}(N)$ with an antisymmetric and four fundamental flavors in 4d.
This theory for $N=2n$ is conjectured to have various self-dual phases, and we provided a proof of this fact for $n=2$, in terms of tensor deconfinement.
Generalizing the approach of such proof to generic $\mathrm{SU}(N)$ we found that there is a self-dual description between the first and the last quiver of Figure \ref{figdec1su}, where the dual phase is equipped with a non-trivial superpotential given in formula \eqref{thirdspot} for $N=2n$ and in formula \eqref{thirdspot2np1} for $N=2n+1$. 
This self-duality is crucial for our analysis, because, upon deforming the electric superpotential through a dangerously irrelevant baryonic deformation, we have shown that the dual picture gets Higgsed to $\mathrm{USp}(2m)$ with either $m=n$ or $m=n-1$ or $m=n-2$ depending on the electric deformation, with an antisymmetric and eight  fundamentals, interacting with a series of flippers.
In this way we have constructed new SU/USp dualities\footnote{Avatars of such dualities were previously discussed in \cite{Razamat:2017wsk}.}. We corroborated our results by studying the Higgsing at the level of the superconformal index, finding the exact identities that represent the dualities proposed from the field theoretical analysis. Furthermore, we provided an alternative proof of such dualities, by using a different tensor deconfinement, by trading the antisymmetric tensor involved in the baryonic superpotential with a symplectic gauge group.  
We also studied the existence of an interacting fixed point for the dualities under investigations, observing that by increasing $N$ an increasing amount of gauge invariant operators in the chiral operators hits the bound of unitarity, and it requires an intricate structure of flippers that need to be added on the electric sides of the dualities.
Then, we have reduced the 4d dualities to 3d, by using the ARSW \cite{Aharony:2013dha} prescription, first considering the effective dualities on $S^1$, where the electric and the magnetic superpotential acquire a further contribution associated to the addition of a KK monopole, and then flowing to ordinary dualities, where the effects of the KK monopole are lifted by opportune real mass flows. 
Remarkably, we obtained also the 3d confining gauge theory associated to $\mathrm{SU}(2n)$ with four fundamentals and an antisymmetric flavor found in \cite{Nii:2019ebv}. In this way we provided the 4d “parent" of this confining gauge theory (see \cite{Amariti:2025jvi} for a similar observation in the 4d/2d reduction of dualities).
Lastly we applied on the effective dualities on $S^1$ the duplication formula for the hyperbolic Gamma functions, by freezing the mass parameters in the squashed three-sphere partition function opportunely.
In this way we read new identities that are interpreted as SU/USp and SU/SO dualities, where in the electric side we have a symmetric and a conjugated antisymmetric and in the dual phases we have an adjoint.
In each case we showed how to obtain such dualities by tensor deconfinement, providing a physical proof of the new dualities in terms of other known and “more ordinary" dualities.

Various generalizations of our analysis are possible. 
First, it should be interesting to connect the $\mathrm{E}_7$ and the $\mathrm{D}_6$ enhancements for the $\mathrm{USp}(2n)$ and the $\mathrm{SU}(2n)$ studied here. This may also give rise to a geometric interpretation of the dualities discussed here, for which a brane description is absent so far.
Furthermore, it should be interesting to increase the number of flavors on the electric side and in addition to consider also other possible baryonic deformations.
Motivated by the relation with the 3d (and 2d) dualities, one could also consider $\mathrm{SU}(N)$ theories with two antisymmetric tensors (i.e. without conjugation) in addition to fundamentals and antifundamentals (consistently with the requirement from the anomaly freedom). New dualities in such case may emerge in presence of 
a non-trivial electric superpotential.
Another issue that we did not discuss, but that certainly deserves a further analysis, regards the existence of a conformal window for the 3d dualities found here. In such cases one should study possible violations of bounds of unitarity by F-maximization and mimic the 4d analysis based on a-maximization.
A last direction that should be interesting to explore regards the matching of other indices for the 3d dualities studied here.

\section*{Acknowledgments}
 The work of A.A., S.R. and A.Z. has been supported in part by the Italian Ministero dell'Istruzione, Università e Ricerca (MIUR), in part by the Istituto Nazionale di Fisica Nucleare (INFN) through the “Gauge Theories, Strings, Supergravity” (GSS) research project. The work of F.M. is funded by the Deutsche Forschungsgemeinschaft (DFG, German Research Foundation) – SFB 1624 – "Higher structures, moduli spaces and integrability" –506632645. The work of S.R. has been partially supported by the MUR-PRIN grant No. 2022NY2MXY.
 
\bibliographystyle{JHEP}
\bibliography{refE7D6.bib}

\end{document}